\documentclass[prx,notitlepage,nofootinbib,twocolumn,longbibliography,superscriptaddress]{revtex4-1}

\usepackage{amsmath}
\usepackage{amsthm, amssymb}

\usepackage{graphicx}

\usepackage{bbold}

\usepackage[usenames,dvipsnames]{color}
\usepackage[colorlinks=true,citecolor=blue,linkcolor=magenta]{hyperref}

\usepackage{dsfont}
\usepackage{changepage}
\usepackage{array}

\newcommand{\bit}{\begin{itemize}}
\newcommand{\eit}{\end{itemize}}

\newcommand{\f}{\frac}
\renewcommand{\>}{\right\rangle}
\newcommand{\<}{\left\langle}
\newcommand{\ba}{\begin{align}}
\newcommand{\ea}{\end{align}}
\newcommand{\be}{\begin{equation}}
\newcommand{\ee}{\end{equation}}
\newcommand{\bi}{\begin{itemize}}
\newcommand{\ei}{\end{itemize}}
\newcommand{\lf}{\left(}
\newcommand{\ri}{\right)}
\newcommand{\dd}{\mathrm{d}}

\newcommand{\Tr}{\operatorname{Tr}}

\def\a{\alpha}
\def\b{\beta}
\def\s{\sigma}

\graphicspath{{entanglementfigures/}}

\begin{document}

\newcommand{\bra}[1]{\< #1 \right|}
\newcommand{\ket}[1]{\left| #1 \>}

\title{Quantum Entanglement Growth Under Random Unitary Dynamics}

\author{Adam Nahum}
\author{Jonathan Ruhman}
\affiliation{Department of Physics, Massachusetts Institute of Technology, Cambridge, Massachusetts 02139, USA}
\author{Sagar Vijay}
\affiliation{Department of Physics, Massachusetts Institute of Technology, Cambridge, Massachusetts 02139, USA}
\affiliation{Kavli Institute for Theoretical Physics, University of California Santa Barbara, CA 93106, USA}
\author{Jeongwan Haah}
\affiliation{Department of Physics, Massachusetts Institute of Technology, Cambridge, Massachusetts 02139, USA}
\date{\today}

\begin{abstract}
\noindent
Characterizing how entanglement grows with time in a many-body system, for example after a quantum quench, is a key problem in non-equilibrium quantum physics. We study this problem for the case of random unitary dynamics, representing either Hamiltonian evolution with time--dependent noise or evolution by a random quantum circuit. Our results reveal a universal structure behind noisy entanglement growth, and also provide simple new heuristics for the `entanglement tsunami' in Hamiltonian systems without noise. In 1D, we show that  noise causes the entanglement entropy across a cut to grow according to the celebrated Kardar--Parisi--Zhang (KPZ) equation. The mean entanglement grows linearly in time, while fluctuations grow like $(\text{time})^{1/3}$ and are spatially correlated over a distance $\propto (\text{time})^{2/3}$. We derive KPZ universal behaviour in three complementary ways, by mapping  random entanglement growth to: (i) a stochastic model of a growing surface; (ii) a `minimal cut' picture, reminiscent of the Ryu--Takayanagi formula in holography; and (iii) a hydrodynamic problem involving the dynamical spreading of operators. We demonstrate KPZ universality in 1D numerically using simulations of random unitary circuits. Importantly, the leading order time dependence of the entropy is deterministic even in the presence of noise, allowing us to propose a simple `minimal cut' picture for  the entanglement growth of {generic} Hamiltonians, even \textit{without} noise, in arbitrary dimensionality. We clarify the meaning of the `velocity' of entanglement growth in the 1D `entanglement tsunami'. We show that in higher dimensions, noisy entanglement evolution maps to the well-studied problem of pinning of a membrane or domain wall by disorder. 
\end{abstract}

\maketitle

\section{Introduction}

The language of quantum entanglement ties together condensed matter physics, quantum information and high energy theory. The von Neumann entanglement entropy is known to encode universal properties of quantum ground states and has led to new perpectives on the AdS-CFT correspondence. But the dynamics of the entanglement  are far less understood. The entanglement entropy is a highly nonlocal quantity, with very different dynamics to energy or charge or other local densities. Traditional many-body tools therefore do not provide much intuition about how entanglement spreads with time, for example after a quantum quench (a sudden change to the Hamiltonian).  We need to develop simple heuristic pictures, and simple long-wavelength descriptions, for entanglement dynamics.

If a many-body system is initialized in a state with low entanglement, the dynamics will typically generate entanglement between increasingly distant regions as time goes on. This irreversible growth of entanglement --- quantified by the growth of the von Neumman entropy --- is important for several reasons. It is an essential part of thermalization, and as a result has been addressed in diverse contexts ranging from conformal field theory \cite{CardyCalabrese2005EntanglementEvolution, CardyCalabrese2009EntanglementFieldTheory,Asplund} and holography \cite{HubenyRangamaniTakayanagi2007,Abajo-Arrastia2010,Hartman2013,LiuSuh2014,LiuSuh20142,CasiniLiuMezei2015} to integrable \cite{FagottiCalabrese2008, PeschelEisler2009, Buyskikh2016}, nonintegrable \cite{LauchliKollath2008, KimHuse2013, ho2015entanglement}, and strongly disordered spin chains \cite{BardarsonPollmanMoore2012,SerbynPapicAbanin2013,HuseNandkishoreOganesyan2014,ChandranLaumann2015,LuitzLaflorencieAlet2016}.  Entanglement growth is also of practical importance as the crucial obstacle to simulating quantum dynamics numerically, for example using matrix product states or the Density Matrix Renormalization Group (DMRG) \cite{VerstraeteMurgCircac2008}.  The entanglement entropy, and even its time dependence, is also beginning to be  experimentally measurable in cold atom systems \cite{GreinerMeasuringEntanglement2015,Daley2012}.  In a very different context, black holes have motivated studies of how fast quantum systems can scramble information by dynamically generating entanglement \cite{HaydenPreskill2007, SekinoSusskind2008,Dankert2009, Roberts2015}. Simple quantum circuits --- quantum evolutions in discrete time ---  serve as useful toy models for entanglement growth and scrambling \cite{BrandaoHarrowHorodecki}.

This paper gives a new perspective on entanglement growth by studying quantum dynamics that are spatially local, and unitary, but random both in time and space. Physically, this class of problems includes closed, many-body systems whose Hamiltonian $H(t)$ contains noise, and also quantum circuits in which qubits are acted on by randomly chosen unitary gates. The latter can be viewed either as as discrete approximations of continuous time dynamics or as  toy models for quantum computations.

The motivation for studying noisy systems is twofold. First,  random dynamics  are a `least structured' model for quantum dynamics. This theoretical laboratory leads us to remarkably simple heuristics for entanglement growth and the so-called `entanglement tsunami' \cite{LiuSuh2014}. As discussed below, we conjecture that many of these heuristics remain valid for Hamiltonian dynamics without noise.  Second, noisy entanglement growth exhibits a remarkable long-wavelength description in its own right, with an  emergent universal structure. This structure has  surprising connections to paradigmatic models in  \textit{classical} non-equilibrium  statistical mechanics. Fluctuations and spatial correlations in the entanglement entropy are characterized by universal scaling exponents, expected to be independent of the details of the microscopic model.

For systems in one spatial dimension (1D), we argue that the critical exponents for noisy entanglement growth are those of the Kardar---Parisi---Zhang (KPZ) equation, originally introduced to describe the stochastic growth of a surface with time $t$ \cite{kpz}. In the simplest setting, we find that the `height' of this surface at a point $x$ in space is simply the von Neumann entanglement entropy $S(x,t)$ for a bipartition which splits the system in two at $x$. The average entanglement grows linearly in time, while fluctuations are characterized by non-trivial exponents. We support this identification with analytical arguments and numerical results for discrete time quantum evolution (unitary circuits).

\begin{figure}[t]
 \begin{center}
  \includegraphics[width=0.98\linewidth]{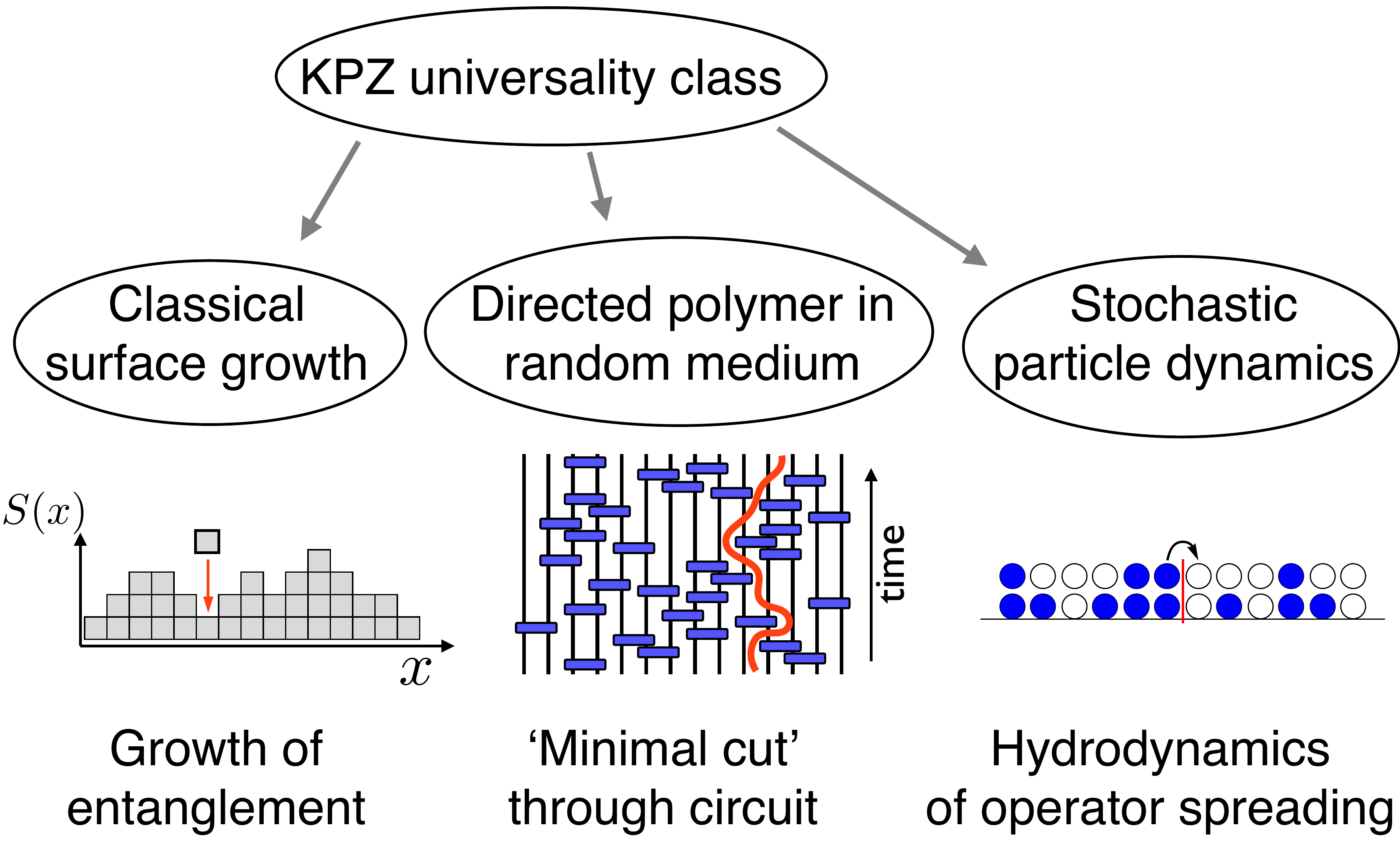}
 \end{center}
\caption{The KPZ `triumvirate' is made up of three very different problems in classical statistical mechanics which all map to the KPZ universality class. As we will discuss, each of them can be usefully related to entanglement in 1+1D.}
 \label{triumvirate}
\end{figure}

The KPZ universality class also includes two other classical problems besides surface growth \cite{kpz, HuseHenleyFisher1985respond}, as summarized in Fig.~\ref{triumvirate}. We show that each one provides a useful  perspective on entanglement dynamics. They are the statistical mechanics of a directed polymer in a disordered potential landscape \cite{HuseHenley1985}, and 1D hydrodynamics with noise (the noisy Burgers equation \cite{ForsterStephenNelson}). Remarkably,  entanglement growth  can be related to all three of the classical problems in in Fig.~\ref{triumvirate}, which are sometimes referred to as the `KPZ triumvirate' \cite{Halpin-Healy1998}. 

In the quantum setting, the directed polymer is related to the `minimal cut', a curve in space-time  which bisects the unitary circuit representing the time evolution. This is reminiscent of the Ryu-Takayanagi prescription for calculating  the entanglement entropy of conformal field theories in the AdS-CFT correspondence, which makes use of a minimal surface in the bulk space \cite{RyuTakayanagi2006}, and analogous results for certain tensor network states \cite{Swingle2012, Pastawski2015, HaydenEtAl2016}. Here however the cut lives in spacetime rather than in space, and its shape is random rather than deterministic. (For a different use of the idea of a minimal cut in spacetime, see Ref.~\cite{CasiniLiuMezei2015}.) This picture is more general than the surface growth picture, as  it allows one to consider the entropy for any bipartition of the system. It also allows us to generalize from 1D to higher dimensions. In $d+1$ spacetime dimensions  the minimal cut becomes a $d$-dimensional membrane pinned by disorder. This picture allows us to pin down approximate critical exponents for noisy entanglement growth in any number of dimensions.   

This picture also leads to  a conjecture for entanglement growth  in systems \textit{without} noise, both in 1D and higher dimensions, as we discuss below. According to this conjecture, the calculation of the entanglement in higher dimensions reduces to a \textit{deterministic} elastic problem for the `minimal membrane' in spacetime. 

The third member of the triumvirate in Fig.~\ref{triumvirate} is a noisy hydrodynamic equation describing the diffusion of interacting (classical) particles in 1D. We show that this can be related to the spreading of quantum operators under the unitary evolution, giving a detailed treatment of the special case of stabilizer circuits.  Note that while the minimal cut picture generalizes to higher dimensions, the KPZ and hydrodynamic pictures are special to 1D.

We propose that noisy dynamics are a useful toy model for quantum quenches in generic (non-integrable, non-conformally--invariant) systems, even without noise.  The logic of our approach is  to pin down the universal behaviour of noisy systems (Secs.~\ref{surface growth section}---\ref{numerical results section}), to establish simple heuristics capturing this behaviour (Secs.~\ref{directed polymer section},~\ref{hydrodynamics section}), and then to draw conclusions that are likely to be true even without noise (Secs.~\ref{entanglement velocity section},~\ref{higher dimensions section}).  While the detailed physics of entanglement fluctuations certainly relies on noise, the coarser features of the dynamics --- i.e. the leading order time dependence of the entanglement entropy and mutual information --- is in fact deterministic. We conjecture that this leading order behaviour (as captured by the directed polymer and hydrodynamic pictures) carries over to  Hamiltonian dynamics without noise. On the basis of this we address (Sec.~\ref{entanglement velocity section}) some features of entanglement growth that have previously been unclear. We argue that in generic 1D  systems the entanglement growth rate {can} be interpreted as a well-defined \textit{speed} $v_E$, but that  this speed is \textit{smaller} than another characteristic speed, which is the speed  $\tilde v$ at which quantum operators spread out under the dynamics. This difference is related to the failure of pictures for entanglement growth in terms of independently spreading operators. We discuss the meaning of $v_E$. In Sec.~\ref{higher dimensions section} we discuss the geometry-dependence of the dynamical entanglement in higher-dimensional systems.

\tableofcontents

\section{Surface growth in 1D}
\label{surface growth section}

We begin by studying entanglement growth under random unitary dynamics in one dimension.  After summarizing the KPZ universal behavior, we derive this behaviour analytically in a solvable model, using a mapping to a classical surface growth problem. In the following sections we provide  more general perspectives on this universal behaviour by relating the `minimal cut' bound on the entanglement to the classical problem of a directed polymer in a random environment, and by relating the spreading of quantum operators to a 1D hydrodyamics problem.

\begin{figure}[t]
 \begin{center}
  \includegraphics[width=0.85\linewidth]{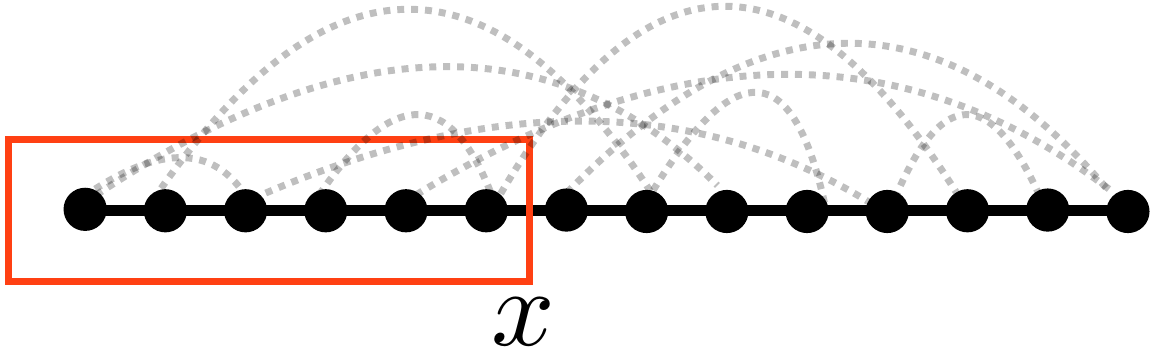}
 \end{center}
\caption{Spin chain with open boundary conditions. $S(x)$ denotes the entanglement entropy (von Neumann or Renyi depending on context) between the part of the chain to the left of bond $x$, indicated by the box, and the part to the right.}
 \label{spin chain fig}
\end{figure}

Consider a chain of quantum spins with local Hilbert space dimension $q$ (for example spin--1/2s with $q=2$). We take open boundary conditions, and label the $\textit{bonds}$ of the lattice by $x=1,\ldots L$. We consider only unitary dynamics, so the full density matrix $\rho = \ket{\Psi}\bra{\Psi}$ represents a pure state. For now we consider the entanglement across a single cut at position $x$; we will generalize to other geometries later in the paper. The reduced density matrix $\rho_x$ is defined by splitting the chain into two halves at $x$ and tracing out the left-hand side (Fig.~\ref{spin chain fig}).  The $n$th Renyi entropy for a cut at $x$ is defined as
\be
S_n(x) = \f{1}{1-n} \log \lf  \Tr \rho_x^n \ri.
\ee
Logarithms are taken base $q$. In the limit $n\rightarrow 1$ the Renyi entropy becomes the von Neumann entropy,
\be
S_\text{vN}(x) = - \Tr \rho_x \ln \rho_x.
\ee
A basic constraint on the von Neumann entropy is that neighboring bonds can differ by at most one\footnote{This follows from subadditivity of the von Neumann entropy.}:
\be\label{von neumann constraint}
\big| S_\text{vN}(x+1) - S_\text{vN}(x) \big| \leq 1.
\ee

In this section we focus on the growth of the bipartite entropies $S(x,t)$ with time, starting from a state with low entanglement.
(Here $S(x,t)$, without a subscript, can denote any of the Renyi entropies with $n>0$.) For simplicity we take the initial state to be a product state, but we expect the same long-time behaviour  for any initial state with area-law entanglement.\footnote{The setup with area-law entanglement in the initial state is analogous to a quantum quench which starts in the ground state of a non-critical Hamiltonian. We briefly consider initial states with non-area-law entanglement in Sec.~\ref{outlook section}.} We will argue that for  noisy unitary dynamics, the universal properties of $S(x,t)$ are those of the Kardar--Parisi--Zhang equation:
\be
\label{kpz equation}
\f{\partial S}{\partial t} = \nu\, \partial_x^2 S - \f{\lambda}{2} (\partial_x S)^2 + \eta(x,t) + c.
\ee
This equation was introduced to describe the stochastic growth of a 1D surface or interface with height profile $S(x)$  \cite{kpz}. It captures an important universality class which has  found a wealth of applications in classical nonequilibrium physics, and its scaling properties have been verified in high-precision experiments \cite{Takeuchi2012, Takeuchi2011}. The constant $c$ in Eq.~\ref{kpz equation} contributes to the positive average growth rate, while $\eta(x,t)$ is noise which is uncorrelated in space and time.  The $\nu$ term describes diffusive smoothing of sharp features. The nonlinear term, with coefficient $\lambda$, describes how the average growth rate depends on the slope; the negative sign is natural here, as discussed in Sec.~\ref{properties of solvable model section} (and implies that $B$ in Eq.~\ref{mean growth} below is positive).

KPZ scaling is characterized by an exponent $\beta$ governing the size of  fluctuations in the  interface height, an exponent $\alpha$ governing spatial correlations, and a dynamical exponent $z$ determining the rate of growth of the correlation length  ($z=\alpha/\beta$ by a scaling relation). These are known exactly \cite{kpz}:
\ba
\beta & = 1/3, &
\alpha & = 1/2, &
z & = 3 /2.
\end{align}
In our context the height of the surface is the bipartite entanglement $S(x,t)$. This is a random quantity which depends on the  realization of the noise in the quantum dynamics. The mean height/entanglement grows linearly in time, with a universal sub-leading correction:
\be\label{mean growth}
h(x,t) \equiv  \<S(x,t) \> = v_E t + B t^\beta.
\ee
Angle brackets denote an average over noise. The fluctuations grow as
\be \label{width growth}
w(x,t) \equiv \langle\langle S(x,t)^2 \rangle\rangle^{1/2} = C t^\beta.
\ee
We will refer to $w$ as the `width' of the surface. The ratio $C/B$ is universal (the constants $v_E$ and $B$ are not). The spatial correlation length grows with time as
\be \label{correlation length}
\xi(t) = t^{1/z},
\ee
and the equal time correlation function has the scaling form
\ba\label{spatial correlator}
G(r) \equiv  \big\langle \left[S(x,t)  - S(x+r,t)\right]^2 \big\rangle^{1/2}
&=  r^{\alpha} g\lf r/ \xi(t)  \ri.
\end{align}
On length scales $1\ll r\ll \xi(t)$, the surface profile $S(x)$ resembles the trace of a 1D random walk: this is consistent with the exponent $\alpha = 1/2$.

At short times the entanglement growth is affected by initial conditions, while on  very long timescales, of the order of the system size, the entanglement saturates. The formulae (\ref{mean growth})---(\ref{spatial correlator}) apply  \textit{prior} to this saturation. In a finite  system the asymptotic $\<S(x)\>$ profile is that of a pyramid, with a maximum at height $x=L/2$, whose height is $L/2$, minus an $O(1)$ correction \cite{Page1993, NadalMajumdarVergassola2011}. This profile is reached at a time
\be
t_\text{saturation} \simeq \f{L}{2 v_E},
\ee
with bonds closer to the boundary saturating sooner (Secs.~\ref{directed polymer section},~\ref{entanglement velocity section}). Saturation is also captured in the surface growth description, once we note that there are Dirichlet boundary conditions on the entropy: $S(0,t) = S(L+1,t) = 0$.

Note that the scaling described in Eqs.~\ref{mean growth},~\ref{correlation length} implies the existence of  two distinct diverging lengthscales during entanglement growth. The fact that $\<S(x,t)\>$ is of order $t$ implies that spins are entangled over distances of order $t$. In fact we will show in Sec.~\ref{entanglement velocity section} that $v_E t$ is a sharply defined  lengthscale. But prior to saturation, the relevant lengthscale for spatial variations in $S(x,t)$ is parametrically smaller than $v_E t$, namely $\xi(t) \sim t^{2/3}$.

Before  deriving KPZ  for entanglement, let us briefly consider the status of this equation. At first sight we might try to justify this description of $S(x,t)$ simply on grounds of symmetry and coarse graining. \textit{If} we were describing classical surface growth, we would  appeal to translational symmetry in the growth direction ($S\rightarrow S+ \text{const.}$) in order to restrict the allowed terms, and would note that the right-hand-side includes the lowest--order terms in $\partial_x$ and  $\partial_x S$. But for entanglement we cannot rely on this simple reasoning. First, the transformation $S\rightarrow S+\text{const.}$ is {not} a symmetry  (or even a well-defined transformation) of the quantum system. More importantly, it is not clear \textit{a priori}  that we can write a stochastic differential equation for $S(x,t)$ alone, since the full quantum state contains much more information than $S(x,t)$. Despite these differences from simple surface growth, we will show that the above equation does capture the universal aspects of the entanglement dynamics.

In the next section we exhibit a solvable quantum model which maps  to a classical surface growth problem that is manifestly in the KPZ universality class.   Then in the two following sections we  give heuristic arguments for more general systems by making connections with the other members of the KPZ triumvirate. Together with the results for the solvable model, these arguments suggest that KPZ exponents should be generic for entanglement growth in any quantum system whose dynamics involves time-dependent randomness.  In Sec.~\ref{numerical results section} we perform numerical checks on KPZ universality for quantum dynamics in discrete time.

\subsection{Solvable 1D model}

\label{haar protocol}

\begin{figure}[t]
 \begin{center}
  \includegraphics[width=0.8\linewidth]{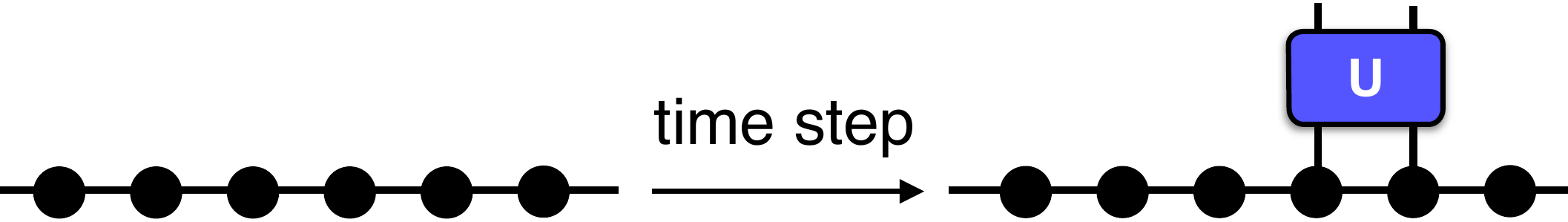}
 \end{center}
\caption{Dynamical update in the solvable model: application of a random unitary $U$ to a randomly chosen pair of adjacent spins.}
 \label{updatefig}
\end{figure}

We now focus on a  specific quantum circuit model for the dynamics of a spin chain with strong noise. We take random unitaries to act on pairs of adjacent spins (i.e. on bonds) at random locations and at random times, as illustrated in Fig.~\ref{updatefig}. For simplicity we discretize time and apply one unitary per time step. (Dynamics in continuous time, with unitaries applied to the links at a fixed rate in a Poissonian fashion, are equivalent.) We choose the initial state to be a product state, with $S_n(x) = 0$ for all $n$ and $x$. We choose the unitaries from the uniform (Haar) probability distribution on the unitary group for a pair of spins, $\mathrm{U}(q^2)$. This model is solvable in the limit of large local Hilbert space dimension $q$.

\subsubsection{Dynamics of Hartley entropy}

A useful starting point is to consider the $n\rightarrow 0$ limit of the Renyi entropy, $S_0$.
This is known as the Hartley entropy, and quantifies (the logarithm of) the number of nonzero eigenvalues of the reduced density matrix. Equivalently, the Hartley entropy determines the necessary value of the local \textit{bond dimension} in an exact matrix product  representation \cite{VerstraeteMurgCircac2008, Perez-Garcia2007} of the state:
\begin{align}
S_0 (x) = \log\lf  \textit{bond dimension at $x$} \ri.
\end{align}
Like the von Neumann entropy, the Hartley entropy of neighboring bonds can differ by at most one:
\begin{align}
\big| S_{0}(x+1) - S_0(x) \big| \leq 1.\label{hartley constraint}
\end{align}
Recall that logarithms are base $q$. For the present we keep $q$ finite.

For the random dynamics described above (Sec.~\ref{haar protocol}),
the Hartley entropy obeys an extremely simple dynamical rule.  In a given time step, a unitary is applied at a random bond, say at $x$.
Applying this unitary may change the Hartley entropy across the bond $x$; the entropy remains unchanged for all {other} bonds.
The rule for the change in $S_{0}(x)$ is that, with probability one, it increases to the \textit{maximal} value allowed by the general constraint (\ref{hartley constraint}):
\begin{align}
S_0(x, t+1) = \min \{ S_0(x-1,t) , S_0(x+1,t) \} + 1.\label{hartley dynamics}
\end{align}
This `maximal growth' of $S_0$ occurs with probability one when all unitaries are chosen randomly. Fine-tuned  unitaries (e.g. the identity) may give a smaller value, but these choices are measure zero with respect to the Haar distribution. 

We present a rigorous proof of  Eq.~\ref{hartley dynamics} in Appendix~\ref{hartley entropy growth}. The appendix also gives a heuristic parameter-counting argument which suggests the same result, but as explained there the more rigorous argument is  necessary.

The dynamical rule in Eq.~\ref{hartley dynamics} defines a simple but  nontrivial stochastic process. Before discussing its properties, we use Eq.~\ref{hartley dynamics} as a starting point  to show that in the limit of large Hilbert space dimension the von Neumann entropy (and in fact all the higher Renyi entropies) obeys the same dynamical rule. The von Neumann entropy is of more interest than $S_0$, since the latter  behaves pathologically in many circumstances\footnote{This is because it simply counts up all the (nonzero) eigenvalues in the spectrum of $\rho_x$, regardless of how small they are. For example,  Hamiltonian dynamics in continuous time --- as opposed to unitary circuits like the above --- will generally give an infinite growth rate for  $S_0$, in contrast to the finite growth rate for $S_\text{vN}$ and the higher Renyi entropies.}.

\subsubsection{Limit of large Hilbert space dimension}
\label{large q section}

The present quantum circuit dynamics lead to a solvable model in the limit of large local Hilbert space dimension, $q\rightarrow\infty$. In this limit \textit{all} the Renyi entropies obey the dynamical rule in Eq.~\ref{hartley dynamics}.

To show this we consider the reduced density matrix for a cut at $x$, where $x$ is the bond to which we are applying the unitary in a given time step. We may write $\rho_x(t+1)$ in terms of $\rho_{x-1}(t)$ and the applied unitary matrix.  Averaging $\Tr \rho_x^2$ \textit{over the choice of this unitary}, we then obtain:
\be\notag
\< \Tr \rho_x(t+1)^2 \>_\text{Haar} = \f{q}{q^2 + 1} \lf
\Tr \rho_{x-1}(t)^2 +
\Tr \rho_{x+1}(t)^2
\ri.
\ee
See App.~\ref{haar appendix} for details. In terms of the second Renyi entropy $S_2$ this is:
\be\label{purity average}
\< q^{-S_2(x,t+1)}  \>_\text{Haar} =
\f{q^{-S_2(x-1,t)-1} +
q^{-S_2(x-1,t)-1}}{1 + 1/q^2}.
\ee
The general constraint $S_2 \leq S_0$ allows us to write
\be
S_2(x,t) = S_0(x,t)-\Delta(x,t)
\ee
with $\Delta\geq 0$. We now use Eqs.~\ref{hartley dynamics},~\ref{purity average} to show that $\Delta$ is infinitesimal at large $q$. Rewriting Eq.~\ref{purity average}  in terms of $\Delta$, and substituting Eq.~\ref{hartley dynamics},  immediately shows
\be
\< q^{\Delta(x,t+1)}\>_\text{Haar}
< \,
q^{\Delta(x-1,t)} + q^{\Delta(x+1,t)}.
\ee
For a simple bound\footnote{For a large system, this bound on $\<q^{\Delta_\text{max}}\>$ will be far from the tightest possible since we have not exploited the large size of the system.}, define $\Delta_\text{max}(t)$ to be the maximal value of $\Delta(x,t)$ in the entire system. The  equation above implies
\be
\< q^{\Delta_\text{max}(t+1)} \>_\text{Haar}  < 2 q^{\Delta_\text{max}(t)}
\ee
We may iterate this by averaging over successively earlier unitaries:
\be
\< e^{(\ln q) \Delta_\text{max}(t)} \>_\text{Haar}  < 2^t.
\ee
This shows that as $q\rightarrow \infty$ at fixed time $t$, the probability distribution for $\Delta$ concentrates on $\Delta = 0$, so that $S_2$ and $S_0$ become equal.

This implies that the entanglement spectrum is flat, so in fact \textit{all} the Renyi entropies obey Eq.~\ref{hartley dynamics} for the application of a unitary across bond $x$.

\subsubsection{Properties of the solvable model}
\label{properties of solvable model section}

\begin{figure}[t]
 \begin{center}
  \includegraphics[width=0.97\linewidth]{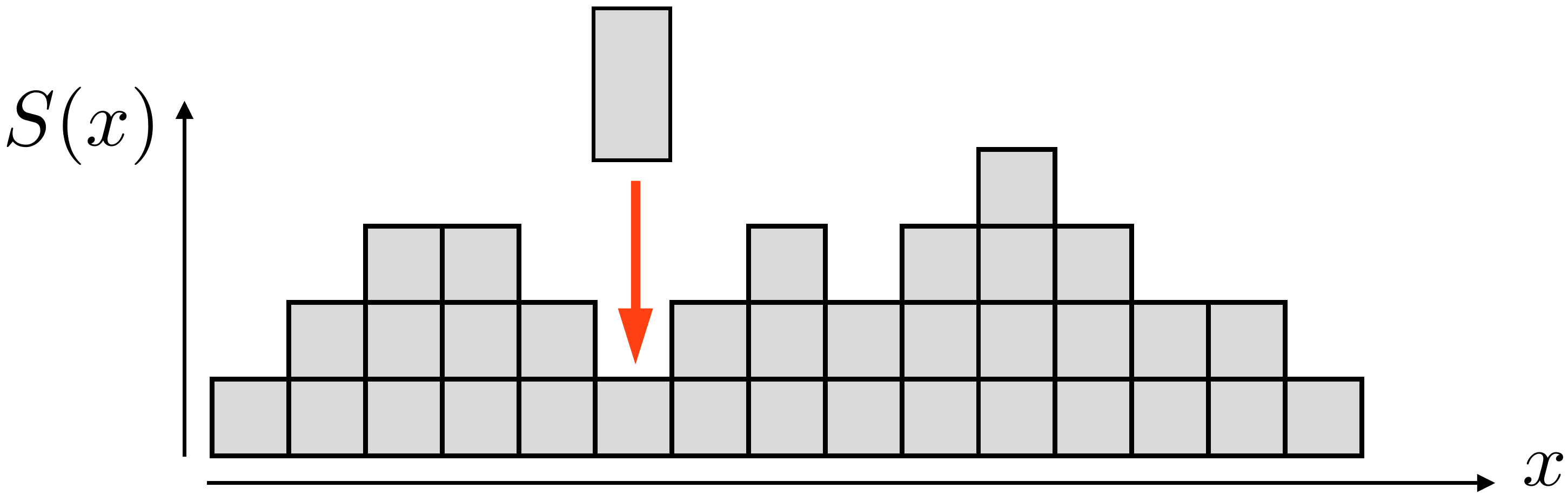}
 \end{center}
\caption{Surface growth model for entanglement $S(x,t)$ across a cut at $x$, in the large $q$ limit. Applying a unitary to bond $x$ can increase the height of the surface locally (Eq.~\ref{large q dynamics}), corresponding to dropping a `block' of height $\Delta S = 1$ or $\Delta S = 2$.}
 \label{surface figure}
\end{figure}

\begin{figure}[t]
 \begin{center}
  \includegraphics[width=0.8\linewidth]{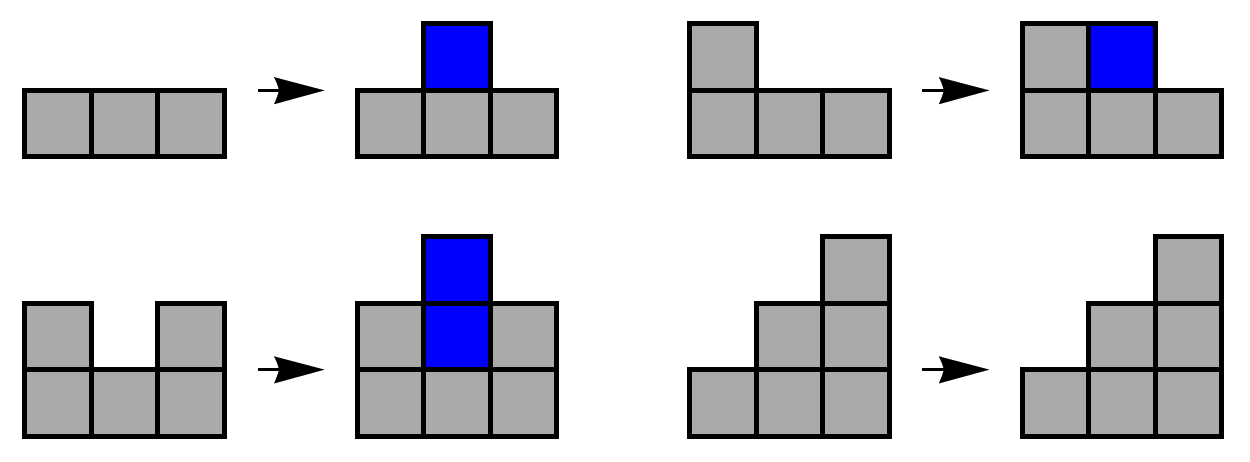}
 \end{center}
\caption{Entanglement growth in the large $q$ model:  Effect of applying a random unitary to the central bond, for four choices of the  initial local entropy configuration of three adjacent bonds.}
 \label{growth examples}
\end{figure}

The dynamical rule we have arrived at for the bipartite von Neumann and Renyi entropies at large $q$,
\be
\label{large q dynamics}
S(x, t+1) = \min \{ S(x-1,t) , S(x+1,t) \} + 1
\ee
defines a stochastic surface growth model in which $S(x,t)$ is always an integer-valued height profile (Fig.~\ref{surface figure}). The remaining randomness is in the choice of which bond is updated in a given time step. At each time step, a bond $x$ is chosen at random, and the `height' $S_0(x)$ is increased to the maximal value allowed by the neighbors.  Fig.~\ref{growth examples} gives examples of local configurations before and after the central bond is updated.

This model is almost identical to standard models for surface growth \cite{MeakinRamanlalSanderBall1986, KimKosterlitz1989}. It is in the KPZ universality class (it is straightforward to simulate the model and confirm the expected KPZ exponents) and some non-universal properties can also be determined exactly. Note that the boundary conditions $S=0$ on the right and the left, and the restriction ${|S(x+1)-S(x)|\leq 1}$, imply that the entanglement eventually saturates in the expected pyramid profile.

When we move to the continuum (KPZ) description of the interface (\ref{kpz equation}) the nonlinear  $\lambda$ term  appears with a negative sign, meaning that entanglement growth is slower when the coarse-grained surface has a nonzero slope. This is natural given the microscopic dynamics: if the slope is maximal in some region, local dynamics cannot increase the entropy there.

In the present model the difference in height between two adjacent bonds is either $\Delta S = \pm 1$ or  $\Delta S = 0$. At early stages of the evolution both possibilities occur. However one may argue that the `flat points', where $\Delta S = 0$, become rarer and rarer at late times.\footnote{Flat points can disappear by `pair annihilation' (Fig.~\ref{growth examples}, top left), and can diffuse left or right (Fig.~\ref{growth examples}, top right), but cannot be created. As a result their density decreases with time.}  At late times the model therefore becomes equivalent to the well-known `single step' surface growth model \cite{MeakinRamanlalSanderBall1986}, in which $\Delta S = \pm 1$ only. An appealing feature of this model is that, for a certain choice of boundary conditions,\footnote{The solvable case corresponds to choosing periodic BCs in the classical problem. (These BCs are useful for understanding the classical model, but they do not have an interpretation in terms of entanglement.) In this setting  the mean height grows indefinitely, but the probability distribution for the height fluctuations reaches a well-defined steady state.} the late-time probability distribution of the growing interface can be determined exactly \cite{MeakinRamanlalSanderBall1986}. This shows that on scales smaller than the correlation length (and prior to  saturation) the interface looks like a 1D random walk with uncorrelated  $\Delta S = \pm 1$ steps. In addition to confirming the expected KPZ exponent $\alpha = 1/2$, this allows the mean growth rate of the surface to be calculated. After rescaling time so that one unit of time corresponds to an average of one unitary per bond,  this gives an entanglement growth rate (\ref{mean growth}) of $v_E = 1/2$.

The mapping to  surface growth  gives us a clean derivation of universal entanglement dynamics in a solvable model. However this surface growth picture is restricted to the entropy for a single cut (as opposed to the entropy of a region with multiple endpoints) and to one dimension. It will be useful to find a more general language which extends the above results. To do this we now make a connection with the second member of the KPZ triumvirate (Fig.~\ref{spin chain fig}), the statistical mechanics of a polymer in a random environment.

\section{Directed polymers \& min--cut}\label{directed polymer section}

\begin{figure}[t]
 \begin{center}
    \includegraphics[width=0.8\linewidth]{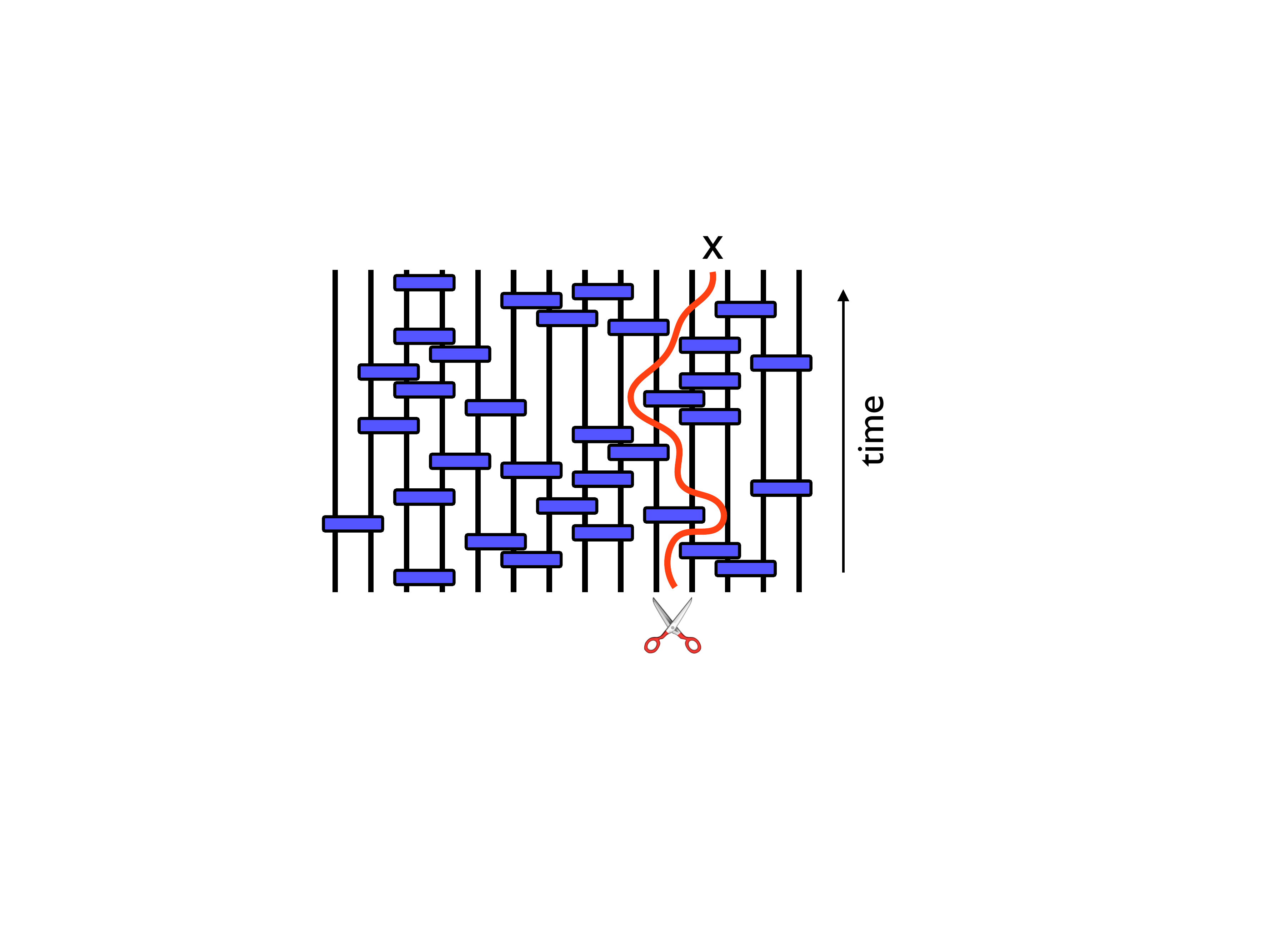}
 \end{center}
\caption{Any cut through the unitary circuit which separates the legs to the left and right of $x$ (on the top boundary) gives an upper bound on $S(x,t)$. The best such bound is given by the minimal cut (note that the cut shown in the figure is not the minimal one). Finding the  minimal cut in a random network is akin to finding the lowest energy state of a polymer in a random potential landscape.}
 \label{cut fig}
\end{figure}

In this section we relate the dynamics of $S(x,t)$ to the geometry of a `minimal cut' through the quantum circuit which prepares the state (Fig.~\ref{cut fig}). This provides an alternative perspective on the exact result (\ref{large q dynamics}) for the solvable model, and also a useful heuristic picture for noisy quantum dynamics in general.  This line of thinking reproduces KPZ behaviour in 1D. Importantly, it also allows us to generalize to higher dimensions and to more complex geometries.

Our starting point is the minimal cut bound for tensor networks. This very general bound has been related to the Ryu-Takayanagi formula for entanglement in holographic conformal field theories \cite{RyuTakayanagi2006, Swingle2012, Pastawski2015, HaydenEtAl2016}, and has also been applied to unitary networks as a heuristic picture for entanglement growth \cite{CasiniLiuMezei2015}.

Consider again a random quantum circuit in 1+1D, and a  curve like that  in Fig.~\ref{cut fig} which bisects the circuit and divides the physical degrees of freedom into two at position $x$. \textit{Any} such curve gives an upper bound on the entanglement: all the Renyi entropies satisfy $S(x) \leq S_\text{cut}$, where $S_\text{cut}$ is  the number of `legs' that the curve passes through. (This is  because the rank of the reduced density matrix $\rho_x$ is at most $q^{S_\text{cut}}$.\footnote{This can be seen by writing the density matrix  in terms of a sum over the indices on the cut bonds.}) The \textit{best} bound of this type is given by the minimal cut, which passes through the smallest number of legs. We denote the corresponding estimate for the entropy $S_\text{min-cut}(x)$. If the geometry of the circuit is random,  $S_\text{min-cut}(x)$ and the corresponding curve are also random.

In the solvable large $q$ model, $S_\text{min-cut}(x)$ in fact gives the von Neumann entropy exactly. This follows straightforwardly from the results of the previous section (see below).  In a typical microscopic model, on the other hand, $S_\text{min-cut}$ is only a bound on the true entropy.  Nevertheless we conjecture that the following is generally valid as a \textit{coarse-grained} picture: i.e. that it  correctly captures the {universal} properties of the entanglement dynamics. This conjecture is equivalent to the  applicability of the KPZ description to generic noisy systems; further evidence for the latter is in Secs.~\ref{hydrodynamics section},~\ref{numerical results section}.

The problem of finding the minimal curve is a version of a well studied problem in classical statistical mechanics, known as the directed polymer in a random environment or DPRE \cite{HuseHenley1985, KardarComment1985}. Here the polymer is the curve which bisects the circuit, and its energy $E(x)$ is equal to $S_\text{cut}(x)$, the number of legs it bisects.  The spatial coordinate of the polymer's upper endpoint is fixed at $x$, while the lower endpoint is free.  Finding $S_\text{min-cut}(x)$ is equivalent to finding  the minimal value of the polymer's energy.  This corresponds to the polymer problem at zero temperature; however the universal behaviour of the DPRE is the same at zero and at nonzero temperature.\footnote{For any finite temperature, the DPRE flows under  renormalization to a zero temperature fixed point at which temperature is an irrelevant perturbation.}

DPRE models with short-range-correlated disorder are in the same universality class as the KPZ equation \cite{kpz}. Let $E(x,t)$ be the minimal energy of the polymer in a sample of height $t$. We may increase $t$ by adding an additional layer to the top of the sample. $E(x,t+\delta t)$ can then be expressed recursively in terms of $E(y, t)$ for the various possible values of $y$. In the continuum limit, this leads to an equation for $E(x,t)$ which is precisely the KPZ equation \cite{kpz}. The KPZ exponents given in Sec.~\ref{surface growth section} may therefore be applied to the energy of the polymer.  The exponent $z=3/2$ also determines the lengthscale for transverse fluctuations of the polymer:
\be\label{transverse wandering} 
\Delta x \sim (\Delta t)^{2/3}.
\ee

Since in our case the minimal $E$ is simply $S_\text{min-cut}$, we find that the latter executes KPZ growth. In the light of the previous section, this is not surprising. In fact in our solvable model, $S_\text{min-cut}$ is exactly equal to the true entanglement entropy (in the large $q$ limit). This follows from the fact that the recursive construction of $E(t)$ described above precisely matches the large $q$ dynamics of Eq.~\ref{large q dynamics}. Examples of non-unitary tensor networks in which the minimal cut bound becomes exact are also known \cite{Pastawski2015}, including a large-bond-dimension limit similar to that discussed here \cite{HaydenEtAl2016}.

The utility of the DPRE picture is that it is far more generalizable than the surface growth picture, which is restricted to the entropy across a single cut in 1D. As noted above, the value of $S_\text{min-cut}$ in a given microscopic model is typically not equal to any of the physical entropies $S_n$ with $n>0$. Nevertheless we conjecture that the DPRE  and  KPZ pictures are valid universal descriptions for all noisy models, so long as they are not fine tuned or nonlocal. This includes Hamiltonian dynamics in continuous time; we discuss this case further in the Outlook section.

\subsection{Saturation in the minimal cut picture}
\label{saturation polymer section}

Eq.~\ref{transverse wandering} shows that the coarse-grained minimal cut is essentially vertical (prior to saturation of the entropy): the lengthscale for transverse fluctuations is negligible in comparison with $t$. This leads to an extremely simple deterministic picture for the leading-order behaviour of the entanglement, which we will discuss in more detail in Sec.~\ref{entanglement velocity section}. Here we briefly consider the saturation of the entanglement entropy, reproducing behaviour known from other contexts \cite{LiuSuh2014, CasiniLiuMezei2015}.

The definition of the entanglement growth rate implies that the `energy' $E$ of such a vertical cut is $v_E t$ to leading order. The entanglement in a finite system grows at this rate until time $t_\text{saturation} =x/v_E$, when it reaches its saturation value $S = x$. (Here we are neglecting subleading terms, and assuming $x < L-x$.) After this time a vertical cut is no longer favourable: instead the minimal cut  exits the circuit via the left-hand side. Its shape is no longer unique, but it can be taken to be horizontal, and it  has `energy' $E=x$.   This picture corresponds to a simple scaling form (again, neglecting subleading terms)
\be\label{scaling form saturation main text}
S(x,t) = v_E t \, f(x / v_E t),
\ee
with 
\be\label{scaling form saturation 2}
f(u) = 
\left\{
\begin{array}{c}
 u \quad \text{for} \quad u < 1  \\
 1 \quad \text{for} \quad u\geq 1
\end{array}
\right.
\ee
For a finite interval of length $l$ in an infinite system we have instead $S(t) = 2 v_E t \, f(l/2v_E t)$. 

These scaling forms are our first confirmation that $v_E$ is really a \textit{speed}, as well as a growth rate for the  entanglement. We will give an independent derivation of this fact for Clifford circuits in the following section, and we will test the above scaling form numerically in Appendix~\ref{appendix finite size effects}. We will discuss the interpretation of $v_E$ further in Sec.~\ref{entanglement velocity section}. Note that fluctuations have dropped out of  Eq.~\ref{scaling form saturation main text}, as a result of considering only the leading order behaviour of $S(x,t)$.  These scaling forms agree with the results for holographic CFTs \cite{LiuSuh2014} and with an application of the minimal cut formula to a regular tensor network \cite{CasiniLiuMezei2015}.

It is straightforward to generalize the directed polymer picture to the entanglement or mutual information of arbitrary regions, and even to higher dimensions.  We propose in  Secs.~\ref{entanglement velocity section},~\ref{higher dimensions section} that this picture in terms of a \textit{coarse-grained} minimal cut is the simplest way to understand the basic features of the `entanglement tsunami' for generic many-body systems, with or without noise.

\section{Hydrodynamics of operator spreading} \label{section: Burgers}
\label{hydrodynamics section}

An alternative way to think about the quantum dynamics is in terms of the evolution of local operators $\mathcal{O}_{i}$. For example, a Pauli operator initially acting on a single spin (e.g. $\mathcal{O}_{i} \equiv Y_{i}$; we denote the Pauli matrices by $X$, $Y$, $Z$) will evolve with time into an operator $\mathcal{O}_i(t)$ which acts on many spins. Operators typically grow ballistically \cite{Roberts2015}, in the sense that the number of spins in the support of $\mathcal{O}_i(t)$ grows linearly with $t$.

In this section we relate the growth of the bipartite entanglement to the spreading of operators. We focus on the special case of unitary evolution with Clifford circuits (defined below), but we expect the basic outcomes to hold for more general unitary dynamics. We find that the entanglement growth rate is not given by the rate at which a single operator grows, but is instead determined by collective dynamics involving many operators. Remarkably, in 1D these collective dynamics have a long wavelength hydrodynamic description.

\begin{figure}[t]
 \begin{center}
  \includegraphics[width=0.9\linewidth]{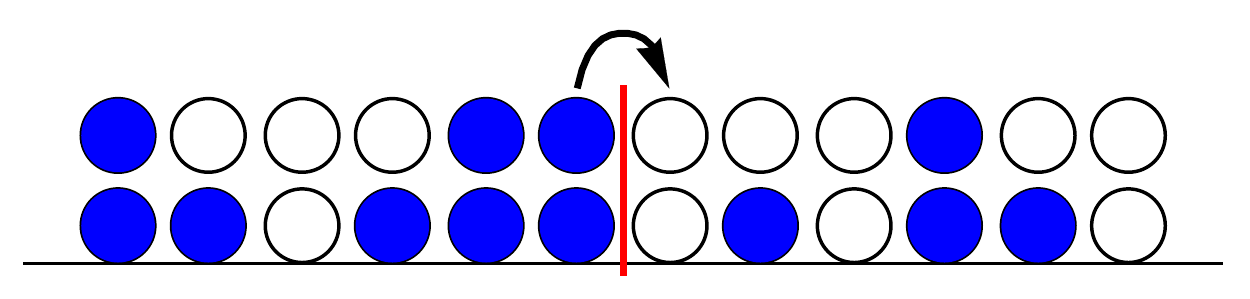}
 \end{center}
\caption{Spreading of stabilizer operators defining the quantum state (Sec.~\ref{hydrodynamics section}). Each blue particle marks the right \textit{endpoint} of some stabilizer (the rightmost spin on which it acts). Blue particles hop predominantly to the right. Whenever a particle enters the right-hand region (A) the entanglement $S_A$ increases by one bit. The particle density is described by the noisy Burgers equation, which maps to KPZ. A `hole' (empty circle) marks the left-hand endpoint of some stabilizer.}
 \label{asep picture}
\end{figure}

This hydrodynamic description turns out to be the noisy Burgers equation, which is related to the KPZ equation by  a simple change of variable and is the final member of the KPZ triumvirate shown in Fig.~\ref{triumvirate}. In the present case the hydrodynamic mode is the density of certain fictitious `particles', shown in blue in Fig.~\ref{asep picture}. The quantum state is defined by a set of operators (Sec.~\ref{stabilizer section}) which spread out over time, and the particles are markers which show how far these operators have spread.  We will derive their coarse-grained dynamics in Sec.~\ref{coarse grained operator dynamics section} after introducing the necessary operator language.

This picture shows that there is a well-defined velocity at which entanglement spreads (at least for the present class of dynamics) and that this is \textit{distinct} from the velocity $\tilde v$ governing the growth of operators (Sec.~\ref{entanglement velocity section}).

\subsection{Stabilizer operators}\label{stabilizer section}

It will be convenient to use the language of `stabilizer' operators to describe the entanglement dynamics. We may define the initial state $\ket{\Psi_{0}}$ by specifying $L$ stabilizers under which it is invariant (in this section we take the number of sites to be $L$). These operators, denoted $\mathcal{O}_i$ ($i=1,\ldots,L$) satisfy
\ba
\mathcal{O}_{i} \ket{\Psi_{0}} & = \ket{\Psi_{0}}.
\end{align}
For example, if the spins are initially polarized in the $y$ direction we may take $\mathcal{O}_{i} = Y_i$. At a later time, the above equation still holds, with each stabilizer $\mathcal{O}_{i}$ replaced with the time-evolved stabilizer $\mathcal{O}_{i}(t) = U(t)\mathcal{O}_{i}U^{\dagger}(t)$, where $U(t)$ is the unitary operator that evolves the initial state to the state at time $t$.

In the following, we focus on evolution of the initial state with unitary gates in the \emph{Clifford group} \cite{gottesman1998heisenberg}. Such gates have recently been used in toy models for many-body localization \cite{ChandranLaumann2015}.  The defining feature of Clifford unitaries is that they have a simple action on Pauli operators:  single-spin Pauli operators are mapped  to \textit{products} of Pauli operators. 

Any product of Pauli matrices can be written as a product of $X$ and $Z$ matrices, so to follow the dynamics of a given stabilizer $\mathcal{O}_{i}(t)$, we need only keep track of which $X_i$ and $Z_i$ operators appear in this product. Furthermore, the overall sign of the stabilizer $\mathcal{O}_{i}(t)$ does not affect the entanglement properties of a system undergoing Clifford evolution, so we do not keep track of it.  By writing $\mathcal{O}_{i}(t)$ as
\be\label{sigma string}
\mathcal{O}_{i}(t) \propto
 X_1^{v_{1x}}
  Z_1^{v_{1z}}
\ldots
X_L^{v_{Lx}}
  Z_L^{v_{Lz}},
\ee
we may specify any stabilizer by a binary vector $\vec{v}$ with $2L$ components:
\be
\vec v = (v_{1x}, v_{2x}, \ldots, v_{Lx}, v_{Lz}).
\ee
For example, the first component of the vector $v_1 = 1$ if $X_1$ appears in the product, and $v_1 = 0$  otherwise.   The binary vector corresponding to a stabilizer $\mathcal{O}_{i} =Y_i$ is
\be
\vec{v} \equiv (0, \ldots, 0, 1, 1, 0, \ldots, 0),
\ee
where the locations of the nonzero elements correspond to $X_i$ and $Z_i$.

We consider the dynamics in two stages. First we consider the evolution of a single operator. Then we will generalize this to understand the dynamics of the state.

How does a single stabilizer $\mathcal{O}_{i}(t)$ evolve? Applying a one or two-site Clifford unitary to $\mathcal{O}_{i}(t)$ corresponds to applying simple local updates to the string $\vec {v}$. Although the precise details of these updates will not be crucial,  we now give some explicit examples of gates which we encounter again in the numerical simulations.

As single-site examples, consider the  Hadamard and Phase gates. The Hadamard is a rotation on the Bloch sphere\footnote{The rotation is by $\pi$ around the  $(1,0,1)$ axis.} which exchanges the $X$ and $Z$ axes,
\be
R_H = {1\over \sqrt{2}} \left( X+ Z\right) \label{RH}
\ee
so applying a Hadamard to site $i$ updates the string by ${v_{ix} \longleftrightarrow v_{iz}}$.  The Phase gate is a rotation around the $Z$ axis which maps $X_i$ to $Y_i = i X_i Z_i$
\be
R_{P} = \sqrt{Z} \,. \label{RP}
\ee
This means that an additional $Z_i$ is generated whenever $X_i$ is present in the string, or equivalently ${v_{iz}\rightarrow v_{iz} + v_{ix} \quad {(\text{mod 2})}}$.
For a two-site example, consider the left and right controlled-NOT (CNOT) gates acting on the leftmost spins in the chain. In the $Z$ basis, the action of these operators is to flip the `target' spin iff the `control' spin is down:
\begin{align}
\operatorname{CNOT}^{\text{(L)}}  &= {1\over 2} \left[  (1+Z_1) + (1-Z_1) X_{2}   \right],
\label{CNOT} \\
 \operatorname{CNOT}^\text{(R)} &= {1\over 2} \left[  (1+Z_2) + (1-Z_2)X_1 \right].\nonumber
 \end{align}
Conjugating the Pauli matrices by $\operatorname{CNOT}^{\text{(L)}}$ yields:
\ba\notag
X_1 & \rightarrow X_1 X_2, &
Z_1 & \rightarrow Z_1, &
X_2 & \rightarrow X_2, &
Z_2 & \rightarrow Z_1 Z_2.
\end{align}
We see that the operator $X_2$ is added to the string if $X_1$ is present (and similarly for $Z_1$ and $Z_2$). Applying $\operatorname{CNOT}^{\text{(L)}}$ therefore updates $\vec v$ by
\ba\notag
v_{2x} & \rightarrow v_{2x}+v_{1x} \, \, (\text{mod 2}), &
v_{1z} & \rightarrow v_{1z} + v_{2z} \,\, (\text{mod 2}).
\end{align}
$\operatorname{CNOT}^{\text{(R)}}$ acts similarly with the roles of the spins reversed.

It is simple to argue that random application of such operations causes the region of space in which $\vec v$ is nonzero to grow ballistically. This  corresponds to the operator spreading itself over a region of average size $2\tilde v t$, where $\tilde v$ is the \textit{operator spreading velocity}.  This operator spreading velocity is the analogue, for the present dynamics, of \textit{both} the Lieb Robinson velocity and the recently discussed `butterfly velocity' \cite{RobertsSwingle2016} in the context of non-noisy systems. The value of $\tilde v$ depends on the precise choice of dynamics, but it is the same for all initial operators so long as the dynamics (the probability distribution on gates) is not fine-tuned. Further, one may argue that  the interior of the region where  $\vec v$ is nonzero is  `structureless'. Within the interior, $\vec v$ rapidly `equilibrates' to become a completely random binary string.\footnote{Consider the late time dynamics of an operator, or equivalently its string $\vec v$, in an $L$-site system. Random application of Clifford gates gives random dynamics to $\vec v$. It is easy to see that the \textit{flat} probability distribution on $\vec v$ is invariant under the dynamics, regardless of the probabilities with which the gates are applied.  By standard properties of Markov processes, this is the unique asymptotic distribution to which the system tends, so long as the choice of Clifford gates is not fine-tuned to make the process non-ergodic. (If the gate set includes each gate and its inverse with the same probability, detailed balance is also obeyed, but this is not necessary.) We expect $\vec v$ to equilbrate locally to this structureless state on an $O(1)$ timescale, and similarly for the internal structure of operators smaller than $L$.}

Now consider the dynamics of a quantum state. Once the sign information in Eq.~\ref{sigma string} is dropped, the relevant information in the state $\ket{\Psi(t)}$ is contained in binary vectors $\vec v^1, \ldots, \vec v^L$ corresponding to the $L$ stabilizers. We may package this information in a $2L\times L$ matrix:
\be
\Psi(t) = \lf \vec v^{1\top}, \ldots, \vec v^{L\top} \ri.
\ee
Each \textit{column} corresponds to a stabilizer, and each \textit{row} to a spin operator $X_i$ or $Z_i$. The dynamical updates correspond to row operations (with arithmetic modulo two) on this matrix. For example, a Hadamard gate exchanges the rows corresponding to $X_i$ and $Z_i$.

A crucial point is that there is a large \textit{gauge freedom} in this definition of the state. This gauge freedom arises because we can redefine stabilizers by multiplying them together. For example if a state is stabilized by $\{ X_1, Z_2 \}$, then it is also stabilized by $\{ X_1 Z_2, Z_2\}$, and vice versa. This freedom to redefine the stabilizers corresponds to the freedom to make column operations on $\Psi$, or equivalently the freedom to add the vectors $\vec v^i$ modulo two. Note that by making such a `gauge transformation' we may be able to reduce the size of one of the stabilizers, giving a more `compact' representation of the state.

The final fact we need is an expression for the entropy $S_A(t)$ of a region $A$ in terms of the stabilizers. Heuristically, this  is given  by  the number of stabilizers that have spread into region $A$ from outside. More precisely, define $I_A$ as the number of stabilizers that are
\textit{independent} when restricted to region $A$.\footnote{We {truncate} all  stabilizers to region $A$ by throwing away all the spin operators acting outside $A$. In this process some of the stabilizers become trivial, and some become redundant: i.e., equal to products of the others.  $I_A$ is the number of stabilizers that are {independent} when truncated to $A$. Equivalently, $I_A$ is the rank of the matrix $\Psi$ after the rows corresponding to the complement of $A$ have been deleted; this is how we calculate the entropy numerically for Sec.~\ref{section:numerics:clifford}.}  (Independence of the stabilizers corresponds to linear independence of the vectors $\vec v^i$, with arithmetic modulo two, once they are truncated to region $A$.) The entropy is equal to  \cite{HammaIonicioiuZanardi2005,AliosciaHammaZanardi2005Ground}
\be\label{entropy formula}
S_A(t) = I_A - |A|,
\ee
where $|A|$ is the number of sites in $A$. See Appendix~\ref{stabilizer state entropy} for a simple derivation of Eq.~\ref{entropy formula}. For Clifford dynamics all Renyi entropies are equal, so we omit the Renyi index on $S$. The maximal value of $I_A$ is $2|A|$, so $S_A$ is bounded by $|A|$ as expected.

This formula has a simple interpretation. In the initial product state we may take one stabilizer to be localized at each site, so $I_A = |A|$ and the entanglement is zero. As time goes on, stabilizers that were initially localized outside of $A$ grow and enter $A$. Each time a new {independent} operator appears in $A$, the entanglement $S_A(t)$ increases by one bit. The linear independence requirement in the definition of $I_A$ is crucial, as it leads to effective interactions between the stabilizers which we discuss in the following subsection.

From now on we take $A$ to consist of the spins to the \textit{right} of the bond $x$, and revert to the notation $S_A = S_x$ used in the rest of the text for the entanglement across a cut at $x$.

\subsection{Coarse-Grained Operator Dynamics}
\label{coarse grained operator dynamics section}

\begin{figure}[t]
 \begin{center}
  \includegraphics[width=0.95\linewidth]{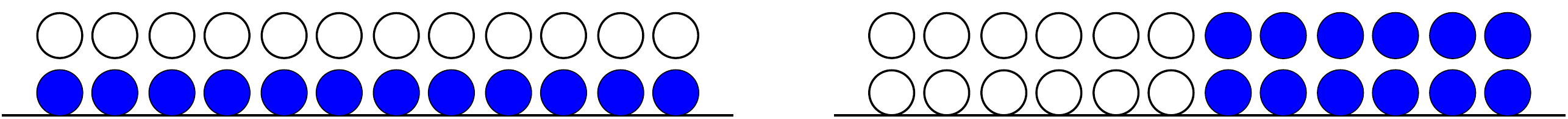}
 \end{center}
\caption{Left: the initial product state represented in terms of the fictitious particles. Right: a state with maximal $S(x)$.}
 \label{initial state fig}
\end{figure}

Each stabilizer $\mathcal{O}_i(t)$ (labelled $i=1,\ldots L$) has a left and a right endpoint $l_i$ and $r_i$, marking the extremal spins included in the stabilizer. We view $l_i$ and $r_i$ as the positions of two fictitious particles of type $l$ and $r$, represented in white and blue respectively in Figs.~\ref{asep picture},~\ref{initial state fig}. There are $L$ of each type of particle in total.

In the initial product state, $\mathcal{O}_i(0)$ is a single Pauli operator on site $i$, say $Y_i$. This means that each site has one $l$ particle and one $r$ particle (since $l_i = r_i = i$) as shown in Fig.~\ref{initial state fig} (left). As time increases the $r$ particles will typically move to the right and the $l$ particles to the left.

The nature of this motion depends on how we define the stabilizers. At first sight the obvious choice is to define $\mathcal{O}_{i}(t)$  as the unitary time evolution of the initial stabilizer, $\mathcal{O}_{i}(t)=U(t) Y_i U^\dagger(t)$. But in fact it is useful to exploit the gauge freedom in the choice of stabilizers to impose a different `canonical' form. One result of this is that the stabilizers effectively grow more slowly than the velocity $\tilde v$ (mentioned in the previous subsection) for the spreading of an operator considered in isolation.

Let $\rho_l(i)$ and $\rho_r(i)$ be the number of particles of each type at site $i$. The constraint that we impose is:
\be\label{density constraint}
\rho_l(i) + \rho_r(i) = 2.
\ee
To see that we can impose this constraint, consider the situation  $\rho_l(i) = 3$, so that there are three stabilizers that start at $i$. The initial element of each string can be either $X$, $Y$, or $Z$. If $\rho_l(i) = 3$, it is impossible for all  three initial elements to be independent. We can then redefine one of the stabilizers, by multiplying it by one or both of the others, in such a way that its length decreases by one.\footnote{By choosing the longer stabilizer we avoid adding length at the right-hand side.} Making reductions of this kind wherever possible guarantees that $\rho_{l}(i)\leq 2$, and also that if $\rho_l=2$, the initial elements of the two stabilizers are distinct. (And similarly for $\rho_r$.) With this convention it also follows that $\rho_l(i) + \rho_r(i) \leq 2$: otherwise the operators involved could not commute, which they must.\footnote{Consider the case where $\rho_r(i)=1$: for example let the corresponding stabilizer read $\mathcal{O} =\ldots X_i$.  Any stabilizer contributing to $\rho_l(i)$ must be of the form $X_i \ldots$ in order to commute with $\mathcal{O}$. By the rule imposed in the text this means that $\rho_l(i)\leq 1$.} (The initial stabilizers commute, and this is preserved  by the unitary dynamics and the redefinitions of the stabilizers.)  Since there are a total of $2L$ particles which all have to live somewhere, we have Eq.~\ref{density constraint}.

With this convention, the dynamics of the bipartite entropy $S(x)$ is simply related to the hopping dynamics of the particles. By Eq.~\ref{density constraint} it suffices to consider only the $r$ particles: an $l$ particle is just an $r$ `hole'. We will write the density $\rho_r$ of $r$ particles as  $\rho$. See Fig.~\ref{asep picture} for a typical configuration in a subregion of the system.

The utility of the canonical form (\ref{density constraint}) is that the independence requirement becomes trivial. One can easily check that all the operators which have spread into $A$ (the region to the right of $x$) are independent.\footnote{Consider the stabilizers which act in region $A$, i.e. the stabilizers with  $r_i>x$. We may argue by contradiction that they remain independent after truncation to subsystem $A$. If not, this means  there is some product of the truncated stabilizers which equals one. Let the rightmost spin appearing in any of these stabilizers be $j$. But by our convention for `clipping' the stabilizers, it is impossible for the Pauli matrices acting on spin $j$ to cancel out when they are multiplied together. Therefore the operators must in fact be independent.}  Therefore to find $S(x)$ we need only count the number of $r$ particles to the right of the cut and subtract the number of sites:
\be
S(x) =  \sum_{i > x}  \lf \rho_i -1\ri.
\ee
To reiterate, the entanglement increases by one every time an $r$ particle drifts rightward across bond $x$ (and decreases by one if it drifts across in the other direction).

Now consider the dynamics of the particles. Microscopically, a dynamical time step involves (1) application of a unitary gate, and (2) potentially a `clipping' of stabilizers to enforce the canonical form. Effectively, the particles perform \textit{biased} diffusion, with the restriction that more than two particles cannot share a site,
\be
\rho \leq 2.
\ee
This constraint leads to `traffic jam' phenomena familiar from the  so-called asymmetric exclusion process \cite{HALPINHEALY1995}, and to the same continuum description.  Our essential approximation is to neglect the detailed \textit{internal} structure of the stabilizers, and to treat the dynamics of the endpoints as effectively Markovian. We expect this to be valid at long length and time scales for the reason mentioned in the previous subsection: the internal structure of the operators is essentially featureless, and characterized by finite time scales.

We now move to a continuum description. The coarse-grained density obeys a continuity equation
\be
\partial_t \rho = - \partial_x J
\ee
with $J$ the particle current. Further, there is a symmetry under spatial reflections, which  exchange left and right endpoints ($\rho_l \leftrightarrow \rho_r$). Writing
\be
\rho = 1 + \Delta \rho,
\ee
where $\Delta \rho$ is the deviation from the mean density, the reflection symmetry is
\ba
x & \rightarrow - x, &
\Delta \rho & \rightarrow - \Delta \rho.
\end{align}
To obtain a long wavelength description, we write the current as a power series in $\Delta \rho$ and $\partial_x$. Keeping the lowest order terms that respect the symmetry,
\be
J = c - \nu \partial_x \Delta \rho - \f{\lambda}{2} (\Delta \rho)^2  + \eta.
\ee
These terms have a transparent physical meaning. The drift constant $c>0$ reflects the fact that the average motion is to the right (i.e. operators grow over time). The $\nu$ term is simple diffusion. The noise  $\eta$ reflects the randomness in the dynamics. Most importantly, the nonlinear $\lambda$ term is the effect of the constraint (\ref{density constraint}). It reflects the fact that the current is maximal when the density is close to one. The current evidently vanishes when $\rho=0$, since there are no particles, but also when $\rho=2$ (the particles cannot move if the density is everywhere maximal). Therefore we expect $\lambda >0$.

From the above formulas, the density obeys
\be
\partial_t  \rho = \nu \partial_x^2  \rho + \f{\lambda}{2} \partial_x (\rho-1)^2 - \partial_x \eta,
\ee
known as the noisy Burgers equation \cite{HALPINHEALY1995}. The entanglement $S= \int_x ( \rho-1)$ obeys $\partial_t S = J$, leading to the KPZ equation:
\be
\partial_t S = c + \nu \partial_x^2 S - \f{\lambda}{2} (\partial_x S)^2 + \eta.
\ee
The sign of $\lambda$ is in agreement with that obtained from the surface growth picture in Sec.~\ref{surface growth section} and from the directed polymer picture in Sec.~\ref{directed polymer section}. While we have focussed here on dynamics of a restricted type (Clifford), this derivation of KPZ for entanglement provides  independent support for the arguments in the previous sections.

In the language of the particles, the initial state corresponds to uniform density $\rho=1$. Saturation of the entanglement corresponds (neglecting fluctuations) to all of the $r$ particles accumulating on the right hand side and all of the $l$ particles on the left (Fig.~\ref{initial state fig}), i.e to a step function density.

\begin{figure}[t]
 \begin{center}
  \includegraphics[width=0.9\linewidth]{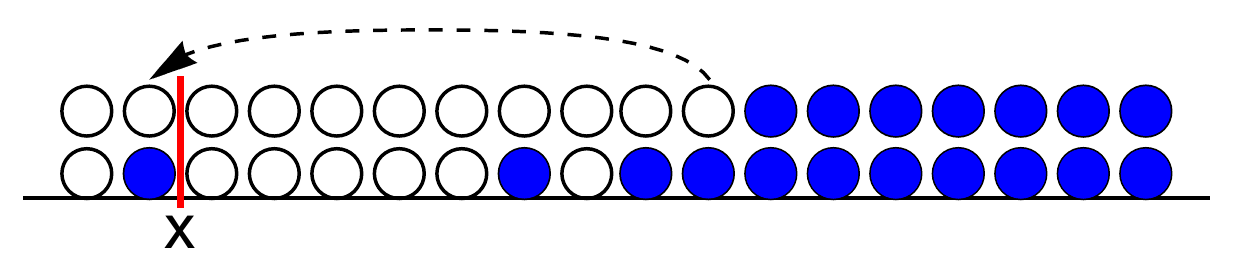}
 \end{center}
\caption{Fluctuations at late times, after saturation of $\<S(x)\>$, in the Clifford case. When $x\ll L/2$ it requires a rare fluctuation (fighting against the net drift) to remove  a particle from region $A$, leading to an exponentially small $S_\text{max}(x) - \<S(x)\>$.}
 \label{instantonfigure}
\end{figure}

As an aside, it is interesting to consider fluctuations in $S(x)$ at late times, i.e. long after the saturation of $\< S(x) \>$. Let us revert to our previous notation, where the system has $L+1$ sites and  bonds are labelled ${x=1,\ldots,L}$. Without loss of generality we take $x \leq L/2$. When fluctuations are neglected, the region to the left of $x$ is empty of $r$ particles, and the entropy is maximal, $S_\text{max} (x) = x$. Fluctuations will reduce the average. But in order for $S(x)$ to fluctuate downward, a blue $r$ particle must diffuse leftward from the right half of the system in order to enter the region to the left of $x$, as in Fig.~\ref{instantonfigure}. This is a fluctuation by a distance $\sim (L/2 - x)$. Such fluctuations are exponentially rare events, because they fight against the net rightward drift for the $r$ particles. Thus when $L/2-x$ is large we expect
\be\label{instanton}
S_\text{max} - \< S(x) \> \sim e^ { - \alpha( L/2 - x )}.
\ee
Our coarse-grained picture does not determine the numerical constants.

The detailed nature of these exponentially small corrections will differ between Clifford circuits and more general unitary circuits.\footnote{For example in the Clifford case $S_n$ is independent of $n$, while in general the corrections will depend on $n$ \cite{NadalMajumdarVergassola2011}.}  Nevertheless the functional form above  agrees with the late time result for generic gate sets, which is simply the mean entanglement in a fully randomized pure state \cite{Page1993, NadalMajumdarVergassola2011}:
\be\label{page}
S_\text{max} - \< S(x) \> \simeq \f{2^{|A|-|\bar A|}}{2 \ln 2 } = \f{4^{-(L/2-x)}}{4\ln 2} .
\ee
$|A|=x$ and $|\bar A|=L-x+1$ are the numbers of sites in $A$ and its complement.

\section{The `entanglement tsunami' and the entanglement speed}
\label{entanglement velocity section}

It is not \textit{a priori} obvious that the rate $v_E$ governing entanglement growth can be viewed as a \textit{speed} in generic systems (see Ref.~\cite{RobertsSwingle2016} for a recent discussion), although this is known to be the case in holographic CFTs \cite{LiuSuh2014}. Our results in the directed polymer picture and in the operator spreading picture suggest that $v_E$ is indeed a well-defined speed in generic systems. As we have seen, there is an appealing visual interpretation of this speed as  the speed at which the operator endpoints move. However, this speed is smaller than the speed  $\tilde v$ which governs the spreading of an operator considered in isolation:  `thermalization is slower than operator spreading'.   

In our formalism the difference between $v_E$ and $\tilde v$ arises because in enforcing Eq.~\ref{density constraint} we `clip' the stabilizers, reducing their rate of growth. We believe this phenomenon to be general and relevant also to non-noisy dynamics. This picture is contrary to that of e.g. Ref.~\cite{ho2015entanglement} where the operator spreading velocity is assumed to determine the entanglement growth rate. In the presence of noise, one may also argue that  a picture of independently spreading operators  underestimates the exponent governing the growth of fluctuations.\footnote{Considering the unitary evolution of a single operator in isolation, its right endpoint executes a biased random walk, traveling an average distance $\tilde v t$ with fluctuations $O(t^{1/2})$. If we were to neglect  the  independence requirement in Eq.~\ref{entropy formula} then the entanglement would be estimated (incorrectly) as the number of independently spreading operators which have reached $A$. The mean of this quantity is $\tilde v t$ and the fluctuations are of order $t^{1/4}$. This is related to the difference between the KPZ universality class of surface growth, which is generic, and the Edwards-Wilkinson universality class which applies when the strength of interactions is fine-tuned to zero \cite{kpz}.}

The language of a `tsunami' is often  used in discussing entanglement spreading, so it is nice to see that --- at least in 1D --- entanglement spreading can be related to a hydrodynamic problem. In higher dimensions the boundary of an operator has a more complicated geometry, so the hydrodynamic correspondence described above does not generalize.

In order to understand the `entanglement tsunami' better, we now return briefly to the directed--polymer--in--a--random--medium picture developed for noisy systems in Sec.~\ref{directed polymer section}. 

\subsection{Simple picture for the entanglement tsunami}

When all length and time scales are large, fluctuations in the entanglement are subleading. Neglecting them is equivalent to saying that the `coarse-grained' minimal cut (prior to saturation) is a straight {vertical} line. This \textit{deterministic} picture generalizes to the entanglement or mutual information of arbitrary regions, and also to higher dimensions (Sec.~\ref{higher dimensions section}). We conjecture that these pictures are valid for the long-time behaviour of entanglement quite generally. The setup relevant to us  in the non-noisy case is a quench, in which the initial state is a ground state of one Hamiltonian, and a different Hamiltonian is  used for the evolution.

In the 1D case, the deterministic scaling form for the entanglement (of an arbitrary region) which results from the leading-order directed polymer picture is rather simple, and is not new --- it agrees with  holographic results \cite{LiuSuh2014, Moosa2015}, and as noted in Ref.~\cite{CasiniLiuMezei2015}, can also be obtained from a more microscopic minimal cut picture in which the geometry of the minimal cut is highly non-unique. (We propose that coarse-graining fixes the geometry of the minimal cut; this will be particularly important in higher dimensions.) Here we consider the scaling of the mutual information, in order to clarify the meaning of the speed $v_E$.

To calculate the entanglement $S_A$ of a region $A$, we must take a  cut, or multiple cuts, with endpoints on the boundary points of $A$ at the top of the spacetime slice. These cuts can either be vertical, in which case they cost an `energy' $v_E t$ (to use the language of Sec.~\ref{directed polymer section}), or they can connect two endpoints, in which case we take them to be horizontal and to have an energy equal to their length. The entanglement $S_A(t)$ is given by minimizing the energy of the cut configuration. It is a continuous piecewise linear function, with slope discontinuities when the geometry of the minimal cut configuration changes.   To generalize the conjecture to systems without noise, we must allow for the fact that the asymptotic value of the entanglement depends on the energy density of the initial state. We therefore  replace the entanglement $S$ in the  formulas with $S / s_\text{eq}$, where $s_\text{eq}$ is the equilibrium entropy density corresponding to the initial energy density \cite{CardyCalabrese2009EntanglementFieldTheory, LiuSuh2014}. This ensures that the entanglement entropy of an $l$-sized region matches the equilibrium thermal entropy when $v_E t \gg l / 2$, as required for thermalization. (Heuristically, $s_\text{eq}$ defines the density of `active' degrees of freedom at a given temperature \cite{RobertsSwingle2016}).

\begin{figure}[b]
 \begin{center}
  \includegraphics[width=0.8\linewidth]{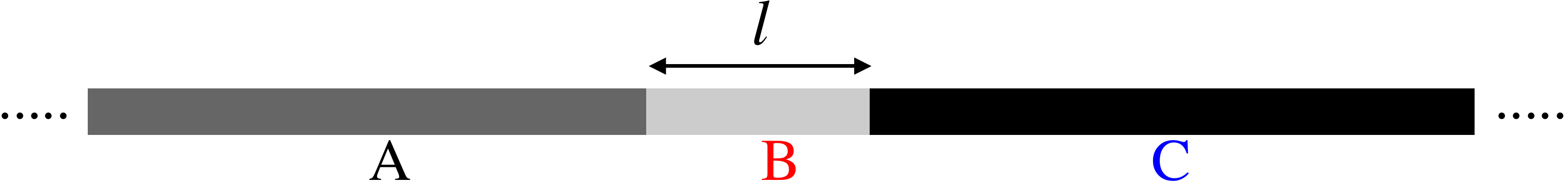}
 \end{center}
\caption{Infinite chain with regions $A$, $B$, $C$ marked. $B$ is of length $l$ while $A$, $C$ are semi-infinite. The mutual information between $A$ and $C$ is nonzero so long as $l < 2 v_E t$: correlations exist  over distances up to  $2 v_E t$, not $v_E t$.}
 \label{infinitechainexample}
\end{figure}

\begin{figure}[t]
 \begin{center}
      \includegraphics[width=0.6\linewidth]{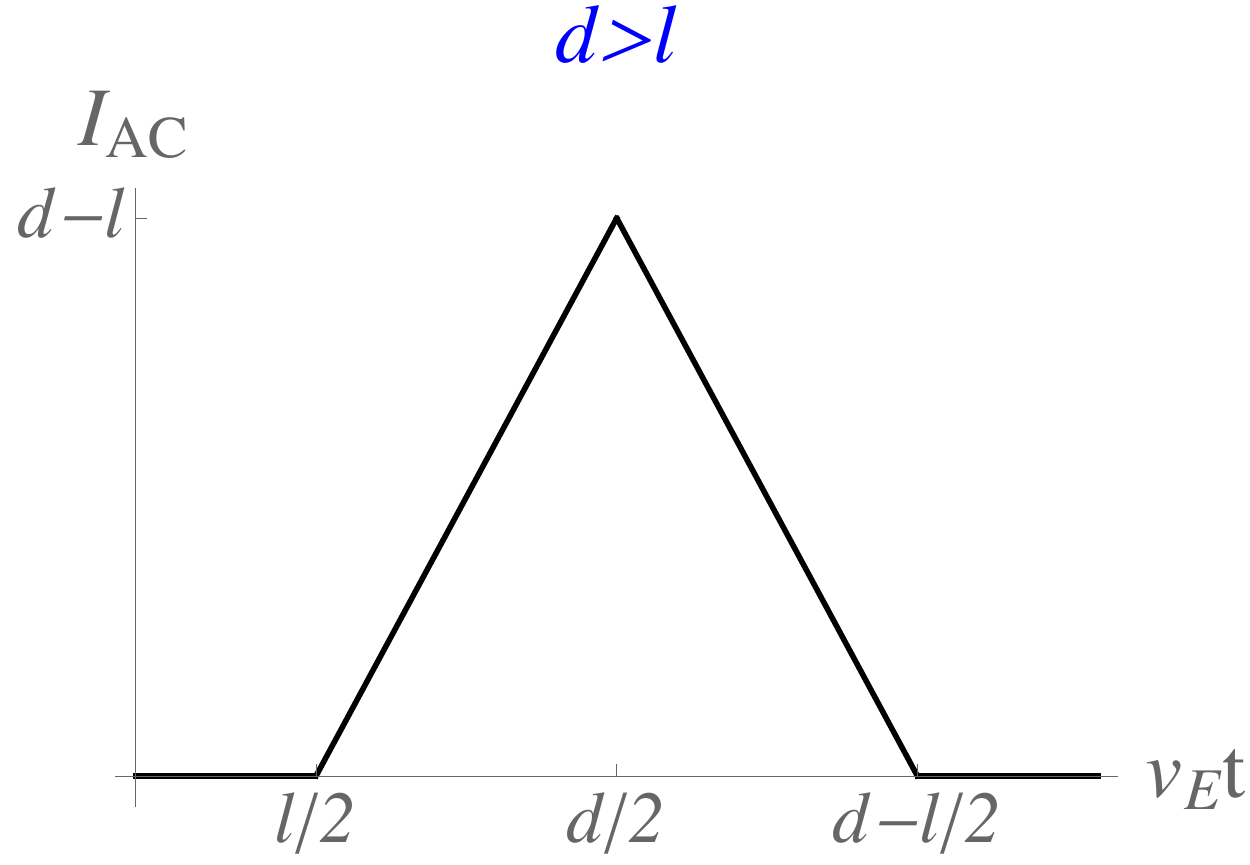}\vspace{3mm}
          \includegraphics[width=0.75\linewidth]{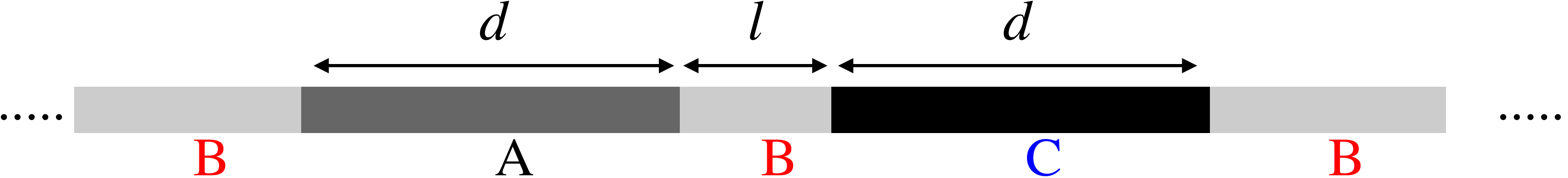}
     \end{center}
\caption{Bottom: Infinite chain with finite regions $A$ and $C$ each of length $d$, separated by distance $l$. Top: The mutual information between $A$ and $C$  in the case $d>l$. In the opposite regime the mutual information vanishes.}
 \label{MIseparatedregions}
\end{figure}

\begin{figure}[]
 \begin{center}
  \includegraphics[width=0.6\linewidth]{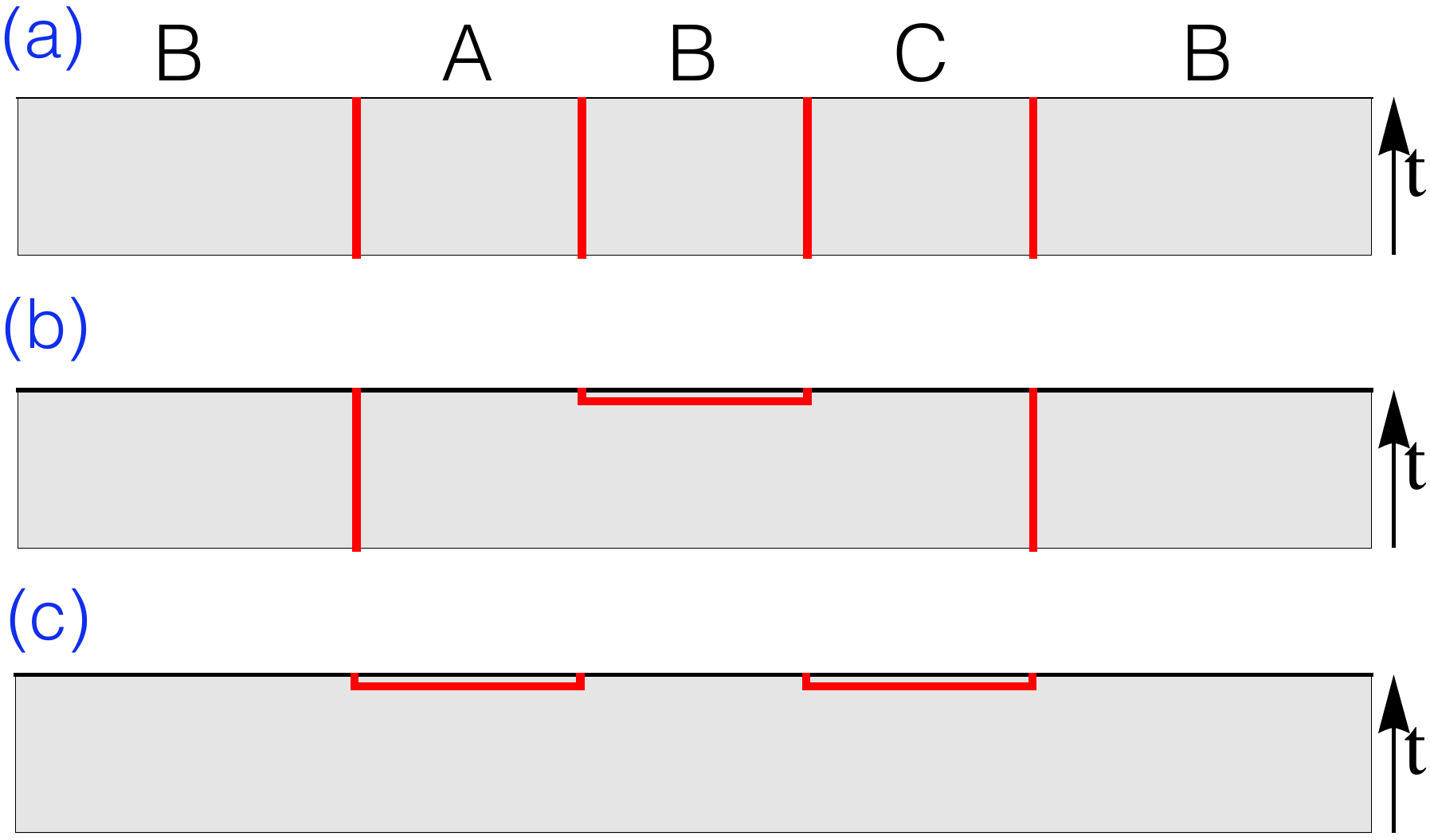}
 \end{center}
\caption{Sequence of minimal cut configurations (red lines) determining the entropy of region $B$ in Fig.~\ref{MIseparatedregions}. $(a)$ gives way to $(b)$ when $2v_E t = l$ and $(b)$ gives way to $(c)$ when $2v_E t + l = 2 d$.}
 \label{min cut sequence}
\end{figure}

\begin{figure}[]
 \begin{center}
   \includegraphics[width=0.49\linewidth]{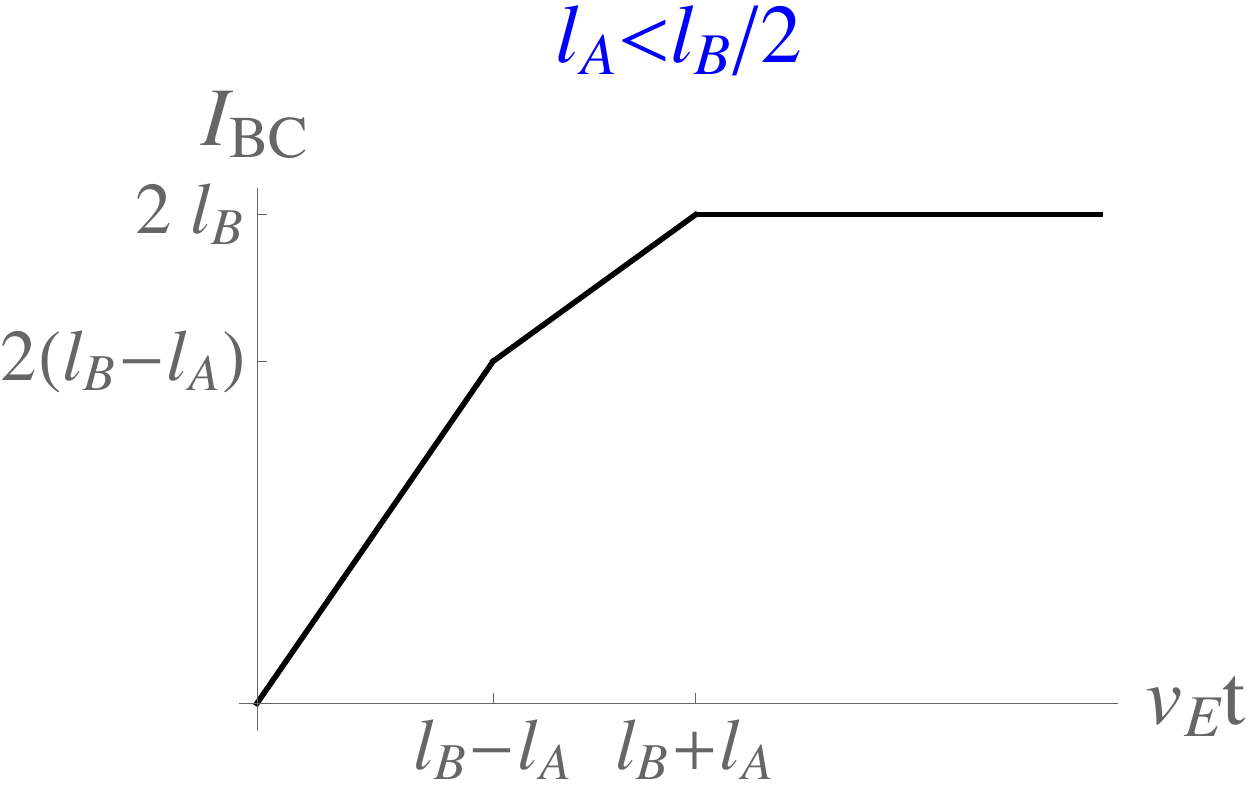}
    \includegraphics[width=0.49\linewidth]{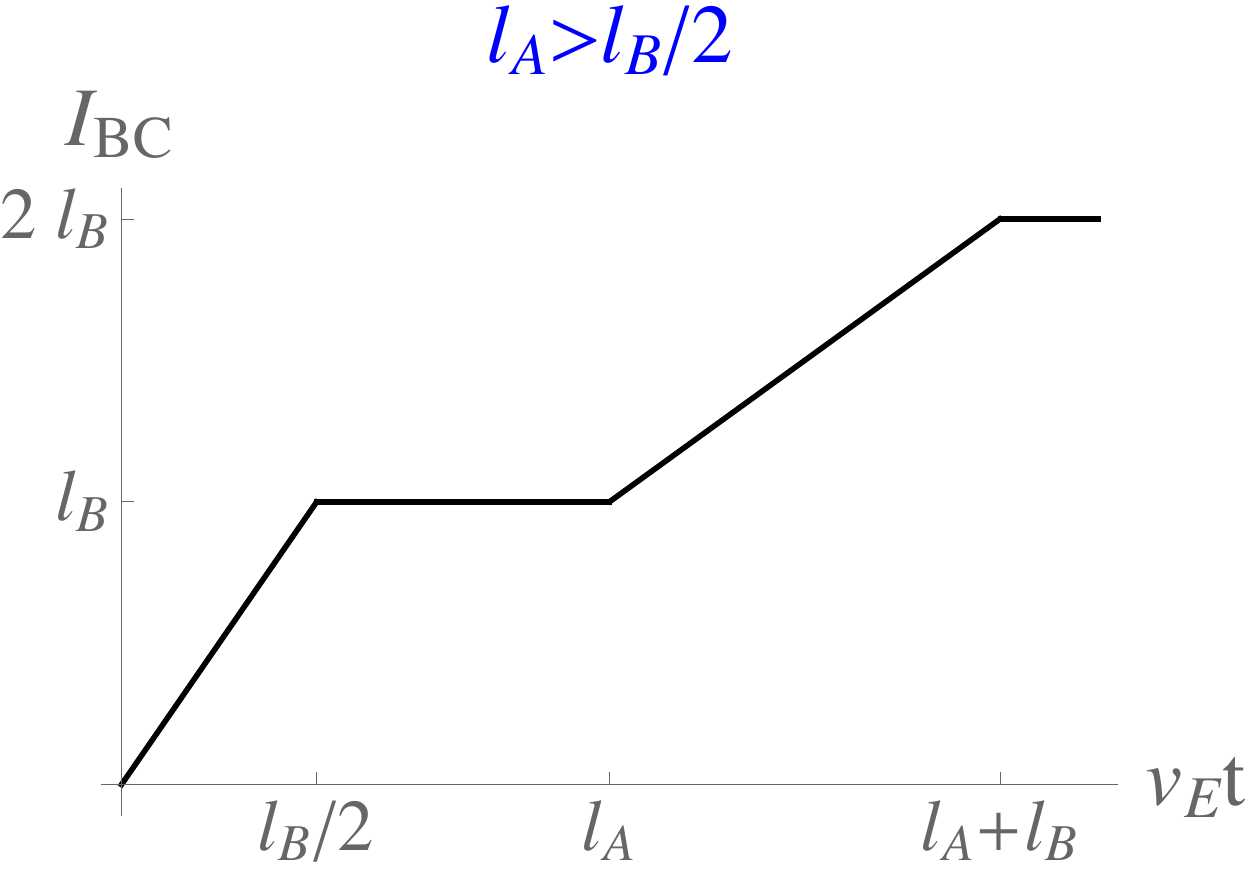}\vspace{3mm}
  \includegraphics[width=0.8\linewidth]{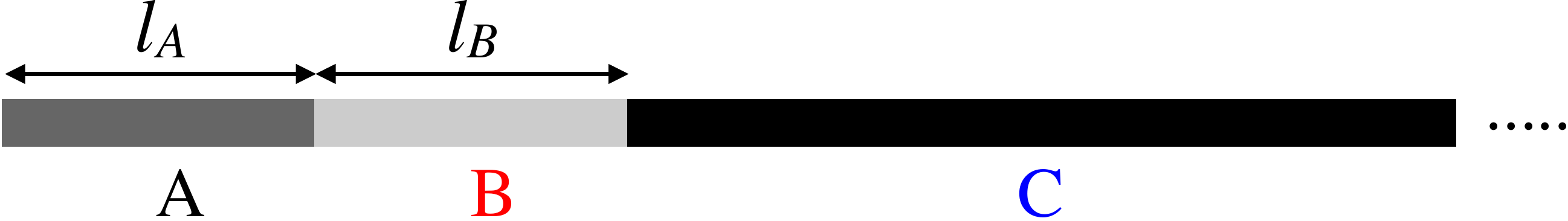}
 \end{center}
\caption{Bottom: Semi-infinite chain with regions $A$, $B$ (length $l_A$, $l_B$ respectively) and $C$ adjacent to the boundary. Top: The mutual information between $B$ and $C$ for this geometry, for the two regimes indicated.}
 \label{finitechainmutualinfoexample1}
\end{figure}

To clarify the meaning of $v_E$, consider the mutual information between two semi-infinite regions that are separated by a distance $l$ (Fig.~\ref{infinitechainexample}). With the labelling of the regions as in the figure, this is given by 
\be
I_{AC} = S_A + S_C - S_{A\cup C} = S_A + S_C - S_B.
\ee
We have $S_A = S_C = v_E t$ for all times, since the appropriate minimal cuts are vertical. If $l > 2 v_E t$, $S_B$ is given by two vertical cuts, so $I_{AC}$ vanishes. When $l< 2 v_E t$, $S_B$ is instead dominated by a horizontal cut, so that  $I_{AC} = 2 v_E t - l$.

The `entanglement tsunami' is sometimes taken to  mean that  at time $t$, a `boundary layer' of width $v_E t$ inside a given region is entangled with the exterior. If this region were maximally entangled with the exterior, this would reproduce the correct value of the entanglement across a  cut ($S=v_E t$). However this picture is not correct: the  result for the mutual information shows that correlations exist over distances up to $2 v_E t$, not $v_E t$. So although $v_E$ is a speed, it should not be thought of as the speed at which the boundary of the entangled region moves. 

Although the rule for calculating the entanglement is almost trivial, the consequences are not always intuitively obvious. First consider the case where the regions $A$ and $C$ above are finite rather than infinite (and embedded in an infinite chain); see Fig.~\ref{MIseparatedregions}. When the length $d$ of the regions $A$ and $C$ exceeds their separation $l$, the time-dependence of the mutual information is as shown in Fig.~\ref{MIseparatedregions}.  [The sequence of minimal cut configurations required for calculating $S_B$ in this case is (a), (b), (c) shown in Fig.~\ref{min cut sequence}.]   By contrast, when the separation $l$ exceeds the length $d$, the mutual information is always zero (or more precisely, exponentially small\footnote{For a simpler example of exponentially small values of the mutual information, consider $\<I_{AC}\>$ at infinite times in a finite system. If the system contains $L$ qubits and $A\cup C$ contains $N$ qubits, the mutual information is exponentially small whenever $N < L/2$, and given by Eq.~\ref{page} as $\<I_{AC}\>\sim(2 \ln 2)^{-1} 2^{-(L-2N)}$.}). [The sequence of cuts for $S_B$ in this case is simply (a), (c)]. 

Finally, consider the effect of a boundary. Take a semi-infinite chain with  regions $A$, $B$, $C$ adjacent to the boundary as in  Fig.~\ref{finitechainmutualinfoexample1} ($C$ is semi-infinite). Consider the mutual information between $B$ and $C$, ${I_{BC}=S_B + S_C - S_A}$. We must distinguish the case $l_A < l_B/2$ from the case $l_A > l_B/2$.\footnote{In the former case the first `event' is that the minimal cut at the boundary of $A$ goes from being vertical to being horizontal; in the latter  the first event is that the two vertical cuts at the boundary of $B$ are replaced by a horizontal one.} The resulting expressions for $I_{BC}$ are plotted in Fig.~\ref{finitechainmutualinfoexample1}.

\section{Numerical results}
\label{numerical results section}

\begin{figure}[t]
 \begin{center}
  \includegraphics[width=0.7\linewidth]{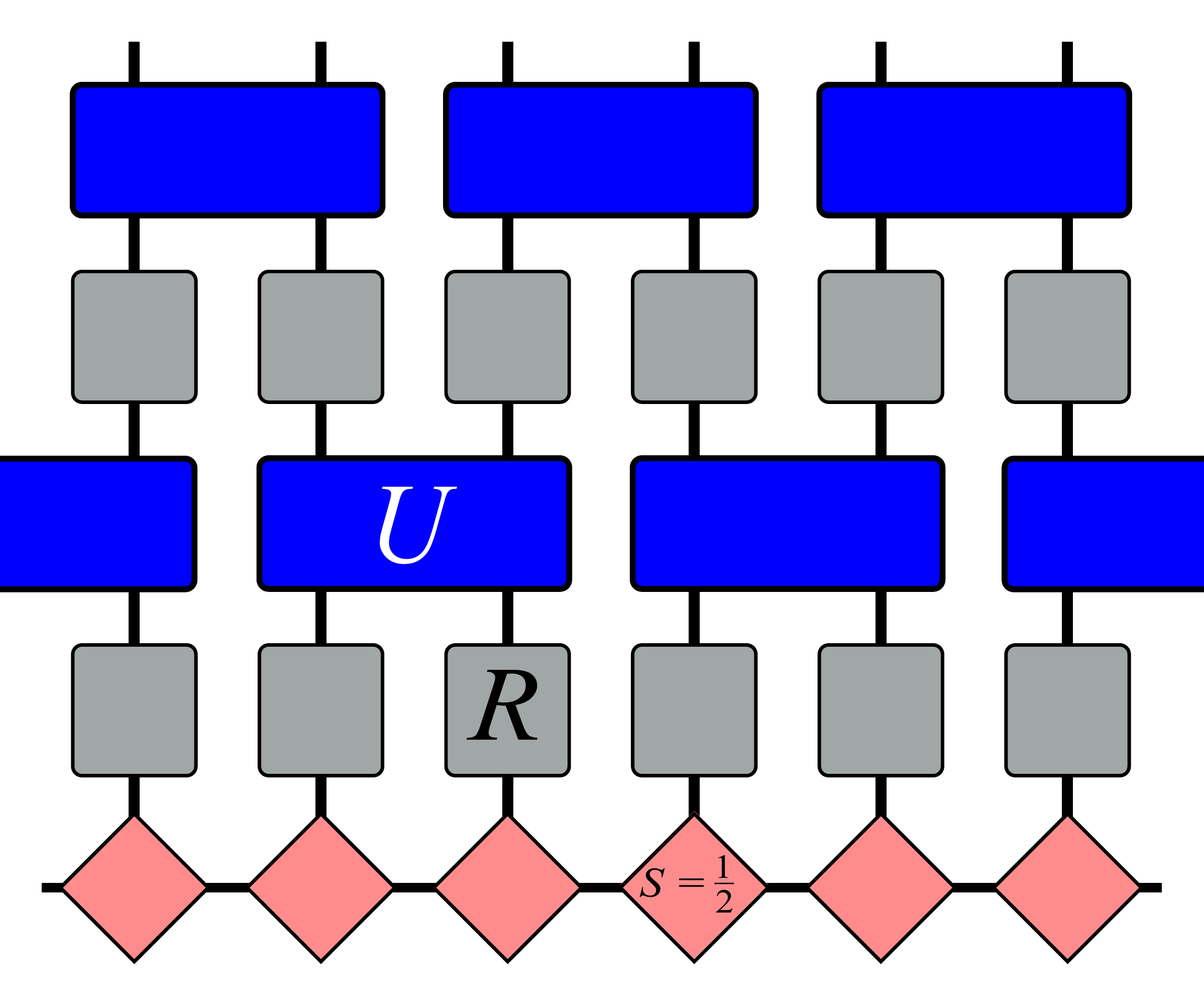}
 \end{center}
\caption{Schematic structure of a layer in the quantum circuits used for simulations.}
 \label{fig:layer}
\end{figure}

We now give numerical evidence that noisy  entanglement growth in 1D is in the KPZ universality class. We study the time evolution of  spin-${1\over 2}$ chains with open boundary conditions, taking the initial state to be a product state with all spins pointing in the same direction (either  the positive $y$ or $z$ direction) and keeping track of the entanglement entropy across each bond during the evolution. The discrete time evolution is  a circuit of one- and two-site unitaries.  Fig.~[\ref{fig:layer}] shows the structure of a single time step: two layers of 2-site unitaries are applied, one layer on odd and one on even bonds, together with single-site unitaries.  Each unitary is chosen independently and randomly (from a certain set specified below). We will use the symbol $R$ to denote a generic single-site unitary, and $U$ to denote a 2-site unitary.

We consider three kinds of dynamics, distinguished by the choice of unitaries.  To begin with we study `\textit{Clifford evolution}' in which the unitaries are restricted to the set of so-called Clifford gates (Sec.~\ref{section: Burgers}). Clifford evolution can be simulated efficiently (in polynomial time) using the stabilizer representation discussed in Sec.~\ref{section: Burgers}. This allows us to access very long times and to pin down KPZ exponents  accurately.
Next we study more general dynamics for which polynomial-time classical simulation is impossible, giving evidence that KPZ behaviour holds more generally. The two types of non-Clifford dynamics studied here are referred to as the `\textit{phase evolution}' and the `\textit{universal evolution}': details are given below. For these dynamics we use a matrix product representation of the state implemented via iTensor~\cite{ITensor}.

The fingerprints of KPZ behaviour that we search for are the two independent critical exponents $\beta$ and $\alpha$ (Sec.~\ref{surface growth section}). We extract $\beta$ both from the fluctuations in the von Neumann entropy and from the corrections to the mean value (Eqs.~\ref{mean growth},~\ref{width growth}), and we extract $\alpha$ from the spatial correlations in the entanglement at distances shorter than the correlation length $\xi(t)$ (Eq.~\ref{spatial correlator}).

\subsection{Clifford evolution}\label{section:numerics:clifford}

Clifford circuits, or `stabilizer circuits', are a special class of quantum circuits which play an important role in quantum information theory. As shown by  Gottesman and Knill, they can be simulated efficiently, even when the entanglement entropy grows rapidly, by representing the quantum state in terms of stabilizers~\cite{Gottesman1996Saturating}: see Sec.~\ref{section: Burgers}. 

The time evolution operator for a Clifford circuit belongs to the Clifford \textit{group}, a subgroup of the unitary group on the full Hilbert space. This group may be generated by a small set of local Clifford gates: the two-site controlled NOT gates (Eq.~\ref{CNOT}) and the single-site Hadamard and Phase gates $R_H$ and $R_P$  (Eqs.~\ref{RH},\ref{RP}). For circuits built from these gates, time-evolving the state on $L$ spins up to a time $t$ takes a computational time of order $Lt$ and measuring the entanglement across a given bond in the final state takes a time of order $L^3$ at most.  This is in sharp contrast to the exponential scaling which is inevitable for more general circuits.

In all our simulations, each \textit{two}-site unitary $U$ in the circuit is chosen with equal probability from three possibilities: the two types of CNOT gate (Eq.~\ref{CNOT}) and the identity matrix:
\be\label{2 site choices}
U \in
\left\{\mathbb{1} , \text{CNOT}^{(L)}, \text{CNOT}^{(R)}\right\}.
\ee
 In this section we discuss the simplest Clifford dynamics, which includes \textit{only} these gates, and no 1-site unitaries ($R = \mathbb{1}$). When the initial state is polarized in the $y$--direction, this set of gates is sufficient to give nontrivial entanglement evolution, with universal properties that turn out to be the same as those for more generic gate sets. We have also studied the `full' Clifford dynamics in which all the Clifford generators are used, choosing the single site unitaries randomly from the three options
 \be
 R\in \{\mathbb{1},\,R_H,\,R_P\}
 \ee
 Results for this case are similar and are given in Appendix.~\ref{appendix: full clifford}.

\begin{figure}[h]
 \begin{center}
  \includegraphics[width=0.9\linewidth]{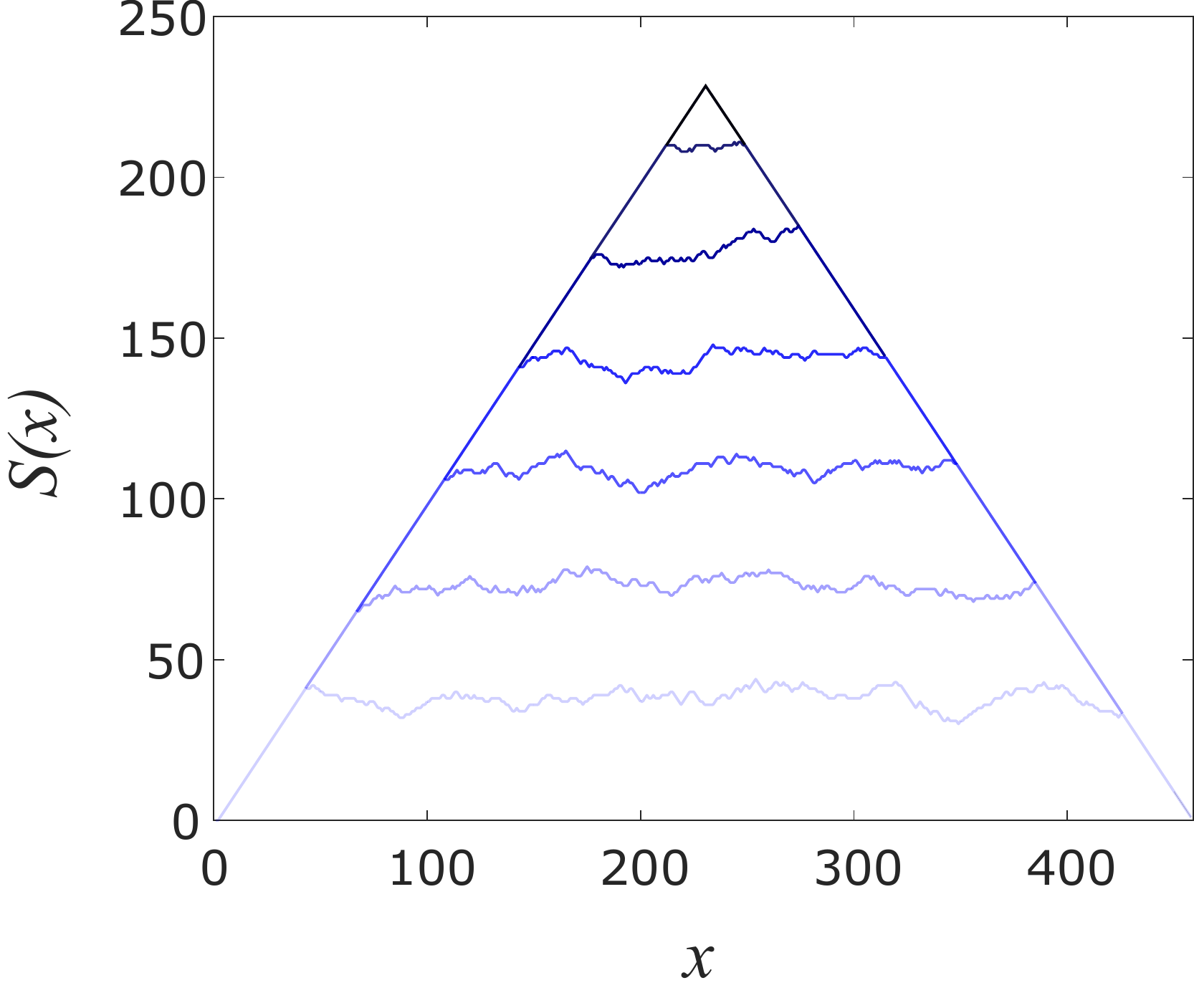}
 \end{center}
\caption{The von Neumann entropy $S(x,t)$ for a system of length $L = 459$, as a function of $x$, for several successive times ($t = 340, 690,1024,1365,1707,2048\; \mathrm{and}\; 4096$), in the Clifford evolution. The figure shows how the state evolves from a product state to a near-maximally entangled one. Prior to saturation the entanglement displays KPZ-like stochastic growth. $S(x,t)$ is in units of $\log 2$.}
 \label{fig:Tower}
\end{figure}

To begin with, Fig.~[\ref{fig:Tower}] shows the evolution of the bipartite von Neumann entropy $S(x)$ (in units of $\log 2$) for a single realization of the noise (i.e. a particular random circuit) in a system of  $L=459$ sites. The  curves show successively later times. Note that the entropy saturates more rapidly closer to the boundary, because the maximum entanglement across a bond is proportional to its distance from the boundary. At very late times $S(x,t)$ saturates to a pyramid-like profile representing close-to-maximal entanglement. Our interest is in the stochastic growth \textit{prior} to saturation, which we will show is KPZ--like. All observables in the following are measured far from the boundary, in order to avoid finite-size effects associated with saturation.

\begin{figure}[h]
 \begin{center}
  \includegraphics[width=0.9\linewidth]{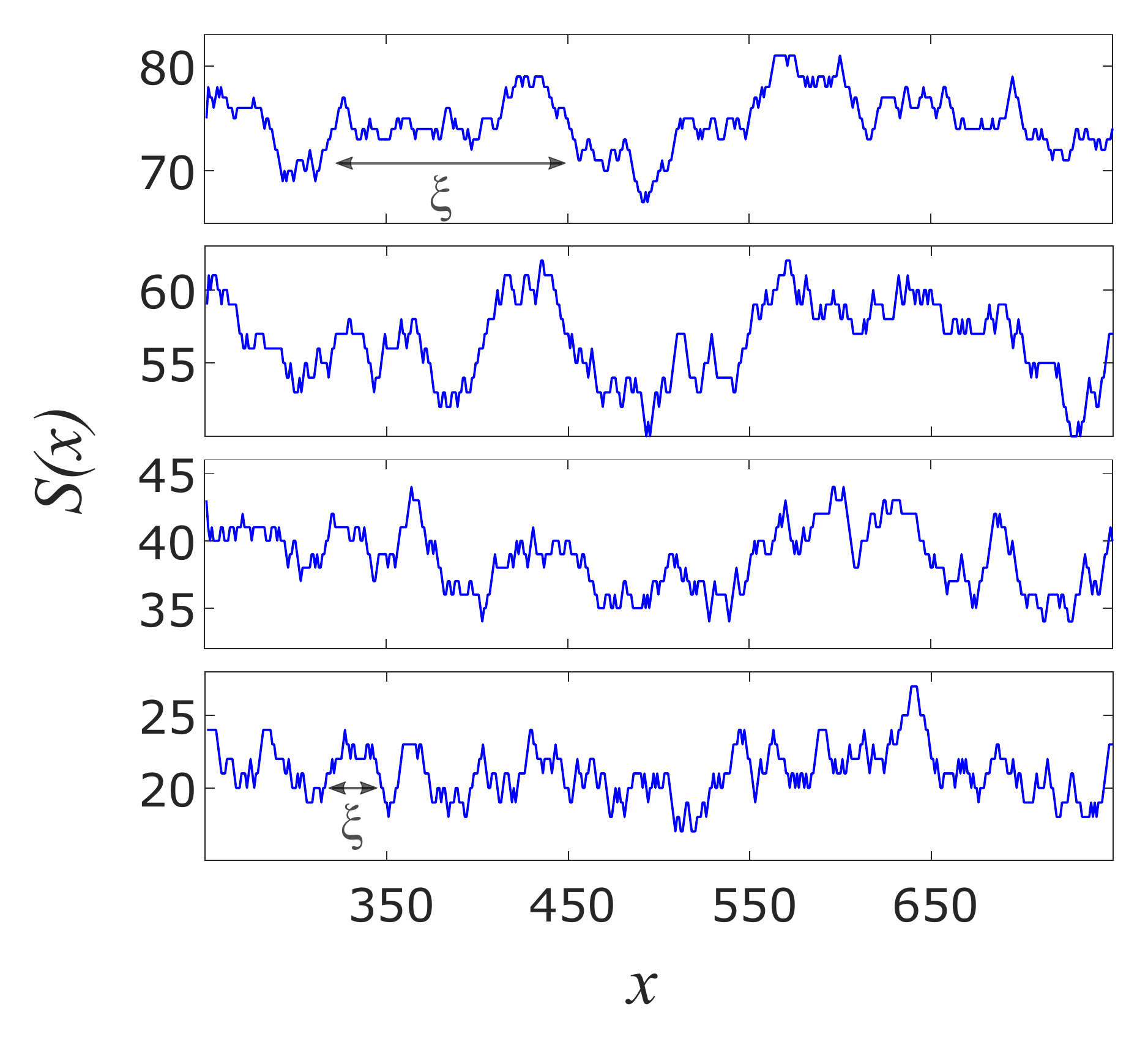}
 \end{center}
\caption{The von Neumann entropy $S(x,t)$ in units of $\log 2$, far from the boundaries, in a system of length $L = 1025$ at various times (from bottom to top $t = 170,340,512\; \mathrm{and}\;682$) evolved with the Clifford evolution scheme. $\xi$ schematically shows the typical correlation length Eq.~[\ref{correlation length}] which grows in time like $t^{1/z}$.  }
 \label{fig:universalevolutionplot}
\end{figure}

Fig.~[\ref{fig:universalevolutionplot}] shows successive snapshots for a subregion of a larger system of  $L = 1025$ bonds (times {$t = 170,340,512,682$}, from bottom to top). The maximal slope that can appear is 1, in accord with Eq.~\ref{von neumann constraint}. Note the gradual roughening of the surface and the growing correlation length.

    \begin{figure}[h]
 \begin{center}
  \includegraphics[width=0.9\linewidth]{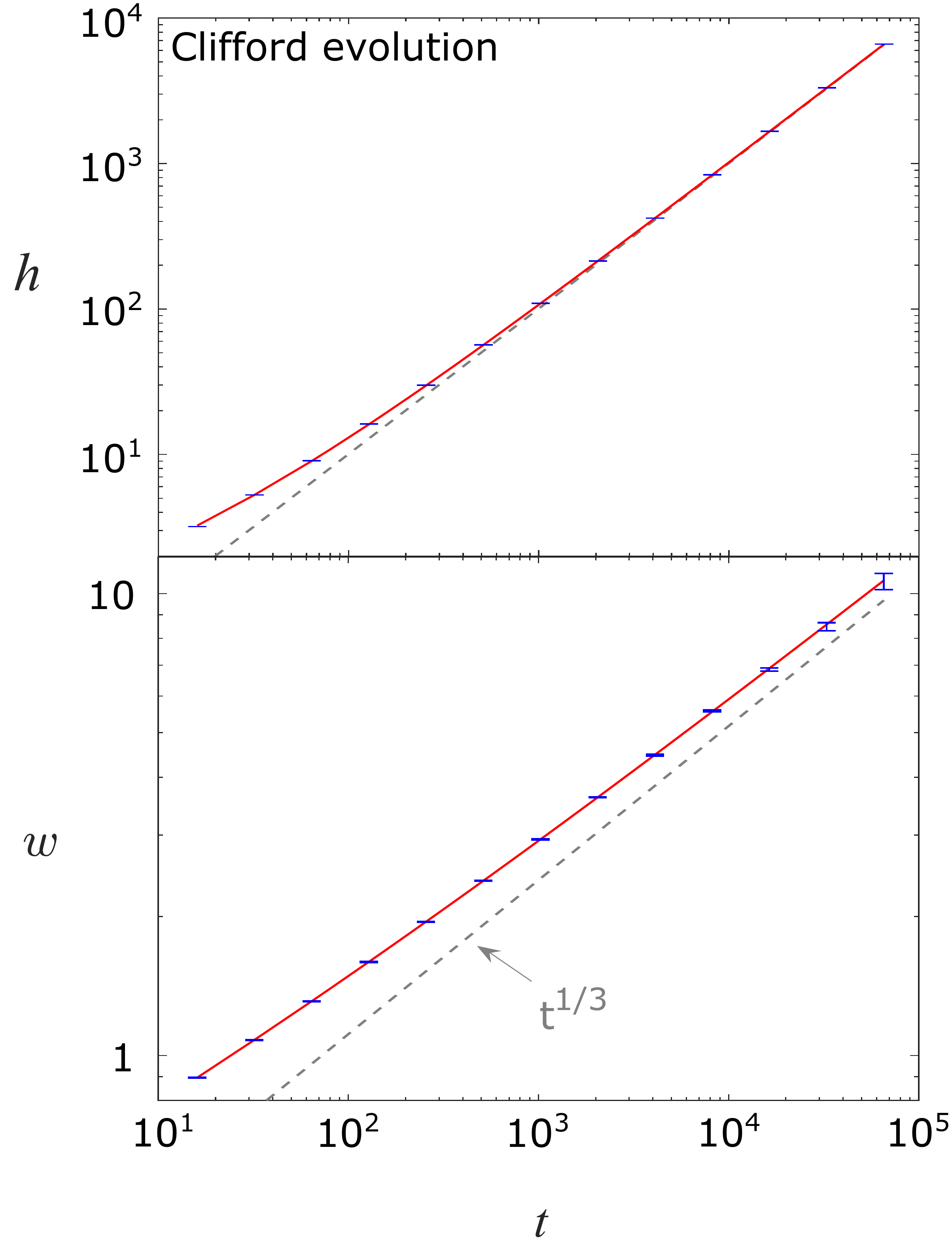}
 \end{center}
\caption{Top: Growth of the mean entanglement with time for the Clifford evolution with only CNOT gates (in units of $\log 2$). The  solid  red curve is a fit using to Eq.~[\ref{fit}]. The exponent $\beta$ is found to be $\beta=0.33 \pm 0.01$, in agreement with the KPZ prediction $\beta=1/3$. Dashed line shows asymptotic linear behaviour. Bottom: Growth in the fluctuations in the entanglement with time. The dashed line shows the expected asymptotic behaviour, ${w(t)} \sim t^{\beta}$ with $\beta=1/3$. The fit includes a subleading correction:  Eq.~[\ref{fit}], with $\b =0.32 \pm 0.02 $. Error bars denote the 1$\sigma$ uncertainty.}
 \label{fig:CNOTheightwidth}
\end{figure}

\begin {table*}
\caption {Summary of all fitting parameters to Eq.~\ref{fit} used in this section. The Errors set the estimated $2\sigma$ uncertainty. } \label{tab:fit}
\begin{center}
    \begin{tabular}{| l | l | l | l | l | l | l | l |}
    \hline
    Evolution type & $v_E$ & $\beta_h$ &  $B$ & $\beta_w$&$C$ & $\eta$&$D$ \\ \hline
    Clifford &$0.1006 \pm 0.0001 $  & $0.33 \pm 0.01$ &$0.66\pm 0.04$& $0.32 \pm 0.02$&$0.28 \pm 0.05 $& $ 0.08 \pm 0.1$&$0.16\pm 0.04$\\ \hline
   Phase & $0.133\pm0.03$ & -- & $0.54\pm0.04$&--& $0.223\pm 0.004$&--&$0.168 \pm 0.03$\\ \hline
    Universal & $0.202\pm0.001$ & -- & $0.09\pm0.005$&  -- &$ 0.14\pm 0.003$&--&$0.36 \pm 0.01$\\
    \hline
    \end{tabular}
    \end{center}
    \end {table*}

Fig.~\ref{fig:CNOTheightwidth} shows the `height' and `width' of the growing surface,
\ba
h(t) &= \langle S_{\mathrm{vN}}(x,t) \rangle,  &
w(t) &= \sqrt{\langle \langle S_{\mathrm{vN}}^2(x,t) \rangle \rangle}
\end{align}
for very long times. These quantities have been averaged over at least $10^5$  realisations. In each realisation only the entanglement across the center bond is used (therefore all data points are uncorrelated) and the system size is $L=a t$, where $a$ is chosen to avoid finite size effects (see Appendix~\ref{appendix finite size effects}). We obtain estimates $\beta_h$ and $\beta_w$ of the exponent $\b$ by fitting the data to the expected forms (cf. Eqs.\ref{mean growth},\ref{width growth}):\begin{align}
 &h(t)  = v_E \,t+B \, t^{\b_h}, &
 &{w}(t) = C\, t^{\b_{\mathrm{w}}}+ D\, t^\eta.\label{fit}
 \end{align}
Here $\eta$ (with $\eta<\beta_w$) captures subleading corrections. We find:
\ba
 \beta_h &= 0.33 \pm 0.01,
 &
 \beta_{{w}} & = 0.32 \pm 0.02.
 \end{align}
Both estimates of $\beta$ are in excellent agreement with the KPZ value $\beta = 1/3$. The solid lines in Fig.~\ref{fig:CNOTheightwidth} show the fits (the fit parameters are in Table~\ref{tab:fit}). The  dashed lines show the slopes corresponding to the expected asymptotic power laws, $h(t)\sim t$ and $w(t)\sim t^{1/3}$.

The analysis in Sec.~\ref{section: Burgers} implies that $v_E$ is a well-defined velocity, and $v_E t$ is a sharply-defined \textit{lengthscale} characterizing the  range of entanglement in the state.  We may confirm this by measuring this lengthscale directly. In Appendix~\ref{appendix finite size effects} we do this by checking the scaling form for the saturation behaviour of the entanglement given in Sec.~\ref{saturation polymer section}.

\begin{figure}[t]
 \begin{center}
  \includegraphics[width=0.9\linewidth]{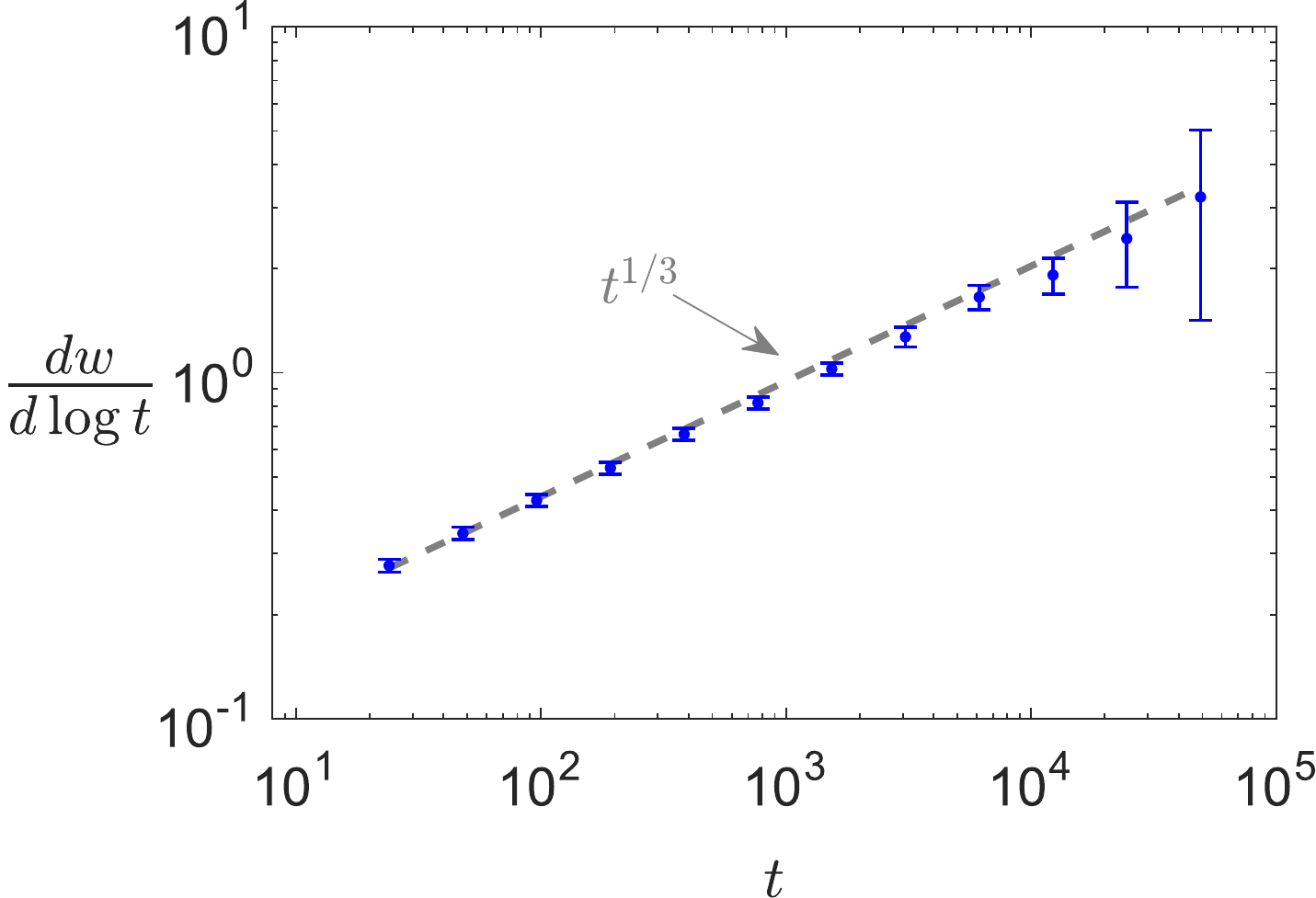}
 \end{center}
\caption{The logarithmic derivative of the width, $ d w / d \log t$, vs. time for the Clifford evolution. The universal behavior with exponent $t^{ 1/3}$ is observed at shorter time scales compared with Fig.~\ref{fig:CNOTheightwidth}.}
 \label{fig:derivative_CNOT}
\end{figure}

Note the small value of the subleading exponent $\eta$ obtained from the fit. This implies that finite time corrections are reduced if we plot the numerical derivative $ d w / d \log t$ rather than $w$ itself (both quantities scale as $t^{1/3}$ at long times). This is done in  Fig.~\ref{fig:derivative_CNOT}. The data fits well to the $t^{1/3}$ law even at short times. This will be useful for the more general dynamics where long times are not available.

\begin{figure}[t]
 \begin{center}
  \includegraphics[width=0.85\linewidth]{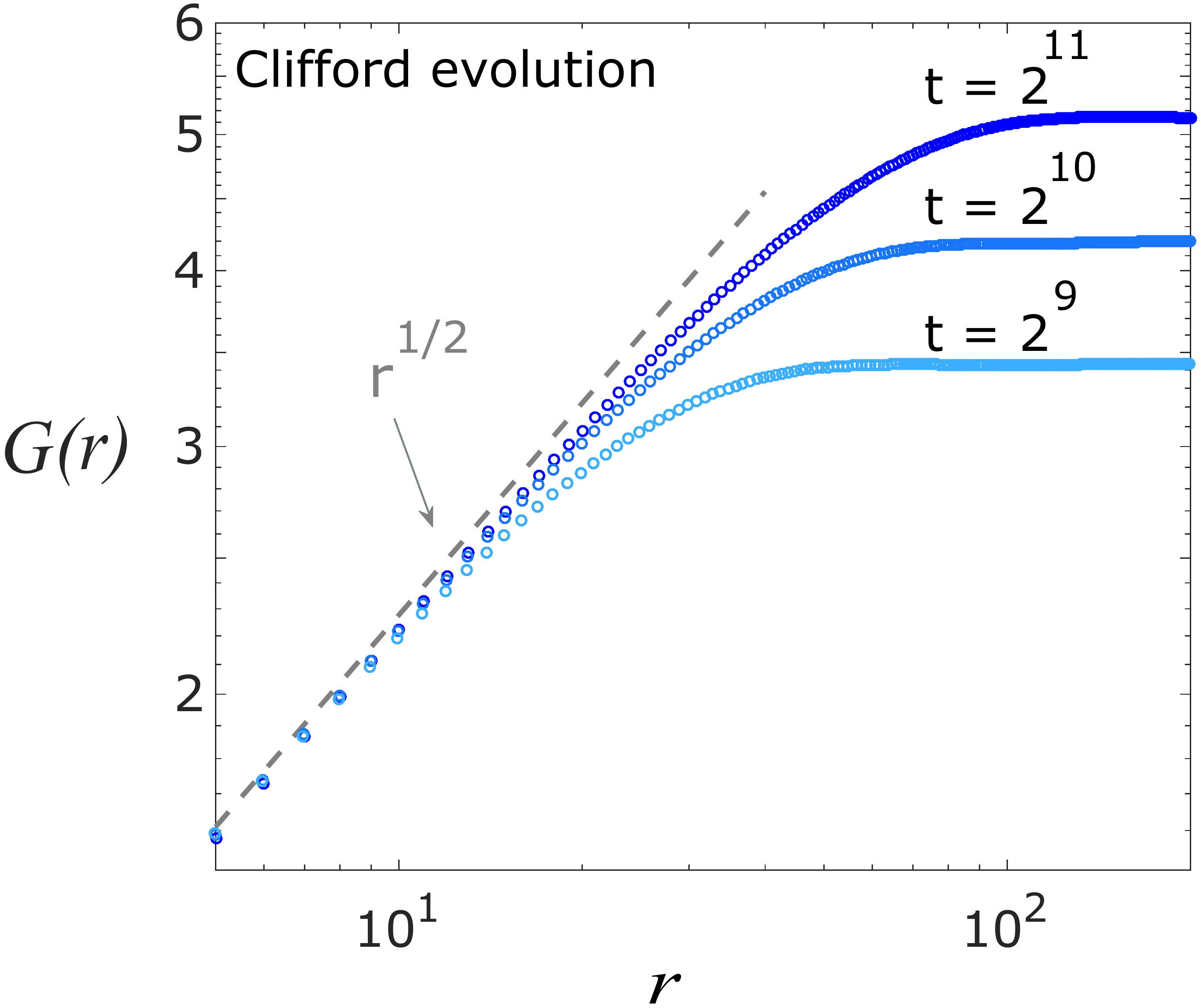}
 \end{center}
\caption{Correlation function $G(r) = \langle [S(r) - S(0)]^2 \rangle^{1/2}$ at time $t=512,1024\; \mathrm{and}\; 2048$ for the Clifford evolution, showing excellent agreement with the KPZ prediction $G(r) \sim r^{\chi}$ with $\chi=1/2$ in the regime $r\ll \xi(t)$.}
 \label{fig:CNOTcorrelator}
\end{figure}

Finally, Fig.~\ref{fig:CNOTcorrelator} shows the spatial correlator $G(r)$ defined in Eq.~\ref{spatial correlator}, as a function of separation $r$, for three successive times.  For small $r$ the correlation grows like a $r^{\a}$ with $\a \simeq 1/2$, in agreement with the KPZ prediction for this exponent. For distances $r \gg \xi(t)$, the correlator saturates to a value proportional to $w(t)$. The figure  gives an idea of the size of the correlation length $\xi(t)$ for these times.

\subsection{Universal and Phase evolution}

The `Phase' and `Universal' dynamics take us outside the Clifford realm, and  cannot be simulated efficiently on a classical computer (in polynomial time). We will give evidence that the correspondence with KPZ continues to hold in this more generic situation.
However, our results will not be as conclusive as in the Clifford evolution as we will not have access to such long times.

The simulations are performed on spin-1/2 chains of length $L = 500$ bonds (501 spins)  using the {iTensor} package \cite{ITensor}. The two types of dynamics are defined as follows. (The 2-site unitaries are always chosen from the set in Eq.~\ref{2 site choices}; the initial state is taken polarized in the $y$-direction.)

{{\it Phase evolution}: Each single-site unitary is  chosen randomly and uniformly from the set of eightfold rotations about the $z$ axis in spin space: ${R = \exp\left(\pi i n \s^z /8 \right)}$ with ${n\in {1,\ldots, 8}}$.}

{{\it Universal evolution}: This set of gates, unlike the others, is `universal' in the quantum information sense (any unitary acting on the full Hilbert space of the spin chain can be approximated, arbitrarily closely, by a product of gates from this set). The single-site gates include the eightfold rotations mentioned above, together with the Hadamard gate $R_H$ (\ref{RH}). $R_H$ is applied with probability 1/2 and the rotations with probability $1/16$ each.}

\begin{figure}[t]
 \begin{center}
  \includegraphics[width=0.8\linewidth]{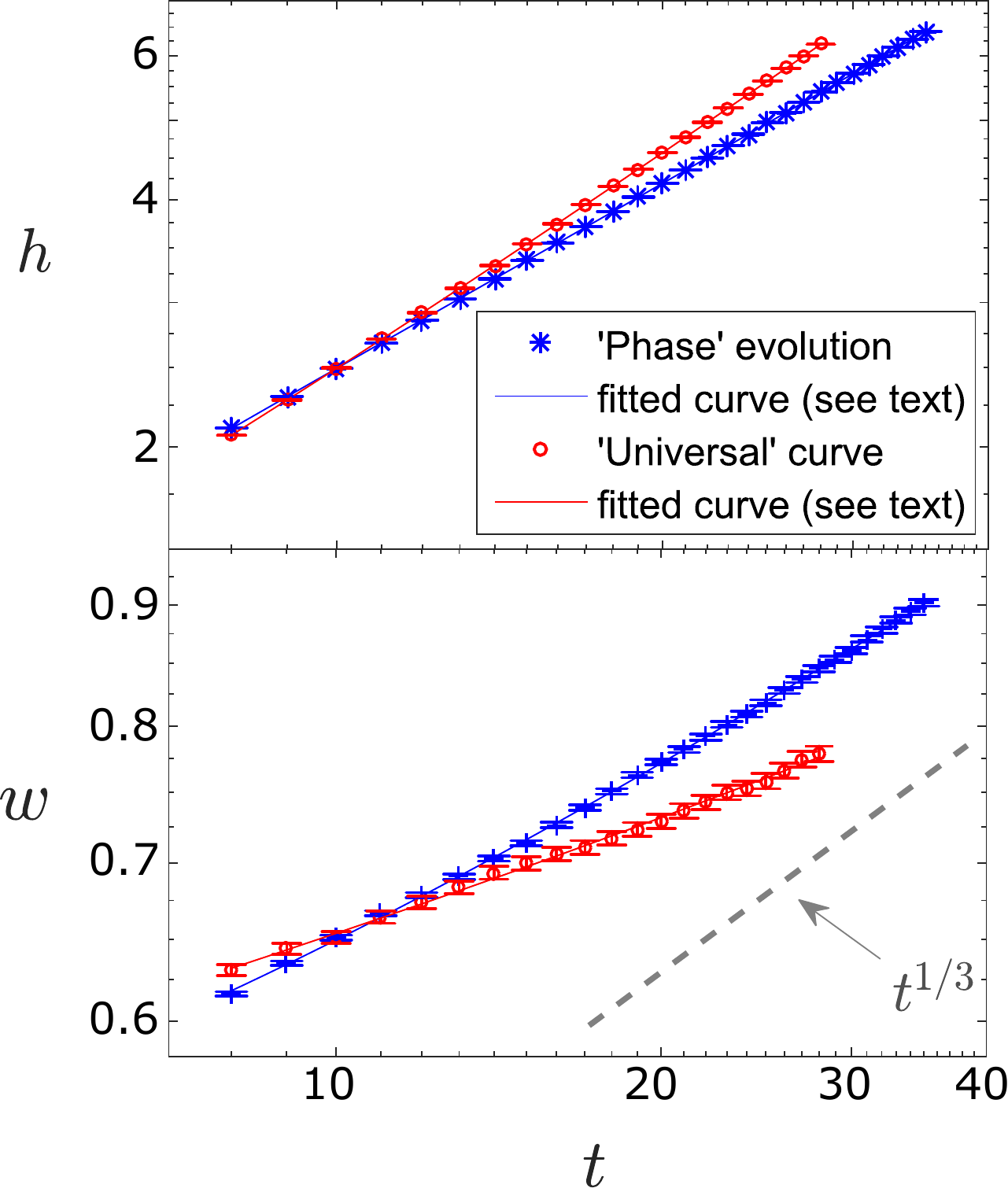}
 \end{center}
\caption{Top: Growth of the mean entanglement as a function of time for the universal and phase gate set fitted to Eq.~\ref{fit} with $\b$ set to $1/3$. The dashed line shows the expected asymptotic behaviour for comparison. Error bars indicate one standard deviation (1$\sigma$) uncertainty.}
 \label{fig:universalheightwidth}
\end{figure}

Fig.~[\ref{fig:universalheightwidth}] shows the height and width $h(t)$ and $w(t)$ for the two protocols (averaged over 380 realisations for the Phase evolution and 200 realisations for the Universal evolution, and over bonds $x$ with $20<x<480$).  The figure shows fits to the forms in Eq.~\ref{fit} with $\beta_h$ and $\beta_w$ \textit{fixed} to the KPZ value and $\eta$ fixed to zero (fit parameters are in Tab.~\ref{tab:fit}). The fits with Eq.~\ref{fit} are consistent with the data. It is not possible to  extract  precise estimates for $\beta$ from the slope of the log-log plot of $w(t)$, although for the Phase evolution the slope at late times is in reasonable agreement with the expected KPZ value, shown by the  dashed grey trendline.

\begin{figure}[t]
 \begin{center}
  \includegraphics[width=1\linewidth]{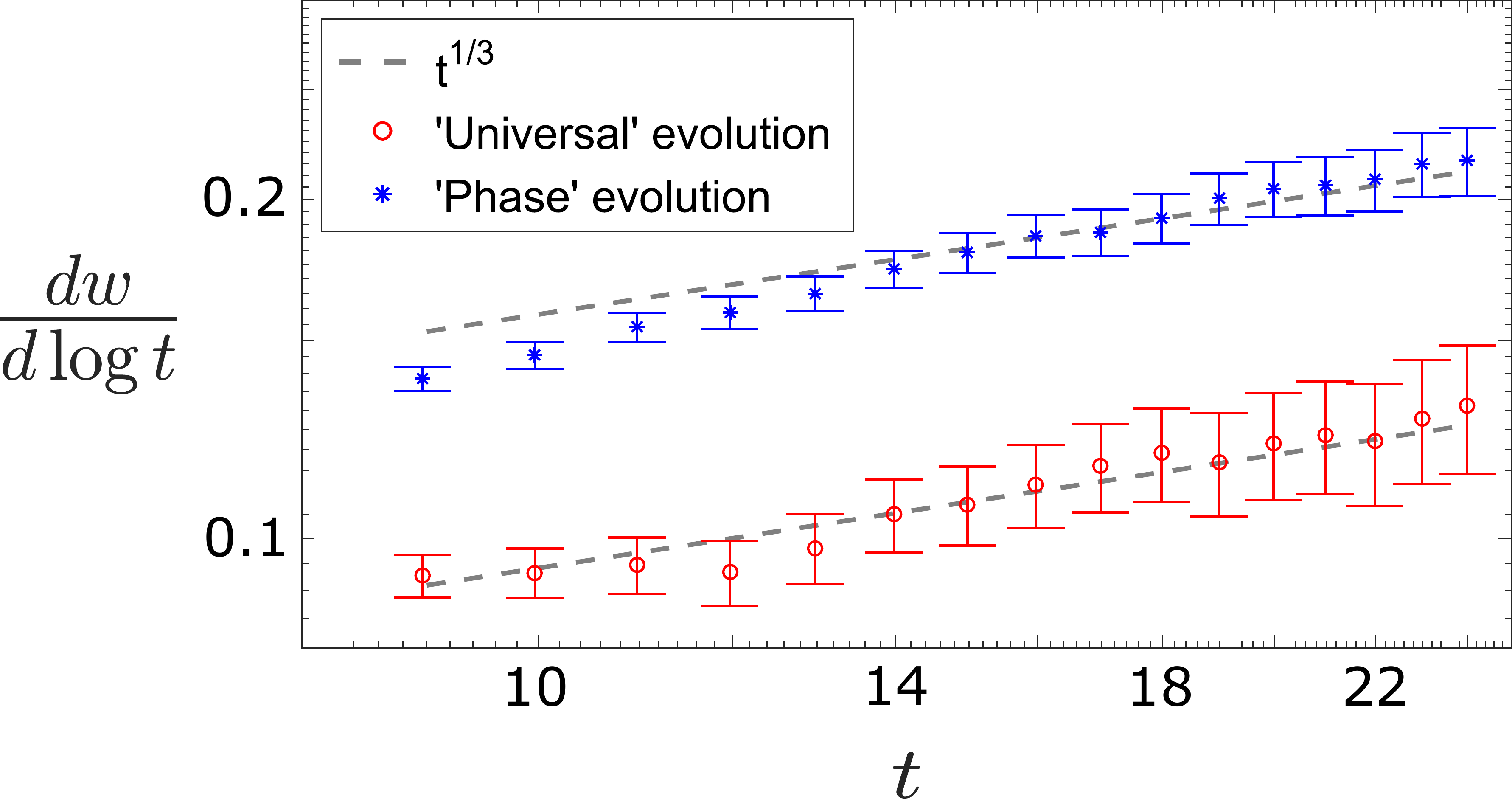}
 \end{center}
\caption{The logarithmic derivative of the width $ d w / d \log t$ vs. time for the Phase and Universal evolution protocols. For comparison we plot the universal behavior with exponent $t^{ 1/3}$ in grey (dashed). (The derivative is calculated using  three data points. Errors are estimated from maximal and minimal slopes obtained within one standard deviation from the averaged data points.)}
 \label{fig:derivative_universal}
\end{figure}

For an alternative attack on $\beta$ we plot the numerical derivative $d w(t) / d \ln t$. We found in the Clifford case that the slope of this quantity (when plotted against time on a log-log plot) had smaller finite size corrections than the slope for $w(t)$ itself. The corresponding plot is shown in Fig.~\ref{fig:derivative_universal}, for times up to $t=25$ (averaging over more than 5000 realisations).  The  dashed grey lines are the $t^{1/3}$ trendlines. Results for both types of dynamics are in good agreement with the expected slope $\beta = 1/3$.

\begin{figure}[t]
 \begin{center}
  \includegraphics[width=0.9\linewidth]{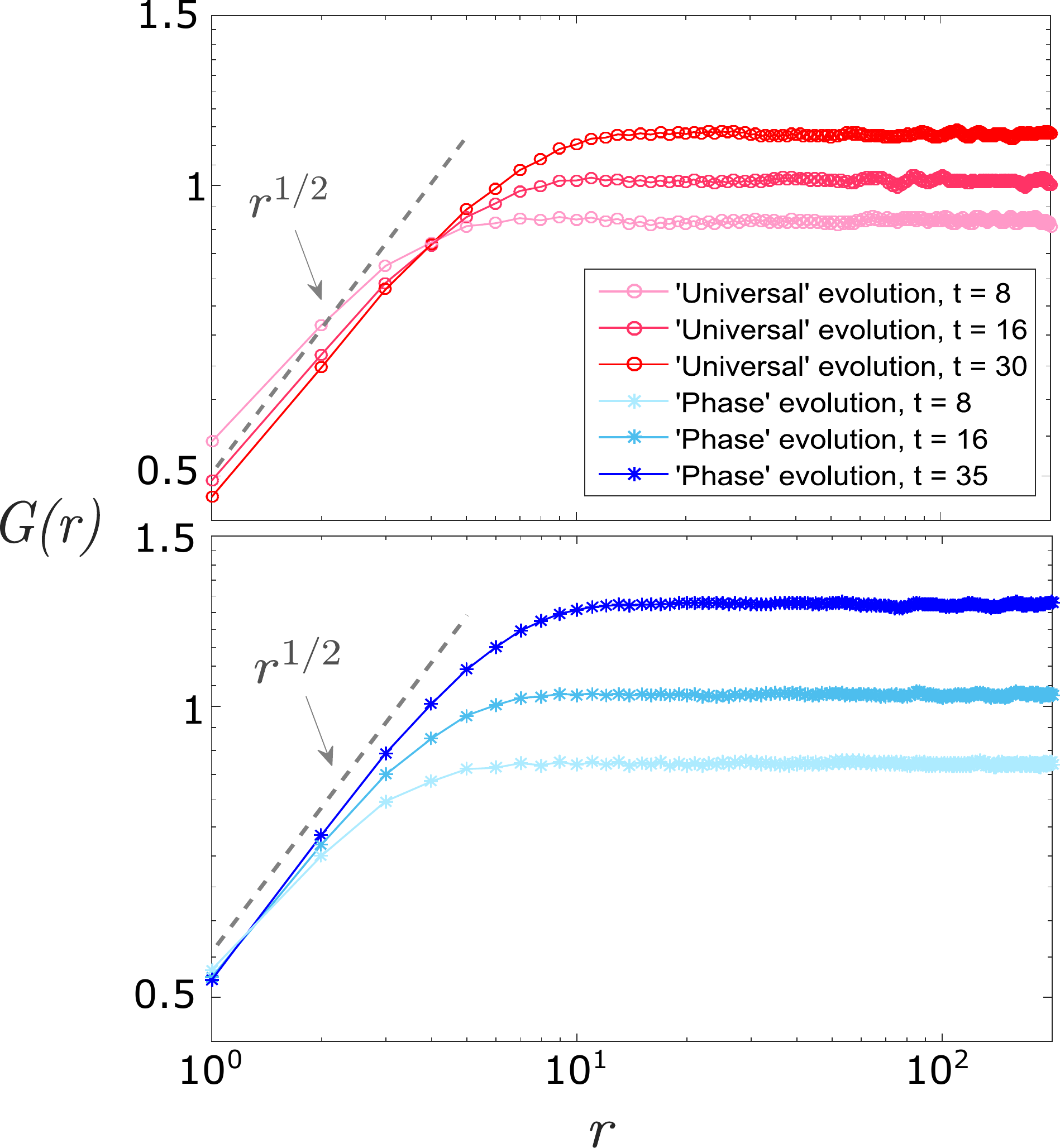}
 \end{center}
\caption{Correlation function $G(r) = \langle [S(r) - S(0)]^2 \rangle^{1/2}$ at three values of the time  for the Phase (top) and Universal (bottom) gate sets, showing good agreement with the KPZ exponent  value $\alpha=1/2$.}
 \label{fig:universalcorrelator}
\end{figure}

Next we examine the spatial correlator (\ref{spatial correlator}) in Fig.~\ref{fig:universalcorrelator}.
 For both types of dynamics, the behaviour for $r \ll \xi(t)$ agrees well with the KPZ exponent value $\a= 1/2$ at the largest available time.

The very long times accessible in the Clifford simulation allowed us to establish KPZ exponents with high accuracy there. For the more generic dynamical rules we cannot reach the same level of precision, but nevertheless the KPZ exponent values are compatible with the data.

Next we briefly discuss a fine-tuned situation in which entanglement dynamics are not KPZ-like: namely when the system is made up of free particles. Then in Sec.~\ref{higher dimensions section} we  move to higher dimensions.

\section{Free fermions are non-generic}
\label{free fermions}

The growth of entanglement in systems of \textit{free} particles is highly non-generic. In the presence of noise the entanglement of a system of free particles on the lattice grows only as $S\sim \sqrt{t}$, in contrast to the behaviour $S\sim t$ of generic systems. The case of spatially homogeneous noise has been discussed recently \cite{Igloi2016}. The basic point is the same when the noise varies in space: the fact that the single-particle wavefunctions spread diffusively in the presence of noise implies that the entanglement cannot be larger than $O(\sqrt t)$ \cite{Igloi2016}.

As a concrete example, consider a short-range hopping Hamiltonian for free fermions,
\be
H(t) = \sum_{ij} H_{ij}(t) c^\dag_i c_j,
\ee
with noisy matrix elements $H_{ij}(t)$. For simplicity, take the initial state to consist of particles localized at sites $i\in S$ for some set $S$; for example we could take $S$ to consist of all the even-numbered sites:
\be
\ket{\Psi(0} = \prod_{i\in S} c_i^\dag \ket{0}.
\ee
Under the evolution, each creation operator evolves into a superposition of creation operators,
\be
c_i^\dag \longrightarrow \sum_j \psi^{(i)}(j,t) c_j^\dag,
\ee
where $\psi^{(i)} (j,t)$ is the solution of the time-dependent Schrodinger equation for a particle initially localized at $i$. In the absence of noise, $\psi^{(j)}$ spreads ballistically, but in the presence of noise it spreads only diffusively. The fact that each creation operator is spread out over only $O(\sqrt{t})$ sites after a time $t$ immediately implies that the mean entanglement is at most of order $\sqrt{t}$. (See also Ref.~\cite{Igloi2016}.) Note however that this argument does not tell us how large the fluctuations are.\footnote{Random unitary evolution of a single wavepacket is discussed in Ref.~\cite{SaulKardarRead1992}.  However we must consider the full  many-body wavefunction, since the formalism of Ref.~\cite{Peschel2003} for the free fermion density matrix shows that the initially occupied orbitals do not simply contribute additively to the entanglement.} 

We have confirmed numerically that $\<S\> \propto \sqrt{t}$  for a noisy 1D hopping model, using the formalism of Ref.~\cite{Peschel2003} to construct the reduced density matrix. This is much slower than the linear-in-time growth of generic interacting models. The $\sqrt{t}$ scaling should apply for free fermions in any number of dimensions. In 1D it  also applies  to certain noisy spin models via the Jordan Wigner transformation: for example the transverse field XY model,
\be
H(t) = \sum_i \lf
J_i(t) \left[ \sigma^x_i \sigma^x_{i+1} +  \sigma^y_i \sigma^y_{i+1}  \right] + h_i(t) \sigma^z_i
\ri.
\ee
However any generic perturbation to the spin chain spoils the free fermion correspondence. We then expect the generic KPZ behavior to reassert itself.

\section{Higher dimensions}
\label{higher dimensions section}

We have discussed several ways of thinking about entanglement growth in 1D. One of these, the directed polymer picture, generalizes naturally to higher dimensions: the polymer is simply replaced by a $d$--dimensional membrane embedded in $(d+1)$--dimensional spacetime.   As in 1D, we think of this membrane as as a coarse-grained version of a minimal cut bisecting a unitary circuit.  The membrane is subject to pinning by `disorder' in space--time arising from the dynamical noise. See Fig.~\ref{min surf random} for the two-dimensional case.

We can explore two kinds of question using this picture. First, we can examine universal properties that are specific to the noisy scenario: as in 1D, fluctuations are governed by universal exponents. Second, we can calculate leading order properties of $S(t)$ that {do not involve fluctuations} and which are therefore likely to be valid even in the absence of noise, i.e. for dynamics with a time--independent Hamiltonian. In higher dimensions the behaviour of $S(t)$ has nontrivial dependence on the geometry of the region for which we calculate the entanglement. We suggest the `minimal membrane in spacetime' as a simple and general heuristic for such calculations. Below we discuss the case of a spherical region (Sec.~\ref{no fluctuations}) and contrast our results with an alternative simple conjecture. For other toy models for entanglement spreading, see Refs.~\cite{ho2015entanglement, CasiniLiuMezei2015}.

Denoting the region for which we wish to calculate the entropy by $A$, and its boundary by $\partial A$, the membrane lives in a spacetime slice of temporal thickness $t$, and terminates at $\partial A$  on the upper boundary of this time slice: see Fig.~\ref{min surf random}. For simple shapes and for times shorter than the saturation time,  the membrane also has a boundary on the lower slice, as shown in Figs.~\ref{min surf random},~\ref{min surf}. In this section we will focus on entanglement growth prior to saturation.

\subsection{Universal fluctuations of $S(t)$}
\label{higher dimensional fluctuations}

\begin{figure}[t]
 \begin{center}
  \includegraphics[width=\linewidth]{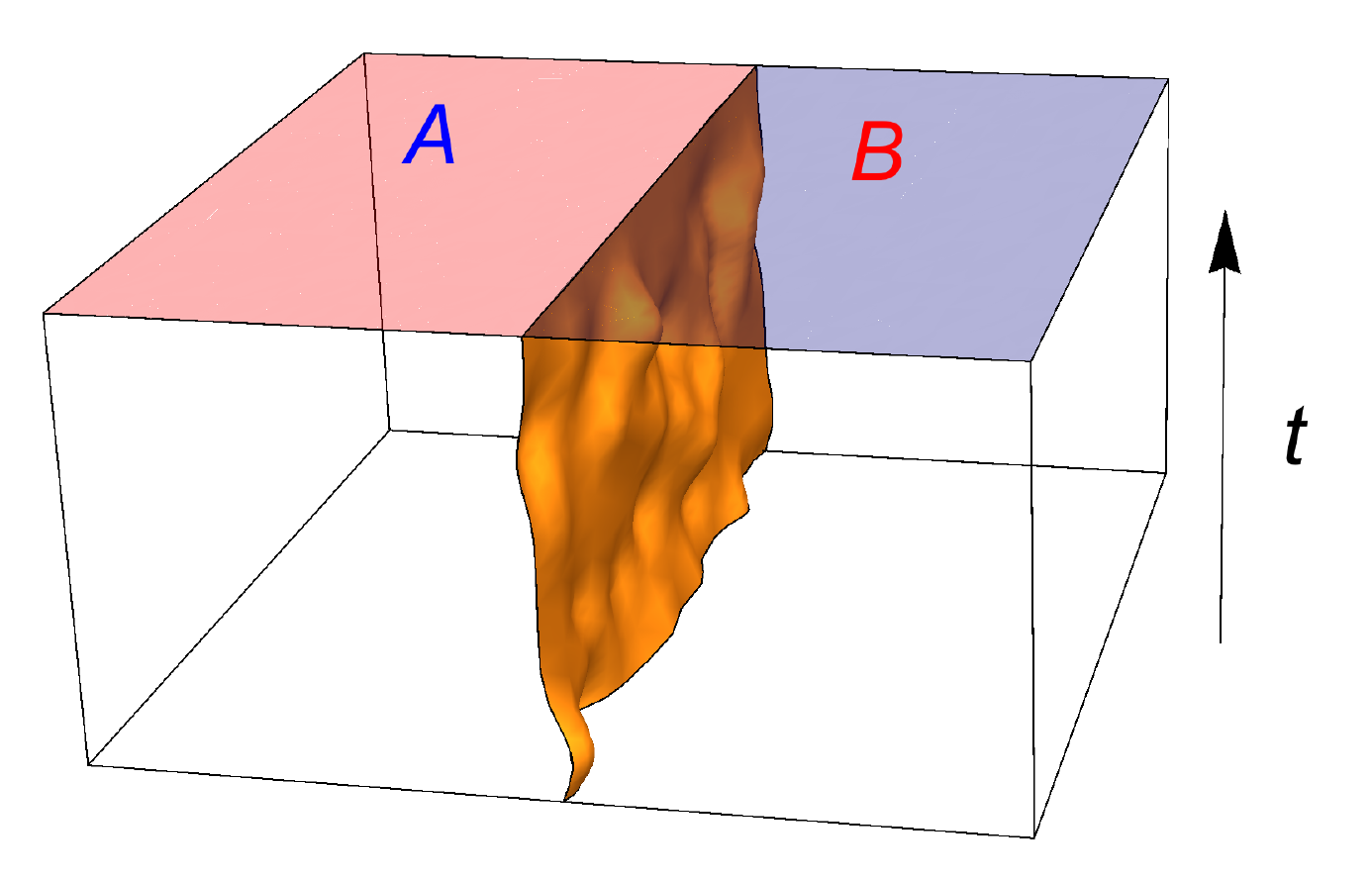}
 \end{center}
\caption{Minimal membrane picture for the entanglement of two regions in $d=2$.}
 \label{min surf random}
\end{figure}

Consider the entanglement $S(t)$ for a region $A$ whose boundary $\partial A$ has length or area $|\partial A|$. In the $d=2$ case, shown in Fig.~\ref{min surf random},  $|\partial A|$ is the length of the spatial boundary. Neglecting fluctuations, the `world volume' of the minimal membrane scales as $|\partial A| \times t$. This gives the leading scaling of the membrane's energy and hence of the entanglement. As in 1D, subleading terms encode universal data. We now consider these terms.

The pinning of a membrane or domain wall by disorder is well studied  \cite{HuseHenley1985, Nattermann1985, Kardar1987, Fisher1986, Middleton1995} (a brief summary is in App.~\ref{membrane details}). Translating standard results into the language of the entanglement in a $d$--dimensional noisy quantum system, we find that in both $d=1$ and $d=2$ there is a unique dynamical phase with nontrivial critical exponents. The same is true for continuum systems\footnote{More precisely, for systems with continuous (statistical) spatial translational symmetry.} in $d=3$. However if a lattice is present, two stable phases (and thus a dynamical phase transition) are possible in $d=3$; one with nontrivial exponents and one with trivial ones. In the trivial phase the membrane is `smooth' and is pinned by the lattice. In the nontrivial phases the membrane is instead pinned by disorder in a `rough' configuration. We will discuss the nontrivial phases (which are the only ones possible in $d<3$ and for continuum systems in $d=3$).

Generally fluctuations have a weaker effect in higher dimensions than in 1D. For simplicity, take a quantum system which is infinite in one direction and of size $L$ in the other $d-1$ directions, and consider the entanglement for a cut perpendicular to the infinite direction. Since $A$ and its complement are both infinite, $S(t)$ will grow indefinitely for this geometry. However there are two regimes, $t \lesssim L$ and $t\gg L$ (here we drop a dimensionful prefactor). For times $t \lesssim L$ (see App.~\ref{membrane details} for details):
\ba\label{mean higher dim}
\< S(t) \> & = L^{d-1} \lf v_E t + B t^{\theta+1-d} + \ldots \ri \\
\label{higher dim subleading correction}
\<\< S(t)^2 \> \>^{1/2} &\propto L^{(d-1)/2} t^{\theta - (d-1)/2},
\end{align}
where the exponent $\theta$ is defined below. This reproduces the 1D result with $\theta = \beta$. Note that when $d>1$, fluctuations are suppressed with respect to the mean by a factor of $|\partial A|^{1/2}$:  distant regions of the boundary give rise to essentially independent fluctuations which add up incoherently. In the opposite regime $t \gg L$, the temporal dimension of the membrane is much larger than its spatial dimensions, so there is a crossover to the 1D directed polymer problem. However, the exponent of the higher dimensional problem appears in the universal $L$ dependence of the growth rate:
\be\label{higher dim growth rate}
S (t) = ( v L^{d-1} + w L^{\theta - 1} ) t + \ldots
\ee
(The higher corrections will include the $t^{1/3}$ term associated with the 1D universality class.) Numerically, the exponent is $\theta = 0.84(3)$ in $d=2$, and $\theta = 1.45(4)$ in $d=3$ \cite{Middleton1995}. The subleading exponent in Eq.~\ref{mean higher dim} is negative for $d>1$, so this correction may be hard to observe numerically.

\subsection{Minimal membrane picture for dynamics without noise}
\label{no fluctuations}

\begin{figure}[t]
 \begin{center}
  \includegraphics[width=0.9\linewidth]{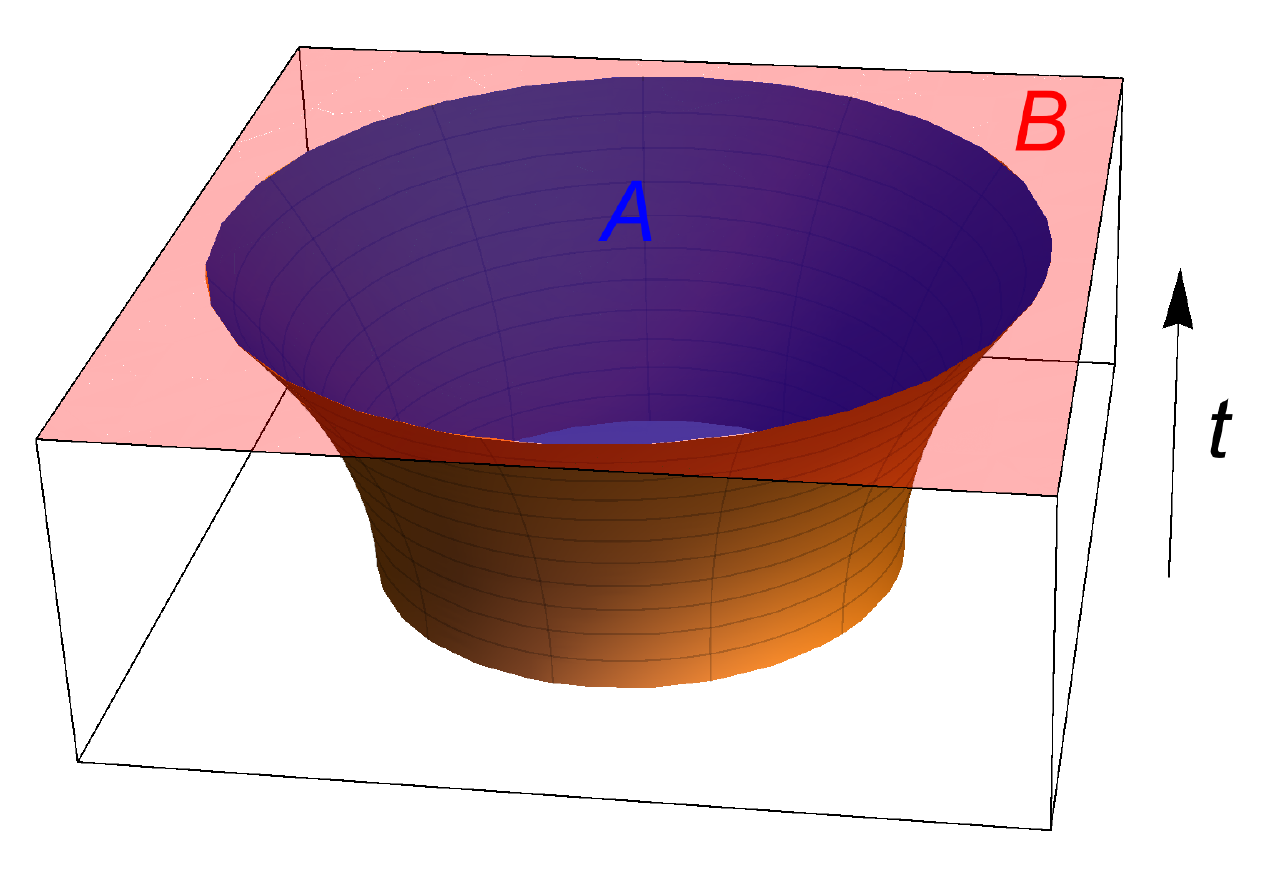}
 \end{center}
\caption{Minimal membrane for a disc-shaped region in $d=2$.}
 \label{min surf}
\end{figure}

In higher dimensions we can ask how $S(t)$ depends on the geometry of region $A$ when this  geometry is nontrivial. Interestingly, the membrane picture makes predictions about this which do not involve the noise-induced fluctuations, and which are likely also to be valid for Hamiltonian dynamics without noise (with the replacement $S\rightarrow S/s_\text{eq}$) discussed in Sec.~\ref{entanglement velocity section}).

As an instructive special case, take $A$ to be a disc-shaped region of radius $R$ in $d=2$. (A ball in higher dimensions is  precisely analogous.) We assume continuous rotational symmetry, at least on average. At short times, the leading scaling of the entanglement is $S(t) \simeq 2\pi v_E R\, t$, since the worldsheet area of the membrane is approximately $2\pi R\times t$. However there are  corrections to this arising from the curvature of $\partial A$.

We consider the limit of large $t$ and large $R$ with a fixed ratio $t/R$. In this regime the effects of fluctuations may be neglected, and instead the energetics of the membrane are determined by deterministic elastic effects. We write the energy of the membrane as
\be
E = \int \dd^2 S \, \mathcal{E},
\ee
where $\dd^2 S$ is the membrane's area element and $\mathcal{E}$ is its `energy' density. $S(t)$ is got by minimising $E$ with appropriate boundary conditions.

Next we Taylor expand $\mathcal{E}$ in terms of local properties of the membrane. For a flat `vertical' membrane (i.e. with  normal perpendicular to the $t$ axis) $\mathcal{E} = v_E$. In general however $\mathcal{E}$ will depend on the angle $\varphi$ by which the surface locally deviates from verticality, as well as, for example, the local curvatures $\kappa_s$ and $\kappa_t$ in the spatial and temporal directions. Using rotational symmetry to parametrise the membrane by the radius $r(t')$,
\be\label{E expansion}
\dd^2 S\, \mathcal{E} = v_E r \dd \theta \dd t \lf 1 + a \dot r^2 +  b \kappa_t^2 + c \kappa_s^2 + c \dot r^4 + \ldots \ri.
\ee
However this simplifies in the limit of interest. We \textit{first} send $t, R\rightarrow \infty$ with $t/R$ fixed. In this limit $\dot r(t')$ remains finite but the curvature terms become negligible (see for example the explicit solution below) so we can write $\mathcal{E} = \mathcal{E}(\dot r)$. Now we make the second approximation that $t/R$ is small, meaning that we can keep only the $O(\dot r^2)$ correction.

The boundary condition at the top of the spacetime slice is $r(t) = R$. We will consider times prior to saturation, so the membrane also has a free boundary on $t=0$. In the relevant limit  its `energy' is
\be
E = 2\pi v_E \int_0^t \dd t' \, r(t') \lf 1 +  a \dot r(t')^2   + \ldots \ri.
\ee
Minimal energy requires the boundary condition ${\dot r(0) = 0}$. When $t/R$ is small we may  expand in $1/R$.
This gives $r(t') \simeq R - (t^2 - t'^2)/(4 a R)$, as illustrated in Fig.~\ref{min surf}. The corresponding entropy is
\be
\label{deterministic S}
S(t) = 2 \pi v_E R \, t \lf 1 - \f{t^2}{12 a R^2} + \ldots  \ri.
\ee
This calculation  generalises trivially to higher dimensions, where the correction is of the same order. Corrections due to fluctuations come in with negative powers of $t$, and are negligible in the limit we are discussing.

Note that the first correction in the brackets in Eq.~\ref{deterministic S} is of order $(t/R)^2$, and not of order $t/R$. This result differs from what one might naively have expected if one guessed that at time $t$ an annulus of width $\tilde v \times t$ inside the disc is entangled with the outside, where $\tilde v$ is a tsunami velocity. This picture gives an entropy proportional to the area of the annulus,
\be\label{area formula}
S(t) \stackrel{?}{\sim} \pi R^2 - \pi (R- \tilde v t)^2 = 2 \pi R \tilde v t \lf 1 - \f{\tilde v t}{R} \ri,
\ee
leading to a negative correction of order $t/R$. The difference between Eqs.~\ref{deterministic S},~\ref{area formula} also indicates that a picture in terms of independently spreading operators is misleading, in agreement with what we found in 1D.

In the regime where $t/R$ is of order one, the full $\dot r$ dependence of $\mathcal{E}(\dot r)$  plays a role. This suggests  that an infinite number of nonuniversal parameters enter the expression for $S(t)$ in this regime, and that  there is no general, universal scaling form for the entanglement of a sphere in $d>1$. However we do expect saturation to remain discontinuous, as in 1D (Eq.~\ref{scaling form saturation 2}), occurring via a  transition between an optimal membrane configuration which reaches the bottom of the spacetime slice and one (with $E = \pi R^2$) which does not.

\section{Outlook}
\label{outlook section}

Quantum quenches generate  complex, highly entangled states whose dynamics cannot usually be tracked explicitly. For this reason,  analytical approaches to quenches have typically relied on  additional structure in the quantum dynamics: for example integrability, or absence of interactions, or conformal invariance. This paper has instead  studied dynamics that are as \textit{unstructured} as possible. We propose that noisy dynamics are a useful toy model for quantum quenches in generic (non-integrable, non-conformally--invariant) systems.

Many of our results are of course specific to noisy dynamics: in particular the emergence of KPZ behaviour at long wavelengths in 1D, and the detailed pictures for entanglement growth afforded by the `KPZ triumvirate'.  But we have suggested that some of our heuristic pictures apply to non-noisy entanglement growth as well (with the replacement $S\rightarrow S/s_\text{eq}$ mentioned above). We proposed a general directed polymer picture/minimal membrane picture for the scaling  of  the entanglement/mutual information (Secs.~\ref{entanglement velocity section},~\ref{no fluctuations}) and we used the operator spreading picture to clarify the meaning of the `entanglement velocity' and its distinction from the operator spreading velocity (Sec.~ \ref{entanglement velocity section}). `Thermalization is  slower than operator spreading' in generic 1D systems: by contrast this is not true in 1+1D CFTs \cite{CardyCalabrese2009EntanglementFieldTheory}, or in certain toy models \cite{ho2015entanglement}.  It would be interesting to make more detailed comparisons with holographic models \cite{LiuSuh2014}.

Many interesting questions remain. First --- within the realm of noisy systems --- an analytical treatment for the regime with \textit{weak} noise would be desirable, i.e. for dynamics of the form 
\be
H(t) = H_0 + \lambda H_1 (t),
\ee
where $H_0$ is a time-independent many-body Hamiltonian, $H_1(t)$ represents noise, and $\lambda$ is small. Our conjecture is that  KPZ exponents apply for any nonzero value of $\lambda$ (unless $H(t)$ is fine-tuned) --- i.e. that there is no universal distinction between continuous time dynamics and quantum circuits. (Note that there is no distinction between these two cases at the level of conservation laws: once noise is added, energy  is not conserved even in the continuous time case.) However our derivations and numerics correspond, roughly speaking, to the large $\lambda$ regime. Perhaps the opposite regime could be addressed using a more explicit RG treatment, although it is not obvious how to set this up.

Such an RG treatment might also shed light on the nature of the entanglement \textit{spectrum}, or equivalently the dependence of $S_n(t)$ on the index $n$.  While we believe that all the Renyi entropies execute KPZ growth in the presence of noise, we have not pinned down the $n$--dependence of the various constants. The solvable models suggest that the leading order behaviour may be independent of $n$ at large times. What is the appropriate scaling form for the spectrum? Limited timescales prevented us from addressing this numerically (except for Clifford circuits, where all the $S_n$ are trivially equal).

It would also be useful to test the higher-dimensional membrane pictures of Sec.~\ref{higher dimensions section}, perhaps exploiting Clifford circuits to reduce the numerical difficulty of higher-dimensional dynamics. 

So far we have only discussed initial states with area law entanglement. We may also ask about initial states with, for example, sub-maximal volume law entanglement. The natural expectation (say in 1D)  is that the directed polymer picture extends to this case if we glue the unitary circuit to a tensor network representation of the  initial state. Then the entropy $S(x,t)$ would include a fluctuating part with KPZ exponents together with a contribution from the initial state.

More speculatively, we may ask whether there exist time-{independent}  Hamiltonians which  show KPZ entanglement growth, despite the absence of explicit noise, in some dynamical regimes. This could arise if chaotic dynamics provided effective noise. This seems unlikely on asymptotically long timescales for a generic system, but may hold on intermediate timescales. We may also ask about the structure of the quantum states generated by the random dynamics: in what ways do they differ from ground states of random Hamiltonians, when the amount of entanglement is similar?

\acknowledgements
\noindent
We thank M. Kardar for useful discussions. AN and JR acknowledge the support of fellowships from the Gordon and Betty Moore Foundation under the EPiQS initiative (grant no. GBMF4303). SV is supported partly by the KITP Graduate Fellows Program and the DOE Office of Basic Energy Sciences, Division of Materials Sciences and Engineering under Award DE-SC0010526. JH is supported by the Pappalardo Fellowship in Physics at MIT.

\appendix

\section{Growth of Hartley entropy $S_0$ in 1D}
\label{hartley entropy growth}

Consider a one-dimensional quantum spin chain of local Hilbert space dimension $q$, prepared initially in a product state,
and apply a sequence of random unitaries
that couple two neighboring spins.
The location of the local unitary at a given time step is arbitrary.
In the following we fix the location of the unitary, but take it to be Haar random.

We prove that in this situation the Hartley entropy $S_0$
generically (i.e. with probability 1)
obeys
\begin{align}
S_0(x,t+1) = \min( S_0( x-1,t), S_0(x+1,t)) + 1,
\label{eq:genericHartleyRule}
\end{align}
if a unitary is applied at the bond $x$. The logarithm is of base $q$.

This formula can be interpreted in matrix-product-state (MPS) language. If  $d_x$ is the minimal value of the local bond dimension required for an exact MPS representation of the state, then $S_0(x) = \log d_x$. A heuristic parameter counting argument for the local bond dimension, given in Sec.~\ref{heuristic parameter counting argument} below,  suggests Eq.~\ref{eq:genericHartleyRule}. 

However a more rigorous proof is necessary as such heuristic arguments can fail. In particular, one might naively conjecture a stronger statement: namely that for \textit{any} state at time $t$,
if the unitary at bond $x$ is Haar random,
then Eq.~\eqref{eq:genericHartleyRule} is true with probability~1.
This conjecture is false;
a counterexample will be given below in Sec.~\ref{counterexample section}.  We now give a proof of Eq.~\ref{eq:genericHartleyRule}.

\subsection{Proof of Eq.~\eqref{eq:genericHartleyRule}}

Our genericity proof consists of two parts.
First, we will show that given locations of unitaries,
there exist certain unitaries such that at each time step,
Eq.~\eqref{eq:genericHartleyRule} is true.
Second, we will show the negation of Eq.~\eqref{eq:genericHartleyRule}
\begin{align}
S_0(x,t+1) < \min( S_0( x-1,t), S_0(x+1,t)) + 1
\label{eq:TooSmallHartley}
\end{align}
happens if and only if
a system of polynomial equations in the entries of the unitaries
is satisfied.
(The inequality ``$>$'' never holds as we noted in the main text.)
By the first part of the proof, the zero locus of these polynomial equations
does not cover the entire set of unitaries.  Therefore it is only a submanifold of strictly smaller dimension,
which implies it has measure zero.

For the first part,
it is sufficient to consider only three types of local unitaries:
the identity $I$, the swap gate $W$,
and a unitary $E$ with the property that  it turns a pair of unentangled polarized spins, $\ket{11}$, into $q^{-1/2} \sum_{i=1}^q \ket{i i}$,
a maximally entangled state. Without loss of generality we may take the initial product state to be the polarized state $\ket{\ldots 1111 \ldots}$. 

We are going to show that using these three types of unitaries at the given locations,
one can construct a state whose entanglement entropy is given by Eq.~\eqref{eq:genericHartleyRule}.
Since Eq.~\eqref{eq:genericHartleyRule} defines the entropy inductively,
we only have to show it inductively too.

At $t = 0$, all the spins are unentangled, so we can simply choose $E$
for every designated location.
Clearly, Eq.~\eqref{eq:genericHartleyRule}
is satisfied.
At later times,
if we do not apply $E$ except for unentangled pair of spins,
then a spin can be either unentangled or maximally entangled with a single other spin.
Therefore, at time $t > 0$
the spin $s_L$ that is immediately left to the bond $x$ can be
\begin{enumerate}
 \item[(i)] unentangled,
 \item[(ii)] entangled with a spin to the left of $s_L$,
 \item[(iii)] entangled with the spin $s_R$ that is immediately to the  right of the bond $y$, or
 \item[(iv)] entangled with a spin to the right of $s_R$.
\end{enumerate}
These are exclusive possibilities, and similarly $s_R$ has four options.
Enumerating all 16 cases,
which in fact reduces to 7 different cases excluding invalid ones and those related by reflection,
one easily checks that there is always a choice among $I,W,E$
that makes Eq.~\eqref{eq:genericHartleyRule} true.
Let us treat three exemplary  cases here.
If $s_L$ and $s_R$ are entangled at time $t$,
then $S(x-1,t) = S(x+1,t)$ and $S(x,t) = S(x-1,t) + 1$,
so one chooses the identity $I$.
If $s_L$ is entangled with a spin on the left of $s_L$
and $s_R$ is entangled with a spin on the right of $s_R$,
then $S(x-1,t) = S(x+1,t) =  1 + S(x,t)$.
One chooses the swap $W$ to obtain $S(x,t+1) = S(x,t) + 2$.
If $s_L$ and $s_R$ are both unentangled, then one applies the entangling unitary $E$
to obtain $S(x,t+1) - 1 = S(x,t) = S(x-1,t) = S(x+1,t)$.

For the second part,
recall that for any bipartite state
\begin{align}
\ket \psi = \sum_{i,j} M_{i,j} \ket i \ket j ,
\end{align}
the number of nonzero Schmidt coefficients
is equal to the number of nonzero singular values of the matrix $M$,
which is nothing but the rank of $M$.
For any positive integer $r$,
the rank of $M$ is \emph{smaller} than $r$ if and only if every $r \times r$
submatrix has determinant zero, i.e., all \emph{$r \times r$ minors} vanish.
Thus, a bipartite state $\ket \psi$ having Hartley entropy ($\log$ of rank of $M$)
strictly smaller than $\log r$ is expressed by a system of polynomial equations
on the coefficients of $\ket \psi$.
If $\ket \psi$ is given by $U_t \ldots U_2 U_1 \ket 0$ where $\ket 0$ is a fixed product state,
then the coefficients are some polynomials of the entries of the unitaries $U_i$,
and hence the equations that expresses vanishing determinants
are polynomial equations in the entries of the unitaries.

Our claim Eq.~\eqref{eq:genericHartleyRule}
completely determines the Hartley entropy
based on the location of unitaries,
and therefore the spatial configuration of the unitaries
tells us which minors we should check.
Namely, the size $r$ of the minors we turn into the polynomial equations
is given by (the exponential of) the right-hand side of Eq.~\eqref{eq:TooSmallHartley}.
In other words, given a spatial configuration of unitaries,
the polynomial equations that express Eq.~\eqref{eq:TooSmallHartley}
are determined.
The polynomial equations are over $tL$ variables,
and the actual number of equations is much larger yet finite.
We do not need explicit expressions for these polynomials,
only the fact of  their existence.
These polynomials might a priori read $0=0$, i.e.,
they could be trivially satisfied.
In that case, the solution to the polynomial equation would be the
entire set of unitaries, and Eq.~\eqref{eq:genericHartleyRule}
could never be satisfied.
However, we just showed in the first part
that this cannot happen because
there exists a choice of unitaries for which Eq.~\eqref{eq:genericHartleyRule}
is satisfied.
This implies that the polynomial equations are nontrivial
and define a measure zero subset of the entire set of unitaries.
This completes the genericity proof.

\subsection{Counterexample to the stronger conjecture}
\label{counterexample section}

We have shown that (\ref{eq:genericHartleyRule}) holds when all unitaries are chosen generically and the initial state is a product state. Naively one might make the stronger conjecture: that the update rule (\ref{eq:genericHartleyRule}) holds whenever a generic unitary $U$ is applied to an arbitrary --- possibly fine-tuned --- state $\ket{\Psi}$. We construct an explicit $\ket{\Psi}$  which is a counterexample to this stronger conjecture.

Consider four degrees of freedom $ABCD$.
The spins $B$ and $C$ have dimension 2 each,
and $A$ and $D$ have dimension 3 each.
(To conform with our consideration of spin chains,
the subsystems $A$ and $D$ should be regarded
as subspaces of two or more spin-$\frac 12$'s.)
The most general form of a quantum state on $ABCD$ is
\begin{align}
 \sum_{a,d = 0}^2 \sum_{b,c = 0}^1 T_{abcd} \ket a \ket b \ket c \ket d .
\end{align}
We consider $T_{abcd} = T'_{abd} \delta_{c,0}$, i.e., $C$ is in $\ket 0$,
where
\begin{align}
 T'_{a0d} &= \begin{pmatrix} 0 & 1 & 0 \\ 1 & 0 & 0 \\ 0 & 0 & 0 \end{pmatrix}_{ad}, &
 T'_{a1d} &= \begin{pmatrix} 0 & 0 & 1 \\ 0 & 0 & 0 \\ 1 & 0 & 0 \end{pmatrix}_{ad}.
\end{align}
(This does not give a normalized state, but we are only concerned about ranks.)

The Hartley entropy for the cut $A/BCD$ is simple to compute.
As remarked in the previous subsection,
it is the rank of the coefficient matrix.
Interpreting this matrix as a linear map,
the rank is the dimension of the image of the map from $BCD$ to $A$.
The image is precisely the linear span of \emph{columns} of $T'_{a0d}$ and $T'_{a1d}$.
They have three linearly independent columns,
implying the Hartley entropy for $A/BCD$ is $\log_2 3$.
Similarly, the rank of the coefficient matrix for $ABC/D$ is
the dimension of the linear span of the \emph{rows} of $T'_{a0d}$ and $T'_{a1d}$,
which reads 3.
That is, the Hartley entropy for $ABC/D$ is $\log_2 3$.

If Eq.~\eqref{eq:genericHartleyRule}
were to be true for generic choice of Haar random unitary on $BC$,
then we should be able to find a unitary on $BC$ such that
\begin{align}
 S_0(AB/CD) = \log_2 3 + 1 = \log_2 6.
\end{align}
We show this cannot hold.
Applying the unitary $U$ on $BC$ the state, we obtain
\begin{align}
 \sum_{b,c} U_{b'c',bc} T_{abcd} = \underbrace{U_{b'c',00}}_{U_0} T'_{a0d} + \underbrace{U_{b'c',10}}_{U_1} T'_{a1d}.
\end{align}
where $U_0$ and $U_1$ are $2 \times 2$ matrices.
The coefficient matrix for the cut $AB/CD$ is then
\begin{align}
 V = U_0 \otimes T'_0 + U_1 \otimes T'_1
\end{align}
whose rank should be 6 if $S_0( AB/CD ) = \log 6$.
Computing all the minors of the $6 \times 6$ matrix $V$ for \emph{arbitrary} matrices $U_0$ and $U_1$,
we find that all $(5 \times 5)$-minors vanish,
implying that $V$ has rank at most 4.
Therefore, for this nongeneric initial state,
\begin{align}
 S_0(x,t+1) \neq \min(S_0(x-1,t), S_0(x+1,t)) + 1.
\end{align}

\subsection{Parameter-counting argument}
\label{heuristic parameter counting argument}

 Consider a 1D state $\ket{\Psi}$ in a matrix product representation. Labelling the states of the qubits (spins) by $\sigma,\sigma'\ldots$ running from $1$ to $q$,
 \be
 \ket{\Psi} =
 \sum_{\{\sigma\}}
 \sum_{\{ a \}}
 \lf
 \ldots A^\sigma_{a_{x-1}, a_{x}} A'^{\sigma'}_{a_{x}, a_{x+1}} \ldots
 \ri
 \ket{\ldots \sigma \sigma'  \ldots}.
 \ee
 Since the state is not translationally invariant, we allow the bond dimension $d_x$ to vary from bond to bond ($a_x = 1,\ldots, d_x$). In an efficient representation, $d_x$ is equal to the rank of the reduced density matrix for a cut at $x$:
 \be
 d_x = q^{S_0(x)}.
 \ee
 We ask how $S_0(x)$ changes when we apply a unitary $U$ to the two spins, $\sigma$ and $\sigma'$, either side of bond $x$. This effects the change (repeated indices summed)
 \be
  A^\sigma_{a_{x-1},a_x}  A'^{\sigma'}_{a_{x},a_{x+1}}  \longrightarrow
 U_{\sigma\sigma', \tau\tau'}
 A^\tau_{a_{x-1},a_x}  A'^{\tau'}_{a_{x},a_{x+1}}.
 \ee
 To update the matrix product representation we must find new matrices $\widetilde A$ and $\widetilde A'$ which satisfy
 \be\label{MPS equations}
 \widetilde A^\sigma_{a_{x-1},a_x}  \widetilde A'^{\sigma'}_{a_{x},a_{x+1}} =
 U_{\sigma\sigma', \tau\tau'}
 A^\tau_{a_{x-1},a_x}  A'^{\tau'}_{a_{x},a_{x+1}}.
 \ee
 In order to solve this equation for $\widetilde A$ and $\widetilde A'$, it will generally be necessary to increase the bond dimension at $x$ to a new value $d'_x$.  Naively,  the necessary value of $d'_x$ will generically be determined by equating the number of independent equations in (\ref{MPS equations})  with the number of degrees of freedom in $\widetilde A$ and $\widetilde A'$.   (However, the previous subsection shows that this expectation  can fail for certain choices of $A$ and $A'$.)
 
 The number of equations is $q^2 d_{x-1} d_{x+1}$, since this is the number of possible values for the external indices in (\ref{MPS equations}). The numbers of degrees of freedom in $\widetilde A$ and $\widetilde A'$ are $q d_{x-1} d'_x$  and  $q d_{x+1} d'_x$ respectively. However, $d_x'^2$ of these are redundant, because the state is unchanged by the transformation $\widetilde A^\sigma \rightarrow \widetilde A^\sigma M$, $\widetilde A'^{\sigma'} \rightarrow M^{-1} \widetilde A'^{\sigma'}$, with $M$ an arbitrary $d_x'\times d_x'$ matrix. Equating the number of equations with the number of independent degrees of freedom gives
 \be
 (d'_x -  q d_{x-1})(d'_x -  q d_{x+1})=0.
 \ee
 Choosing the smallest solution,
 \be
 d_x' = q \times \min \{ d_{x-1}, d_{x+1} \}.
 \ee
 This agrees with Eq.~\ref{eq:genericHartleyRule} since $S_0(x) = \log d_x$.

\section{Haar average for $\Tr \rho_x^2$}
\label{haar appendix}

Let $\rho_x(t)$ be the reduced density matrix for a cut at $x$, obtained by tracing out the spins to the left of the cut. Each index on this matrix labels a configuration of the spins to the right of the cut. Let us temporarily label these spins $1,2,\ldots$, and let the spin immediately to the left of the cut be denoted $0$. The indices on the reduced density matrices are then:
\ba
&\rho_{x-1}(t)^{\sigma_0,\sigma_1,\sigma_2,\ldots}_{\mu_0,\mu_1,\mu_2,\ldots}&
&\rho_x(t)^{\sigma_1,\sigma_2,\ldots}_{\mu_1,\mu_2,\ldots}&
&\rho_{x+1}(t)^{\sigma_2,\ldots}_{\mu_2,\ldots}
\end{align}
After applying a unitary on bond $x$,
\be\notag
\rho_x(t+1)^{\sigma_1,\sigma_2,\ldots}_{\mu_1,\mu_2,\ldots}
=
U_{\tau\sigma_1, \sigma_0'\sigma_1'} U^*_{\tau\mu_1,\mu_0'\mu_1'}
\rho_x(t)^{\sigma'_0,\sigma'_1,\sigma_2,\ldots}_{\mu'_0,\mu'_1,\mu_2,\ldots}
\ee
Let us average $\Tr \rho_x(t+1)^2$ over the choice of unitary, for a fixed initial state:
\ba\notag
 \<\Tr \rho_x(t+1)^2 \>&   =  \rho_{x-1}(t)^{\sigma'_0,\sigma'_1,\sigma_2,\ldots}_{\mu'_0,\mu'_1,\mu_2,\ldots}
\rho_{x-1}(t)^{\mu''_0,\mu''_1,\mu_2,\ldots}_{\sigma''_0,\sigma''_1,\sigma_2,\ldots}\\\notag
&\times
\< U_{\tau\sigma_1, \sigma_0'\sigma_1'} U^*_{\tau\mu_1,\mu_0'\mu_1'}
U_{\nu\mu_1, \mu_0''\mu_1''} U^*_{\nu\sigma_1,\sigma_0''\sigma_1''}\>
\end{align}
The Haar average for four elements of a $\mathrm{U}(d)$ matrix (here $d=q^2$, and each index on $U$ represents a pair of spin indices) is
\ba
\< U_{a,b} U_{a',b'} \right. & \left. U^*_{c,d}   U^*_{c',d'}\>_\text{Haar}
= \\\notag
\f{1}{d^2-1} \bigg(
&\left\{
\delta_{a,c}\delta_{a',c'} \delta_{b,d}\delta_{b',d'}
+
\delta_{a,c'}\delta_{a',c} \delta_{b,d'}\delta_{b',d}
\right\}  \\ \notag
&-
\f{1}{d}
\left\{
\delta_{a,c}\delta_{a',c'} \delta_{b,d'}\delta_{b',d}
+
\delta_{a,c'}\delta_{a',c} \delta_{b,d}\delta_{b',d'}
\right\}
\bigg).
\end{align}
The index contractions give the result in the text,
\ba
\<\Tr \rho_x(t+1)^2\>_\text{Haar} =q(q^2+1)^{-1} \lf
\Tr \rho_{x-1}^2 + \Tr \rho_{x+1}^2 \ri.
\end{align}

\section{Entanglement entropy of stabilizer states}
\label{stabilizer state entropy}

A stabilizer state is a state of an $n$-qubit system
defined by a complete set $\{g_1,\ldots, g_n\}$ 
of commuting tensor products of Pauli matrices
through equations
\begin{align}
g_i \ket \psi = + \ket \psi .
\end{align}
The group generated by $\{ g_1, \ldots, g_n\}$
is naturally called a stabilizer group, and denoted by $\mathcal G$~\cite{Gottesman1996Saturating,CalderbankRainsShorEtAl1997Quantum}.
A trivial example is the all-spin-up state, defined as
\begin{align}
Z_i \ket \psi = + \ket \psi 
\end{align}
for all $i = 1,\ldots, n$.
The condition that $\ket \psi$ is nonzero and unique
is equivalent to the condition that the operator
\begin{align}
\frac{1}{|\mathcal G|} \sum_{g \in \mathcal G} g
\end{align}
is a projector of rank one~\cite{KlappeneckerRotteler2002stabilizer,
LindenMatusRuskaiEtAl2013Quantum}.
Since $\ket \psi$ is in the image of this projector,
we see
\begin{align}
\ket \psi \bra \psi = \frac{1}{|\mathcal G|} \sum_{g \in \mathcal G} g.
\end{align}
Since this is a normalized pure density matrix, its trace is equal to 1.
But a Pauli matrix has the property that it is traceless.
Therefore, only the identity element on the right has nonzero trace:
\begin{align}
1 = \frac{1}{|\mathcal G|} \dim (\mathbb C^2)^{\otimes n}
=\frac{1}{|\mathcal G|} 2^n
\end{align}

From this expression, it is straightforward to obtain expressions for reduced density matrices.
Suppose the $n$-qubit system is partitioned into two complementary regions $A$ and $B$.
Tracing out $B$, we have
\begin{align}
\rho_A = \frac{1}{2^n} \sum_{g \in \mathcal G} \Tr_B (g) .
\end{align}
$\Tr_B(g)$ is nonzero if and only if the tensor component corresponding to $B$ is identity,
in which case 
\begin{align}
\Tr_B(g) = 2^{|B|} g|_A
\end{align}
where $g|_A$ denotes the tensor components of $g$ corresponding to $A$.
The set of all $g|_A$ such that $\Tr_B(g) \neq 0$ 
can be regarded as a subgroup of $\mathcal G$,
which we denote by $\mathcal G_A$.
The formula for $\rho_A$ now reads
\begin{align}
\rho_A = \frac{2^{|B|}}{2^n} \sum_{g \in \mathcal G_A} g = 
\frac{|\mathcal G_A|}{2^{|A|}} \frac{1}{|\mathcal G_A|} \sum_{g \in \mathcal G_A} g.
\end{align}
It is immediate that $\rho_A$ is proportional to a projector
since it is a sum over a group.
It follows that the rank of $\rho_A$ is equal to $2^{|A|} / |\mathcal G_A|$.
In particular, the (R\'enyi or von Neumann) entropy of $\rho_A$ with base-2 logarithm is
\begin{align}
S(\rho_A) = |A| - \log_2 |\mathcal G_A|.
\end{align}
The subgroup $\mathcal G_A$ has period 2,
and therefore $\log_2 |\mathcal G_A|$ is an integer,
which is equal to the number of independent stabilizers
 supported only on $A$.
This expression for the entanglement entropy has also appeared in 
\cite{HammaIonicioiuZanardi2005,AliosciaHammaZanardi2005Ground}.

Now, regard the stabilizer group $\mathcal G$ as a binary vector space $V$ by ignoring the overall phase (sign) factors. Let $\Pi_A$ be the truncation map retaining the components corresponding to the region $A$, and similarly $\Pi_B$ be the truncation map for $B = \bar A$. It is routine to check that $V$ decomposes as $V_A \oplus V_B \oplus V'$ for some subspace $V' \subseteq V$ where $V_A$ and $V_B$ are the spans of stabilizers supported only on $A$ and $B$, respectively. Both the truncation maps are injective on $V'$. It follows that $S_A = |B| - \dim_{\mathbb F_2} V_B = \dim_{\mathbb F_2} (\Pi_A V) - |A|$. This completes the proof of Eq.~\ref{entropy formula}.

\section{Numerics for full Clifford evolution}
\label{appendix: full clifford}

\begin{figure}[b]
 \begin{center}
  \includegraphics[width=0.9\linewidth]{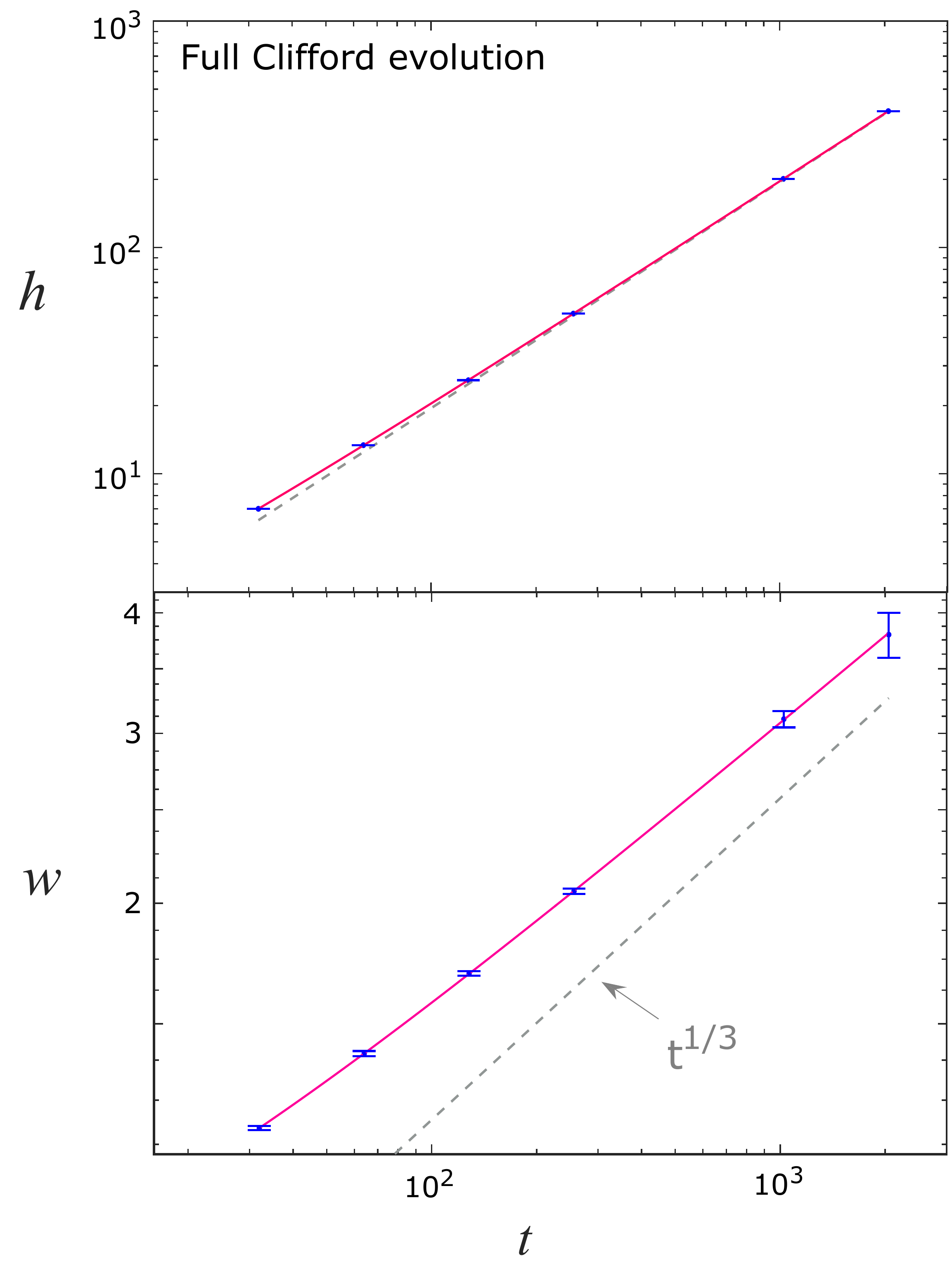}
 \end{center}
\caption{Top: Growth of the mean entanglement in units of $\log 2$ as a function of time for the random Clifford evolution (only CNOT gates). The red solid curve is a fit using the form Eq.~[\ref{fit}]. Dashed line shows asymptotic linear behaviour. Bottom: Growth in the fluctuations in the entanglement with time. The exponent $\beta$ is found to be $\beta_{\mathrm w}=0.3 \pm 0.04$, in agreement with the KPZ prediction $\beta=1/3$. The dashed line shows the expected asymptotic behaviour, $\mathrm{w(t)} \sim t^{\beta}$ with $\beta=1/3$.}
 \label{fig:Cliffordheightwidth}
\end{figure}

In Sec.~\ref{section:numerics:clifford} we have presented numerical results for random unitary evolution using only the CNOT gates Eq.~\ref{CNOT}. Here we present similar analysis using the full set of generators for the Clifford group showing that the additional gates do not modify the universal behavior. The additional single-site gates are the Hadamard and phase gates defined in Eq.~\ref{RH} and Eq.~\ref{RP}, respectively (The Hadamard gate corresponds to row swap between $X$ and $Z$ while the phase gates corresponds to adding the $X$ vector to the $Z$ one).

The von Neumann entropy in units of $\log 2$ and the corresponding width averaged over $\sim2\times 10 ^5$ realisations (except for the last data point where $\sim2\times 10 ^4$ realisations were used for the average) are plotted in Fig.~[\ref{fig:Cliffordheightwidth}]. The fit to the KPZ universal form Eq.~\ref{fit} gives $\b_h =0.2\pm 0.15$ and $\b_\mathrm{w}= 0.3\pm0.04$. We also obtain $v_E  = 0.194\pm0.001$, $B = 0.4\pm0.2$, $C = 0.4\pm0.1$, $D = 0.4\pm0.6$ and $\eta = -0.4 \pm0.8$. These results are consistent with the KPZ universality and with the data presented in Fig.~\ref{fig:CNOTheightwidth}.

\section{Numerical check that $v_E$ is a speed}
\label{appendix finite size effects}

\begin{figure}[]
 \begin{center}
  \includegraphics[width=1\linewidth]{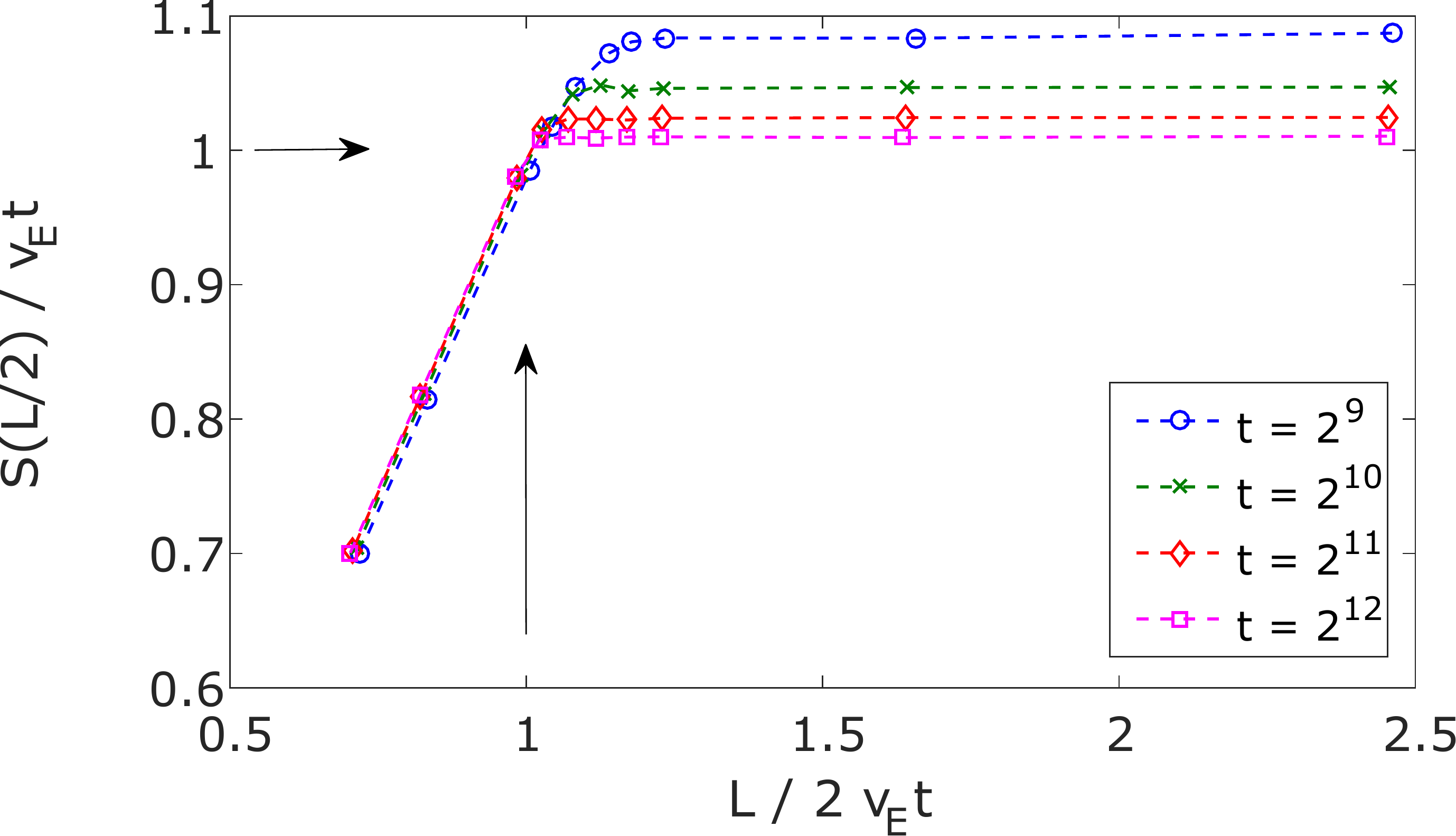}
 \end{center}
\caption{The entropy across the centre of the chain (in units of $\log 2$) divided by $v_E t$ vs. $L/2v_E t$ for various fixed values of $t$.  This plot converges to the scaling form in (\ref{scaling form saturation},~\ref{scaling form saturation 2}), confirming that $v_E$ can be interpreted as a speed in addition to an entanglement growth rate.}
 \label{fig:vE}
\end{figure}

In the main text we have fitted the numerical data points to the form Eq.~\ref{fit}. The parameter $v_E$ quantifies the average entanglement growth rate. We argued in Sec.~\ref{entanglement velocity section} that $ v_E $ can \textit{also} be viewed as a velocity associated with entanglement spreading. (This is the speed of the fictitious particles in Sec.~\ref{hydrodynamics section}.) The simplest manifestation of this is in the saturation behaviour of the entanglement. The analytical arguments imply that to leading order (at large $t$ and $l$) the entanglement across a cut at position $l$ has the simple scaling form
\be\label{scaling form saturation}
S_A = v_E t \, f(l / v_E t),
\ee
with 
\be\label{scaling form saturation 2}
f(u) = 
\left\{
\begin{array}{c}
 u \quad \text{for} \quad u < 1  \\
 1 \quad \text{for} \quad u\geq 1
\end{array}
\right.
\ee
in other words there is no influence of the boundary at times $t< l/v_E$. 

In this section we test this conjecture numerically for the Clifford evolution. We set $l=L/2$ and plot in Fig.~\ref{fig:vE}
\be
\f{S(L/2,t)}{v_E t} \quad\text{vs.} \quad \f{L/2}{v_E t}
\ee
as a function of $L$, for several values of the time ($t = 2^9,2^{10},2^{11}\,\,\mathrm{and}\,\,2^{12}$). Here $v_E = 0.1$ is taken from the fits to Fig.~\ref{fig:Cliffordheightwidth}.  According to  (\ref{scaling form saturation}), this plot should converge for large $t$ to a plot of $f(u)$ against $u$. The results are in excellent agreement with the scaling form, confirming (for the case of Clifford circuits) that $v_E$ is a meaningful velocity.

\section{Details of statistics of membranes}
\label{membrane details}

The exponents governing the membrane problem are traditionally denoted $\theta$ and $\zeta$, and are related by  $2\zeta - \theta = 2 - d$ \cite{HuseHenley1985}. Consider a  patch of the membrane with  linear dimensions scaling as $\ell$. This includes both its  temporal dimension and its internal spatial dimensions: after a rescaling of time, the membrane is statistically isotropic on large scales.  The mean `energy' of this patch of membrane scales as $\ell^d+ \text{const} \times \ell^\theta$, with fluctuations of order $\ell^\theta$. The lengthscale for wandering of the membrane in the transverse direction is of order $\ell^{\zeta}$.   The numerical results quoted in the main text are in good agreement with an epsilon expansion about $d=4$ which gives $\zeta\simeq 0.208(4-d)$ \cite{Fisher1986} (see also \cite{Halpin-HealyManifolds}). The scaling forms for the entanglement discussed in the text are easily found by regarding the membrane as made up of patches of appropriate linear size: size $t$ for Eqs.~\ref{mean higher dim},~\ref{higher dim subleading correction}, and size $L$ for Eq.~\ref{higher dim growth rate}. 

Note that the geometry of the membrane, including the transverse lengthscale (which is $\Delta x \sim t^\zeta$ for the regime $t \lesssim L$) determines the dimensions of the spacetime region around $\partial A$ for which the final entanglement is sensitive to small changes in $H(t)$, i.e. in the history of the noise.


%

\end{document}